\documentclass[12pt]{article}
\pdfoutput=1
\usepackage{jheppub}
\usepackage{amsmath}
\usepackage{graphicx,tabularx}
\usepackage{url}
\usepackage{footmisc}
\usepackage{amsfonts}
\usepackage{cancel}
\usepackage{color}
\usepackage{multirow} 
\usepackage{amssymb}
\usepackage{pifont}
\usepackage{subfigure}
\usepackage{epstopdf}
\usepackage{slashed}
\usepackage{comment}
\usepackage{booktabs} 
\usepackage{natbib}
\usepackage{hyperref}
\usepackage{pstricks}
\usepackage{youngtab}

\allowdisplaybreaks[4]
\newcommand{\beq}{\begin{eqnarray}}% can be used as {equation} or  {eqnarray}
\newcommand{\eeq}{\end{eqnarray}}

\newcommand{\centeron}[2]{{\setbox0=\hbox{#1}\setbox1=\hbox{#2}\ifdim

\wd1>\wd0\kern.5\wd1\kern-.5\wd0\fi
\copy0

\kern-.5\wd0\kern-.5\wd1\copy1\ifdim\wd0>\wd1
                                       \kern.5\wd0\kern-.5\wd1\fi}}
\newcommand{\ltap}{\>\centeron{\raise.35ex\hbox{$<$}}
                               {\lower.65ex\hbox{$\sim$}}\>}
\newcommand{\gtap}{\>\centeron{\raise.35ex\hbox{$>$}}
                               {\lower.65ex\hbox{$\sim$}}\>}

\newcommand\ZZ{\hbox{\zfont Z\kern-.4emZ}}
\font\zfont = cmss10 %scaled \magstep1

\newcommand{\met}{$E_T^{\rm miss}\ $}

\title{Composite Dark Matter and Higgs}
\author[a,b]{Yongcheng Wu,}
\author[c]{Teng Ma,}
\author[a,b]{Bin Zhang,}
\author[d]{Giacomo Cacciapaglia}
\affiliation[a]{Department of Physics, Tsinghua University, Beijing, 100086, China}
\affiliation[b]{Center for High energy Physics, Tsinghua University, Beijing, 100084, China}
\affiliation[c]{CAS Key Laboratory of Theoretical Physics, Institute of Theoretical Physics, Chinese Academy of Sciences, Beijing 100190, China. }
\affiliation[d]{Univ Lyon, Universit\'e Lyon 1, CNRS/IN2P3, IPNL, F-69622, Villeurbanne, France}

\emailAdd{wuyongcheng12@mails.tsinghua.edu.cn}
\emailAdd{mat@itp.ac.cn}
\emailAdd{zb@mail.tsinghua.edu.cn}
\emailAdd{g.cacciapaglia@ipnl.in2p3.fr}

\preprint{
	\begin{flushright}
		LYCEN-2017-03
	\end{flushright}
}

\abstract{
	We investigate the possibility that Dark Matter arises as a composite state of a fundamental confining dynamics, together with the Higgs boson. We focus on the minimal SU(4)$\times$SU(4)/SU(4) model which has both a Dark Matter and a Higgs candidates arising as pseudo-Nambu-Goldstone bosons. At the same time, a simple underlying gauge-fermion theory can be defined providing an existence proof of, and useful constraints on, the effective field theory description. We focus on the parameter space where the Dark Matter candidate is mostly a gauge singlet. We present a complete calculation of its relic abundance and find preferred masses between 500 GeV to a few TeV. Direct Dark Matter detection already probes part of the parameter space, ruling out masses above 1 TeV, while Indirect Detection is relevant only if non-thermal production is assumed. The prospects for detection of the odd composite scalars at the LHC are also established. }

\begin{document}
\titlepage

\maketitle

\newpage

%%%%%%%%%%%%%%%%%%%%%%%%%%%

\flushbottom

%%%%%%%%%%%%%%%%%%%%%%%%%%%

\section{Introduction}

The  discovery of the Higgs boson in 2012~\cite{Aad:2012tfa,Chatrchyan:2012xdj}  by ATLAS  and  CMS   at the  Large  Hadron  Collider (LHC) is  a triumph  of  particle physics.  In fact, this event marks not only the completion of the particle list predicted by the Standard Model (SM),  but also the measurement of a particle of a completely new kind: the first possibly elementary spin-0 particle. However, all elementary scalars are  always  accompanied  by a hierarchy problem  because  the mass  is not protected  by  any symmetry, thus it will be directly sensitive to  higher scales of new physics. In the case of the SM, this fact affects the value of the electroweak (EW) scale versus other Ultra-Violet (UV) scales like the Planck mass and the hypercharge Landau pole. To  stabilize it, therefore, new  physics or  new symmetries  need   to be  introduced in order to push the SM  into  a  near fixed  point. This happens, for instance, in supersymmetric models,  where  a  space-time symmetry  between  scalars (bosons) and  fermions is implemented, yielding a cancellation of divergent quantum corrections to the Higgs  mass. Other examples are Little Higgs~\cite{ArkaniHamed:2002qx,Schmaltz:2002wx}, Twin-Higgs~\cite{Chacko:2005pe}, Maximally Symmetric Composite Higgs ~\cite{Csaki:2017cep} models  where the Higgs  is identified to a pseudo-Nambu-Goldstone boson (pNGB) and its mass  is protected (at least at one loop) by the associated shift  symmetry. In extra--dimensional models~\cite{Manton:1979kb,Fairlie:1979at,Hosotani:1983xw} the Higgs mass is protected by the bulk  gauge  symmetry. Another attractive and time--honoured scenario is Technicolour~\cite{Weinberg:1975gm,Susskind:1978ms,Dimopoulos:1979es,Eichten:1979ah} where the Higgs is associated to a bound  state  of  a  new strong  dynamics, like QCD, and the  EW symmetry is broken  by dynamical condensation.  In the 70s/early 80s, the first versions of Technicolour~\cite{Farhi:128227}  predicted  a  heavy  Higgs (thus leading to an effectively Higgs-less theory) and  induced  very  large  corrections  to  precision  measurements~\cite{Peskin:1990zt}.  A way  to  produce a light  Higgs  is  to  enlarge  the  global  symmetry  such that one of the additional pNGBs  can  play  the  role  of the Higgs~\cite{Kaplan:1983fs,Kaplan:1983sm,Georgi:1984af}. Other attempts include the possibility that a near--conformal dynamics~\cite{Holdom:1981rm,Yamawaki:1985zg,Bando:1986bg,Dietrich:2005jn,Appelquist:2010gy}, or some other dynamical mechanism~\cite{Elander:2010wd,Foadi:2012bb}, may reduce the mass of the Technicolour $0^{++}$ state in non--QCD--like theories~\footnote{For evidence on the Lattice, see Refs.~\cite{Fodor:2012ty,Fodor:2015vwa}.} (even though the effectiveness of this mechanism is still unclear~\cite{Holdom:1986ub,Holdom:1987yu}). In recent years, the AdS/CFT  correspondence~\cite{Maldacena:1997re}  has uncovered that warped  5--dimensional  models~\cite{Randall:1999vf}  show a low energy behaviour  similar to that of  4--dimensional strong near--conformal field theories (CFT), if maximally supersymmetric. The conjecture implies that  the 5--dimensional bulk  gauge  symmetry   corresponds to  the global  symmetry  of the CFT, and that UV  boundary  fields (Kaluza--Klein  modes)     correspond to external  elementary  fields (internal  bound  states of the CFT).  The interest in composite Higgs models was thus rekindled as the composite Higgs can be associated to a bulk gauge state. The  minimal  composite Higgs model,  based on the global symmetry $SO(5)/SO(4)$~\cite{Agashe:2004rs}, loosely based on the AdS/CFT correspondence,   is an  effective theory which  only focuses  on the  low energy properties of the composite dynamics. In this  theory,  the 4--dimensional composite Higgs  field, whose  mass  is protected  by  global symmetries, corresponds  to the additional polarisation of  bulk  gauge  bosons.  Its mass  is, therefore,  protected by  the gauge symmetry in the warped  bulk, while the SM fermion  masses are generated  by  partial  compositeness~\cite{Kaplan:1991dc} (see Refs.~\cite{Contino:2004vy,Cacciapaglia:2008bi} for  AdS/CFT applied to fermions). Models of this kind do not care about the properties of the underlying  dynamics generating the composite states, nor about their UV completion. The physical properties of the models rely on the assumption of a restored conformal symmetry above the scale associated to the heavier resonances, and on the further assumption that large conformal dimensions can push the scale where flavour physics is generated close to the Planck scale.
This  picture, therefore,  is  built  on a set of  rather strong  assumptions and it relies on an effective field theory description. It is, thus, an interesting question to ask how  to  realize it  by specific underlying  dynamics.  In fact,  there  is  no  guarantee  that  there  exists a   suitable  underlying description  for  any low energy effective theory.  Furthermore, providing a definitive UV completion can  constrain  the physics of the associated effective  theory.

Recently  some  work in the literature  has explored possible  underlying  descriptions  of modern composite Higgs models. Underlying models have been  built  based on  a  simple confining  gauge  group  with fundamental  fermions~\cite{Galloway:2010bp,Cacciapaglia:2014uja,Ma:2015gra} (for examples with partial compositeness, see Refs.~\cite{Barnard:2013zea,Ferretti:2013kya,Ferretti:2014qta,vonGersdorff:2015fta,Vecchi:2015fma}). Definite  underlying completions  can  provide a  precise  relation between the components of the underlying theory  and the bound  states  described in the effective theory. Furthermore, in  these fundamental  composite Higgs models,   the global symmetry  breaking  pattern  and the spectrum  of the bound  states can  be  characterized  by use of Lattice simulations. The  minimal Fundamental Composite  Dynamics (FCD)  Higgs  is  realized  by  a  confining  $SU(2)_{\rm FCD}$ gauge group  with  four  Weyl fermions, leading to the global symmetry breaking pattern $SU(4) $ to  $Sp(4)$~\cite{Ryttov:2008xe,Galloway:2010bp,Cacciapaglia:2014uja}. In the  minimal FCD model, the  Higgs  is  a  mixture  of the pNGB  and the Technicolour scalar. Like in most pNGB Higgs models, it  can  provide small corrections to  precision measurements~\cite{Galloway:2010bp,Arbey:2015exa} in a wide  parameter  range. In  this  model  there  is  an EW singlet  pNGB  which has  been studied  as  a composite scalar Dark Matter (DM)  candidate in an effective theory approach~\cite{Frigerio:2012uc,Marzocca:2014msa}, however decays are induced from the Wess-Zumino-Witten (WZW)~\cite{Wess:1971yu,Witten:1983tx,Duan:2000dy} anomaly, that unequivocally present in the underlying fermionic FCD models. In order  to  provide  a natural  scalar DM~\footnote{Composite sectors that contain only a DM candidate, but no Higgs, have been studied, for instance, in Refs.~\cite{Antipin:2015xia,Carmona:2015haa}. Atomic--type composite DM can also arise in mirror models~\cite{Blinnikov:1983gh,Mohapatra:2001sx,Foot:2004pa} or as atomic DM~\cite{Kaplan:2009de}.},  we  propose  a  less  minimal  FCD  model based  on a confining  $SU(N)_{\rm FCD} $ gauge  group  with  four  Dirac fermions~\cite{Ma:2015gra}. This  model has a global  symmetry  $SU(4)_1 \times SU(4)_2 $  broken  to the diagonal  $ SU(4)$, thus containing many more pNGBs than the minimal case: two  Higgs  doublets, two  custodial triplets  and  a  singlet. Nevertheless, precision EW measurements can be passed without increasing the compositeness scale with respect to the minimal case. It has been shown in Ref.~\cite{Ma:2015gra} that the two Higgs doublets are related by a global U(1) transformation contained in SU(4), thus the Higgs vacuum can always be aligned on one of the two doublets without loss of generality. An unbroken parity that protects the second doublet and the two triplets can be defined, which thus prevents the lightest scalar from decaying. 
In addition, there is an unbroken global charge, the Techni--baryon (TB) number U(1)$_{\rm TB}$, which guarantees that the lightest baryon (bound state of $N$ underlying fermions) is stable and may, thus, be a candidate for asymmetric DM~\cite{Nussinov:1985xr,Barr:1990ca,Nussinov:1992he}.

In  this  paper  we  analyze the physical properties of  all  the additional  scalars, odd under the DM parity. We find that the even singlet  scalar has properties very similar to the ones of the singlet in the minimal SU(4)/Sp(4) model. The second Higgs  doublet and  the triplets, odd  under the DM parity,  must  decay to the  lightest scalar which  can  be the  DM candidate.  We  provide a complete analysis of the  phenomenology of the scalar  DM candidate, in the region of the parameter space where it is mainly aligned with the EW singlets contained in the custodial triplet. The reason for this choice is to avoid the strong constraints from Direct Detection in the presence of direct couplings to the EW gauge bosons. We find that the DM candidate  in  this model behaved  differently  from the singlet in the minimal model~\cite{Frigerio:2012uc} because  of its  mixing to EW--charged states which enhances its  annihilation cross  section. Therefore,  the preferred mass  needs  to be  around the TeV scale  to have  enough  relic  density, if  the scalar undergoes thermal freeze--out and needs to saturate the observed DM relic abundance.  We also find that the lightest TB cannot play the role of DM, because its thermal relic abundance is too small and an asymmetry cannot be generated via EW sphalerons as the TB number is exactly conserved in this minimal realization. We thus discuss the possibility that UV generated interactions may break the symmetry and thus generate an asymmetry.
We discuss Direct and Indirect DM searches: if the scalar is the the thermal DM candidate, Direct Detection strongly constraints the parameter space, ruling out larger masses, while Indirect searches are only relevant if the DM scalar is non-thermally produced.
Finally, the LHC reach on the odd scalars is found to be very feeble, due to the small couplings of the scalars which can only be mediated by EW gauge interactions.

The paper is organized as follows: in Section~\ref{sec:model} we recap the main properties of the model introduced in~\cite{Ma:2015gra}, while the phenomenology of the extended scalar sector is characterized in Section~\ref{sec:pheno}. In the following Section~\ref{sec:relic}, we focus on the lightest odd pNGB as a candidate for thermal DM, and we analyse both the relic density and constraints from Direct and Indirect DM searches. Finally, in Section~\ref{sec:LTB} we sketch the properties of the stable Baryons, which could also play the role of DM,  before concluding.

%%%%%%%%%%%%%%%%%%%%%%%%%%%%%%%%%%%%%%%%%%%%%%
\section{The model} \label{sec:model}

We pay our attention to a model of composite Higgs based on the coset SU(4)$\times$SU(4) /SU(4), where the custodial symmetry of the SM Higgs potential, SU(2)$_L \times$SU(2)$_R$, is embedded in the diagonal SU(4). This is the minimal model with global symmetry SU($N_F$)$^2$ which enjoys the possibility of the Higgs arising as a pNGB within custodial invariance~\cite{Bellazzini:2014yua,Mrazek:2011iu}. We shall recall that the SM electroweak symmetry arises as the partial gauging of the custodial symmetry, with U(1)$_Y \subset$ SU(2)$_R$.
The model is particularly interesting as it can arise in the confined phase of a simple underlying gauge theory of fermions based on a gauged SU($N$)$_{FCD}$ with 4 Dirac fermions transforming as the fundamental representation~\cite{Ma:2015gra}. In terms of the EW symmetry, the two underlying fermions can be thought of as transforming as a doublet of SU(2)$_L$ and a doublet of SU(2)$_R$. In the following, we will mainly base our analysis on an effective field theory approach, for which only the symmetry structure matters.

The global symmetry breaking, spontaneously generated by fermion condensation in the underlying theory~\cite{Witten:1983tx}, generates 15 Goldstone bosons transforming as the adjoint of the unbroken SU(4). Here, we will work in the parameterisation where the vacuum of the theory is misaligned to break the electroweak symmetry~\cite{Kaplan:1983fs}, and the misalignment is parameterised by an angle $\theta$~\cite{Cacciapaglia:2014uja} interpolating between a Technicolour-like model and a pNGB Higgs one. This approach differs from the usual parameterisations considered in composite (pNGB) Higgs models (see for instance~\cite{Gripaios:2009pe}), where the Goldstone expansion is operated around the electroweak preserving vacuum and a vacuum expectation value to the pNGB transforming as the Higgs is later applied. The two parameterisations, however, give equivalent physical results, at least to lowest order in a small $\theta$ expansion. The advantage of our approach is that all the derivative couplings respect the Goldstone symmetry, while any explicit breaking of the global symmetry is added in the form of non-derivative potential and/or interactions. We will thus use results already obtained in~\cite{Ma:2015gra}: in the following we will recap the main results useful to understand the remaining of this paper.

The pNGBs of the theory are introduced in terms of a matrix $\Sigma$, transforming linearly under the global symmetry SU(4)$_1 \times$SU(4)$_2$~\footnote{An equivalent method consists in defining Maurer--Cartan one--forms, see for instance~\cite{Marzocca:2012zn}.}:
\beq
\Sigma = \Omega_\theta \cdot e^{i \Pi/f} \cdot \Omega_\theta\,,
\eeq
where
\beq
\Pi = \frac{1}{2} \left( \begin{array}{cc}
\sigma^i \Delta_i + s/\sqrt{2} & - i\ \Phi_H \\
i \ \Phi^\dagger_H & \sigma^i N_i - s/\sqrt{2}
\end{array} \right)
\eeq
parametrises the 15 pNGBs and
\beq
\Omega_\theta = \left( \begin{array}{cc}
\cos \frac{\theta}{2} & \sin \frac{\theta}{2} \\
- \sin \frac{\theta}{2} & \cos \frac{\theta}{2}
\end{array} \right)
\eeq
is an SU(4) rotation matrix that contains the $\theta$-dependent misalignment of the vacuum.
In the pNGB matrix, $\Delta_i$ and $N_i$ transform respectively as a triplet of  SU(2)$_L$ and SU(2)$_R$ ($\sigma^i$ are the Pauli matrices), $s$ is a singlet while $\Phi_H$ is a complex bi-doublet thus describing two Higgs doublets, $H_1$ and $H_2$.
We work in the most general custodial invariant vacuum where, as proven in~\cite{Ma:2015gra}, the vacuum can be aligned with one of the two doublets ($H_1$) without loss of generality.

As  discussed in~\cite{Ma:2015gra}, the mass of the EW gauge bosons is due to the misalignment, thus it is proportional to the angle $\theta$:
\beq
m_W^2 = 2 g^2 f^2 \sin^2 \theta\,, \quad m_Z^2 = \frac{m_W^2}{\cos^2 \theta_W}\,.
\eeq
so that we can identify the relation between the decay constant $f$ and the EW scale
\beq \label{eq:fsm}
2 \sqrt{2} f \sin \theta = v_{\rm SM} = 246\; \mbox{GeV}\,, \qquad \sin \theta=\frac{ v_{\rm SM}}{2 \sqrt{2}  f}\,.
\eeq
Note that the normalisation of $f$ is different from the usual one in Composite Higgs literature~\cite{Marzocca:2012zn} by a factor $2\sqrt{2}$, and that the small number associated with the hierarchy between the EW and compositeness scale is $\sin \theta$. In~\cite{Ma:2015gra} it was also shown that electroweak precision tests require $\theta$ to be small, and the bound can be estimated to be
\beq
\sin \theta \leq 0.2\,.
\eeq
It should be noted, however, that this bound may be released if massive composite states are lighter than the naive expectation: this might be the case for light spin-1 or spin-1/2 states~\cite{Contino:2015mha,Ghosh:2015wiz}, or for a light $\sigma$--like scalar~\cite{Arbey:2015exa} that mixes with the Higgs (there is growing evidence in the Lattice literature that such light scalars may arise if the underlying theory is near-conformal in the UV~\cite{Brower:2015owo}).

We should also remark that we do not assume the presence of light fermionic bound states that mix with the top, and other SM fermions, to give them mass via partial compositeness. Partial compositeness can be nevertheless implemented if the number of flavours is extended by additional coloured fundamental fermions (and for 3 FCD colours), as shown in~\cite{Vecchi:2015fma}: their presence can also explain why the theory runs in a conformal regime at higher energies~\cite{Brower:2015owo}, while a largish mass for the coloured fermions would avoid the presence of additional light coloured scalars.
Partial compositeness, in fact, strongly relies on the presence of large anomalous dimensions for the fermionic operators, which are only possible if the theory is conformal at strong coupling regime, i.e. very close to the lower edge of the conformal window.  However, there is no evidence so far that large anomalous dimensions may arise~\cite{Pica:2016rmv}, and the position of the lower edge in terms of number of flavours, which is expected to lie between 8 and 12~\cite{Appelquist:2009ty}, is disputed~\cite{Aoki:2012eq,Cheng:2014jba,Fodor:2011tu,Lombardo:2014pda}.
In our study of this model we want to be as conservative as possible, thus
we will assume that, if present, the coloured fermions have a mass well above the scale $f$ and can thus be thought of as heavy flavours. If partial compositeness is behind the top mass, the fermionic top partners can be integrated out and their effect can be parameterised in terms of effective couplings of the SM fermion fields to the composite pNGBs. Another possibility would be to couple directly the elementary fermions to the composite Higgs sector via four fermion interactions, even though the issue of generating the correct flavour structures is left to the UV physics leading to such interactions. One possibility may be that masses of the light quarks and leptons are generated at a much higher scale than the top providing enough suppression without the need of flavour symmetries in the composite sector~\cite{Cacciapaglia:2015dsa} (see also~\cite{Matsedonskyi:2014iha,Panico:2016ull}). Finally, we would like to mention the recently proposed mechanism where the masses of the SM fermions are induced thanks to the presence of scalars charged under the FCD dynamics~\cite{Sannino:2016sfx}: while the underlying theory is not natural as it contains fundamental scalars, partial compositeness can be implemented without requiring large anomalous dimensions and the theory is potentially predictive up to high scales.

In the following, we will take the same approach as in~\cite{Ma:2015gra} and assume that the alignment of the vacuum is fixed by the interplay between the contribution of top loops and the effect of an explicit mass term for the underlying fermions. We refer the reader to~\cite{Ma:2015gra} for more details. Here we limit ourselves to notice that the potential has essentially 4 parameters: two are the masses of the underlying fermions, $m_L$ and $m_R$, the other 2 are form factors describing the effect of top and gauge loops ($C_t$ and $C_g$ respectively). Notice that only the average mass, $m_L + m_R$, enters the stabilization of the potential, and it can be traded with the value of the misalignment angle $\sin \theta$, while the mass difference:
\beq \label{eq:delta}
\delta = \frac{m_L - m_R}{m_L + m_R}
\eeq 
will only affect the masses of the pNGBs.
The form factors are potentially calculable, as they only depend on the underlying dynamics: predictions can be obtained either using Lattice techniques~\cite{Arthur:2016dir,Rantaharju:2015nep}, or by employing effective calculation methods borrowed from QCD~\cite{Foadi:2016nbi}. The mass of the Higgs candidate can be predicted as:
\beq 
m_h^2 = \frac{C_t}{4} m_{\rm top}^2 - \frac{C_g}{16} (2 m_W^2 + m_Z^2)\,.
\eeq
We can thus use the above relation to fix the value of $C_t$ to match the experimental value of the Higgs mass at $125$ GeV (which requires $C_t \sim 2$), while $C_g$ is left as an $\mathcal{O} (1)$ parameter. The masses of the other pNGBs can be similarly computed however, before showing this, we will discuss the structure of the effective Yukawa couplings for all SM fermions, thus generalising the results of Ref.~\cite{Ma:2015gra}.

%%%%%%%%%%%%%%%%%%%%%%%%%%%%%%%%
\subsection{Flavour realisation and Dark Matter parity}

Independently on the origin of the quark and lepton masses, at low energy we can write effective Yukawa couplings in terms of the pNGB matrix $\Sigma$ by simply coupling the usual SM fermion bilinears to the components of $\Sigma$ that transform like the Higgs doublets:
\beq \label{eq:Yuk0}
 \mathcal{L}_{\rm Yuk} &=& - f\, (\bar{Q}_{Li}^\alpha u_{Rj}) \left[ \mbox{Tr} [P_{1,\alpha} (y_{u1} ^{ij} \Sigma + y_{u2}^{ij}  \Sigma^\dagger)] + (i\sigma_2)_{\alpha \beta} \mbox{Tr} [P_2^\beta (y_{u3} ^{ij} \Sigma + y_{u4} ^{ij} \Sigma^\dagger)] \right]  \nonumber  \\
& & - f\, (\bar{Q}_{Li}^\alpha d_{Rj}) \left[ \mbox{Tr} [P_{b1,\alpha} (y_{d1} ^{ij} \Sigma + y_{d2} ^{ij} \Sigma^\dagger)] + (i\sigma_2)_{\alpha \beta} \mbox{Tr} [P_{b2} ^\beta (y_{d3} ^{ij} \Sigma + y_{d4} ^{ij} \Sigma^\dagger)] \right]  \nonumber  \\
&& - f\, (\bar{L}_{Li}^\alpha e_{Rj}) \left[ \mbox{Tr} [P_{b1,\alpha} (y_{e1} ^{ij} \Sigma + y_{e2} ^{ij} \Sigma^\dagger)] + (i\sigma_2)_{\alpha \beta} \mbox{Tr} [P_{b2} ^\beta (y_{e3} ^{ij} \Sigma + y_{e4} ^{ij} \Sigma^\dagger)] \right]
  + h.c. \nonumber  \\
\eeq
where $Q_{Li}$ and $L_{Li}$ are the quark and lepton doublets ($\alpha$ is the SU(2)$_L$ index), and $u_{Rj}$, $d_{Rj}$ and $e_{Rj}$ are the singlet quarks and charged leptons. The projectors $P_{1,2}$ and $P_{b1, b2}$ are defined in~\cite{Ma:2015gra}. In the most general case, therefore, one can write 4 independent Yukawa couplings per type of fermion. Expanding $\Sigma$ up to linear order in the pNGB fields, the masses of the fermions can be written as:
\beq \label{eq:Yuk}
& & \mathcal{L}_{\rm Yuk}  = \nonumber \\
& & - \left[  Y_{ui} \delta^{ij} v_{SM} + Y_{ui}\delta^{ij}  \cos \theta\ h_1 + i Y_{uD} ^{ij}  \ h_2 + Y_{uD} ^{ij} \cos \theta\ A_0+ i \frac{Y_{uT} ^{ij}}{\sqrt{2}} \sin \theta\ (N_0 + \Delta_0)\right] (\bar{u}_{Li} u_{Rj}) \nonumber \\
&& - \left[ -i \sqrt{2} \tilde{Y}_{uD} ^{ij} \cos \theta\ H^- + i \tilde{Y}_{uT} ^{ij} \sin \theta\ (N^- + \Delta^-) \right] (\bar{d}_{Li} u_{Rj}) \nonumber  \\
&& - \left[  Y_{di} \delta^{ij} v_{SM} + Y_{di}\delta^{ij}  \cos \theta\ h_1 + i Y_{dD} ^{ij}  \ h_2 - Y_{dD} ^{ij} \cos \theta\ A_0 -i \frac{Y_{dT} ^{ij}}{\sqrt{2}} \sin \theta\ (N_0 + \Delta_0)\right] (\bar{d}_{Li} d_{Rj}) \nonumber \\
&& - \left[ i \sqrt{2} \tilde{Y}_{dD} ^{ij} \cos \theta\ H^+ + i \tilde{Y}_{dT} ^{ij} \sin \theta\ (N^+ + \Delta^+) \right] (\bar{u}_{Li} d_{Rj}) \nonumber \\
&& - \left[  Y_{ei} \delta^{ij} v_{SM} + Y_{ei}\delta^{ij}  \cos \theta\ h_1 + i Y_{eD} ^{ij}  \ h_2 - Y_{eD} ^{ij} \cos \theta\ A_0 -i \frac{Y_{eT} ^{ij}}{\sqrt{2}} \sin \theta\ (N_0 + \Delta_0)\right] (\bar{e}_{Li} e_{Rj}) \nonumber \\
&& - \left[ i \sqrt{2} Y_{eD} ^{ ij} \cos \theta\ H^+ + i Y_{eT} ^{ ij} \sin \theta\ (N^+ + \Delta^+) \right] (\bar{\nu}_{Li} e_{Rj}) \nonumber \\
&&   + h.c.,  \label{eq:genralyukawa}
\eeq
where we have diagonalized the matrices $Y_u$, $Y_d$ and $Y_e$, and 
\beq
\tilde{Y}_{uD/T} =V^\dagger _{CKM} Y_{uD/T}\ , \quad \tilde{Y}_{dD/T} = V _{CKM}  Y_{dD/T} \ , 
\eeq
with  $ V_{CKM}   $ being the standard CKM  matrix~\footnote{We neglect here neutrino masses and the PNMS mixing matrix, which can be introduced in the same way as in the SM.}.
The couplings $Y$ are linear combinations of the couplings in the effective operators in Eq.~(\ref{eq:Yuk0}) and are defined as~\cite{Ma:2015gra}:
\beq
 Y_{f} ^{ij} &=&  \frac{y_{f1} ^{ij} - y_{f2} ^{ij} - (y_{f3} ^{ij} - y_{f4}^{ij} )}{2 \sqrt{2}}\,,  \quad Y_{f0} ^{ij} =  \frac{y_{f1} ^{ij} + y_{f2} ^{ij} - (y_{f3} ^{ij} + y_{f4}^{ij} )}{2 \sqrt{2}}\,, \nonumber \\
Y_{f D} ^{ij} &=&  \frac{y_{f1}^{ij}  - y_{f2} ^{ij} + (y_{f3}^{ij}  - y_{f4}^{ij} )}{2 \sqrt{2}}\,, \quad
Y_{f T}^{ij} = \frac{y_{f1} ^{ij} + y_{f2}^{ij} + (y_{f3}^{ij} + y_{f4}^{ij} )}{2 \sqrt{2}}\,;
\eeq
where $f =u,d,e$. Note that one combinations that we dub $Y_{f0}$ does not appear in the linear couplings in Eq.~(\ref{eq:Yuk}). Also, the expression for the masses allows us to relate
\beq
 Y_f = \frac{m_f}{v_{SM} } .
\eeq

The operator that generates a potential for the vacuum (and masses for the pNGBs) arises from loops of the SM fermions: the leading one, in an expansion at linear order in the pNGBs, reads
\beq
V_{\rm fermions} & = & - 8 f^4 C_t \left\{ \sum_{f=u,d,e} \xi_f \mbox{Tr} [Y_f^\dagger Y_f]\ \left( \sin^2 \theta + \sin (2\theta) \frac{h}{2 \sqrt{2}f} \right) + \right. \nonumber \\
 & & - i \sum_{f=u,d,e}  \xi_f \mbox{Tr} [Y^\dagger_{fD} Y_f - Y_{fD} Y_f^\dagger]\ \sin \theta \frac{h_2}{2 \sqrt{2} f} +  \\
 & & + \left( \mbox{Tr}  [Y^\dagger_{uD} Y_u + Y_{uD} Y_u^\dagger] - \sum_{f=d,e}  \xi_f \mbox{Tr} [Y^\dagger_{fD} Y_f + Y_{fD} Y_f^\dagger] \right)\ \frac{\sin (2 \theta)}{2} \frac{A_0}{2 \sqrt{2} f} + \nonumber \\
 & & - i \left.  \left( \mbox{Tr}  [Y^\dagger_{uT} Y_u - Y_{uT} Y_u^\dagger] - \sum_{f=d,e}  \xi_f \mbox{Tr} [Y^\dagger_{fT} Y_f - Y_{fT} Y_f^\dagger] \right)\ \sin^2 \theta \frac{N_0 + \Delta_0}{4 f} \right\} \,. \nonumber
\eeq
The coefficient $\xi_f$ counts the number of QCD colours, and is defined as 
$$
\xi_{u/d} = 1\,, \qquad \xi_e = \frac{1}{3}\,.
$$
Note that we introduced a single form factor $C_t$ as the loops have the same structure in terms of the underlying fermions (and the FCD is flavour blind). The contribution to the potential for $\theta$ has the same functional form as the one generated by the top alone, thus it suffices to replace
\beq
Y_t^2 \to \sum_{f} \xi_f Y_f^2 = \sum_{q= {\rm quarks}} Y_q^2 + \frac{1}{3} \sum_{l= {\rm leptons}} Y_l^2 
\eeq
in the formulas in Ref.~\cite{Ma:2015gra}~\footnote{In particular, the Higgs mass is given by $m_h^2 = \frac{C_t}{4} \sum_f \xi_f m_{f}^2 - \frac{C_g}{16} (2 m_W^2 + m_Z^2)$, and is dominated by the top contribution.}. As shown in~\cite{Ma:2015gra}, the tadpole for $h_2$ can be always removed by an appropriate choice of phase and thus one can assume without loss of generality that its coefficient vanishes.
The coefficient of the tadpoles for $A_0$ and the triplets, however, are physical and, as shown by the opposite sign of the contribution of the down-type fermions with respect to the up-type ones, violate custodial invariance. One thus needs to impose peculiar conditions on the couplings in order for such tadpoles to vanish, otherwise the vacuum is misaligned along a non-custodial direction~\footnote{In fact, it would be enough to require that the coefficients are small enough to evade bounds from electroweak precision tests, so a strong constraint applies mainly on the top Yukawas.}:
\beq
\mbox{Tr} [ \mbox{Re} (Y_{uD}^\dagger Y_u) - \sum_{f=d,e} \xi_f \mbox{Re} (Y_{fD}^\dagger Y_f) ] &=& 0\,,\nonumber\\ 
\mbox{Tr} [ \mbox{Im} (Y_{uT}^\dagger Y_u) - \sum_{f=d,e} \xi_f \mbox{Im} (Y_{fT}^\dagger Y_f) ] &=& 0\,.
\eeq
From Eq.~(\ref{eq:Yuk}), we see that the couplings $Y_{fT}$ and $Y_{fD}$ generate direct couplings of the triplets and of the second doublet to the SM fermions: in general, therefore, the model will be marred by tree-level Flavour--Changing Neutral Currents (FCNCs) mediated by these pNGBs. The other couplings $Y_{f0}$ appear in couplings with two pNGBs, that we parametrise as (the flavour indices are understood and $\pi_j$ is a generic pNGB field)
\beq \label{eq:NLcoup}
\mathcal{L}_{\pi \pi \bar{f} f} &=& \frac{1}{2 \sqrt{2} f} \left( \frac{1}{2} \xi^N_{u,kl}\ \bar{u}_L u_R \pi^0_k \pi^0_l + \xi^C_{u,kl} \ \bar{u}_L u_R \pi^+_k \pi^-_l  + V_{\rm CKM}^\dagger \cdot \xi^+_{u,kl}\ \bar{d}_L u_R \pi_k^- \pi^0_l +\right.  \nonumber \\
& & \left. + \frac{1}{2} \xi^N_{d,kl}\ \bar{d}_L d_R \pi^0_k \pi^0_l + \xi^C_{d,kl} \ \bar{d}_L d_R \pi^+_k \pi^-_l  + V_{\rm CKM}\cdot \xi^-_{d,kl}\ \bar{u}_L d_R \pi_k^+ \pi^0_l+ \right. \nonumber \\
& & \left. + \frac{1}{2} \xi^N_{e,kl}\ \bar{e}_L e_R \pi^0_k \pi^0_l + \xi^C_{e,kl} \ \bar{e}_L e_R \pi^+_k \pi^-_l  + \xi^-_{e,kl}\ \bar{\nu}_L e_R \pi_k^+ \pi^0_l+ h.c. \right)\,,
\eeq
generated by non-linearities, and listed in Appendix~\ref{app:Yukawa}.

In~\cite{Ma:2015gra} it was shown that there exists a unique symmetry under which some of the pNGBs are odd while being compatible with the correct EW breaking vacuum. Under such parity, that we will call DM-parity in the following, the second doublet and the two triplets are odd. Furthermore, imposing the parity on the effective Yukawa couplings implies that
\beq
Y_{fD} = Y_{fT} = 0
\eeq
for all SM fermions (thus, all the linear couplings of the second doublet and triplets to fermions vanish).
Under these conditions, the scalar sector will contain a DM candidate, being the lightest state of the lot. Furthermore, as a bonus, the dangerous flavour violating tree level couplings of the pNGBS, together with potential custodial violating vacua, are absent!
Imposing the DM-parity, therefore, has a double advantage on the phenomenology of the model: its presence, or not, finally relies on the properties of the UV theory responsible for generating the masses for the SM fermions. 
As it can be seen in Table~\ref{table:topyukawa} in Appendix~\ref{app:Yukawa}, imposing the DM-parity, the system of odd scalars decouples from the even ones and the Higgs and the singlet $s$ do not communicate with the rest of the pNGBs.

Besides the parameters that are fixed by the SM masses and by the vacuum alignment, the free parameters of the model consists of the underlying fermion mass difference $\delta$, defined in Eq.~(\ref{eq:delta}), and the matrices $Y_{f0}^{ij}$. The latter matrices are unrelated to the SM fermions masses, and an eventual CP--violating phase cannot be removed. The first constraint, therefore, that one needs to check is about FCNCs generated at loop level.

%%%%%%%%%%%%%%%%%%%
\subsection{Flavour bounds}

Imposing the DM parity removes the flavour changing tree level couplings of the pNGBs, however, as it can be seen in Eq.~(\ref{eq:NLcoup}), there still exist couplings with two pNGBs proportional to $Y_{f0}$ that are potentially dangerous. Closing a loop of neutral or charged pNGBs, therefore, flavour changing four-fermion interactions are generated as follows:
\begin{multline}
\mathcal{L}_{\rm FCNC}^{\rm 1-loop} = \frac{1}{16 \pi^2} \log \frac{\Lambda^2}{m_\pi^2}\  \sum_{f,f'} \sum_{k,l}\left\{  \frac{\xi^N_{f,kl} \xi^N_{f',kl} +  \xi^C_{f,kl} \xi^{C}_{f',kl} }{16 f^2}\ \bar{f}_L f_R \bar{f}'_L f'_R  + \right.\\
\left.  \frac{\xi^N_{f,kl} \xi^{N,\dagger}_{f',kl} + \xi^C_{f,kl} \xi^{C,\dagger}_{f',kl}}{16 f^2}\ \bar{f}_L f_R \bar{f}'_R f'_L  + h.c. \right\}\,,
\end{multline}
where $f, f' = u, d, e$, and the flavour indices are left understood.
Taking the values of the couplings listed in Appendix~\ref{app:Yukawa}, we found that
\beq
\sum_{k,l} \xi^N_{f,kl} \xi^N_{f',kl} +  \xi^C_{f,kl} \xi^{C}_{f',kl} &=&  9 Y_f Y_{f'} \sin^2 \theta \pm 4 Y_{f0} Y_{f'0}\,, \\
\sum_{k,l} \xi^N_{f,kl} \xi^{N,\dagger}_{f',kl} + \xi^C_{f,kl} \xi^{C,\dagger}_{f',kl} &=& 9 Y_f Y_{f'}^\dagger \sin^2 \theta \pm 4 Y_{f0} Y_{f'0}^\dagger\,,
\eeq
where the positive sign apply to the case where both $f$ and $f'$ are of the same type (up or down), and the negative one when $f$ and $f'$ are of different type.
Flavour changing transitions are thus generated by off-diagonal coefficients of the matrices $Y_{f0}$. We can estimate the bound on these matrix elements by comparing the coefficient of the operators with a generic flavour suppression scale $\Lambda_F \sim 10^5$~TeV (see, for instance, Ref.~\cite{Butler:2013kdw}):
\beq
\frac{\log (4\pi)}{4 \pi^2} \frac{\sin^2 \theta}{v_{\rm SM}^2} \left.Y_{f0} Y_{f'0}\right|_{\rm off-diag} \lesssim \frac{1}{\Lambda_F^2} \Rightarrow \left.Y_{f0} Y_{f'0}\right|_{\rm off-diag} \lesssim \frac{10^{-10}}{\sin^2 \theta}\,;
\eeq
where we have approximated the masses of the pNGBs $m_\pi \sim f$, the cut--off $\Lambda \sim 4 \pi f$, and used the relation between $f$ and the SM Higgs VEV in Eq.~(\ref{eq:fsm}).
We see that a strong flavour alignment is needed, even when the effect of a small $\sin\theta \lesssim 0.2$ required by precision physics is taken into account. In the following, we will ``play safe'' and assume that $Y_{f0}$ is always aligned with the Yukawa couplings $Y_f$: this assumption will not play a crucial role in studying the properties of the DM candidate.

%%%%%%%%%%%%%%%%%%%%%%%%%%%%%%%%%%%%%%%%%%%%%%
\section{Phenomenology of the scalar sector} \label{sec:pheno}

In this section we will explore the properties of the 11 exotic pNGBs predicted by this model in addition to the Higgs and the 3 Goldstone bosons eaten by the massive $W^\pm$ and $Z$.
The lowest order chiral Lagrangian possesses some discrete symmetries which are compatible with the vacuum and with the gauging of the EW symmetry (but are potentially broken by the Yukawa couplings)~\cite{Ma:2015gra}:
\begin{itemize}
\item[-] A--parity, generated by a space--time parity transformation plus an SU(4) rotation, under which the singlet $s$ and the triplets are odd. It is left invariant by the Yukawas if $Y_{fT} = Y_{f0} = 0$, however it is broken by the WZW anomaly term.
\item[-] B--parity, generated by a charge conjugation plus an SU(4) rotation, under which the second doublet and the two triplets are odd. It is left invariant if $Y_{fD} = Y_{fT} = 0$, and it can act as a DM parity.
\item[-] CP, which is only broken by phases present in the Yukawa couplings (in addition to the CKM phase). Namely, the phases of $Y_{fT}$, $Y_{fD}$ and $Y_{f0}$ affect the couplings and mixing of the pNGBs.
\end{itemize}
\begin{table}[tt] \begin{center}
\begin{tabular}{|c|c|ccccc|ccc|}
\hline
  & $h_1$ & $h_2$ & $A_0$ & $s$ & $\Delta_0$ & $N_0$ & $H^\pm$ & $\Delta^\pm$ & $N^\pm$ \\
\hline
CP (real $Y_{f0}$) & $+$ & $-$ & $+$ & $-$ & $-$  & $-$ & $-$ & $-$ & $-$ \\
CP (imaginary $Y_{f0}$) & $+$ & $+$ & $-$ & $-$ & $-$  & $-$ & $-$ & $-$ & $-$ \\
\hline
A & $+$ & $+$ & $+$ & $-$ & $-$  & $-$ & $-$ & $-$ & $-$  \\
B (DM) & $+$ & $-$ & $-$ & $+$ & $-$  & $-$ & $-$ & $-$ & $-$ \\
\hline
\end{tabular}
\caption{Parities assignment of the pNGBs under CP, and A and B: for the charged states, it is left understood that they transform in their complex conjugates (anti-particles) under CP transformation.} \label{tab:BCP}
\end{center} \end{table}
In models where a DM candidate is present, i.e. where the B--parity is preserved, only the phase of $Y_{f0}$ can break CP in the scalar sector. In fact,  there are two cases for CP-conserving scalar sectors: if all the $Y_{f0}$'s are real, then $A_0$ is a CP-even state, while for $Y_{f0}$'s purely imaginary one can redefine the CP transformation such that $h_2$ is CP-even.
The parities of the pNGBs under the discrete symmetries are summarised in Table~\ref{tab:BCP}.

 %%%%%%%%%%%%%%%%%%%%%%
\subsection{Masses of the pNGBs}

The masses of  the  scalars are generated by the interactions that explicitly  break  the global  symmetry: in our minimal scenario, they are the EW gauge interaction, the Techni--fermion masses and the fermion Yukawas.  Complete expressions for the mass matrices can be found in Appendix C of Ref.~\cite{Ma:2015gra}.

We remark that two  scalars do not mix with the others: the Higgs  $h_1$ and the singlet $s$. The mass of the Higgs is given by
\beq
m_{h_1} ^2  = \frac{C_t}{4} \left( \sum_q m_{q}^2 + \frac{1}{3} \sum_l m_l^2 \right) - \frac{C_g}{16} (2 m_W^2 + m_Z^2)\,,
\eeq
where $m_q$ and $m_l$ are the masses of the SM quarks and leptons, respectively.
Thus,  we  can express the parameter $C_t$ as a function of known masses (and $C_g$).
The pseudoscalar $s$ also doesn't  mix  with  other  pNGBs, even when all discrete symmetries (B and  CP)  are  violated: its mass is equal to~\footnote{We use the Techni--fermion mass to stabilise the potential to a small value of $\theta$. In other approaches (with top partners), higher order top loops can do the job. The spectrum will, then, be different from what we use here.}
\beq
m_{s} ^2 = \frac{m_{h_1} ^2 }{\sin^2 \theta}\,,
\eeq
matching the results found in the minimal SU(4)/Sp(4) case~\cite{Galloway:2010bp,Cacciapaglia:2014uja}.

The DM-odd states, on the other hand, mix with each other. In the remaining of the paper, we will work in the CP--conserving case where all $Y_{f0}$ are real, thus $A_0$ is the only CP-even state and it does not mix. We denote the mass eigenstates as follows:
\beq
& \varphi \equiv A_0 \qquad \mbox{(CP even)} \,, & \nonumber \\
& \eta_{1,2,3} \equiv N_0, \Delta_0, h_2 \qquad \mbox{(CP-odd)} \,, & \\
& \eta^\pm_{1,2,3} \equiv N^\pm, \Delta^\pm, H^\pm\,. & \nonumber
\eeq
Note that we renamed $A_0$ in order to avoid confusion with standard 2 Higgs doublet models and supersymmetry, where $A$ indicates a pseudo-scalar.
%Numerical values for the masses, as a function of $\theta$ are shown in Figure~\ref{fig:mass of scalars} for two simple choices of the other parameters. The parameter $\delta = \frac{m_L - m_R}{m_L + m_R}$ describes the difference in mass between the SU(2)$_L$ doublet Techni--fermion and the SU(2)$_R$ one.

The mass of the CP-even scalar is given by
\beq
m_\varphi^2 = m_{s}^2 + \frac{C_g}{16} \left( \frac{4 m_W^2 + m_Z^2}{\sin^2 \theta} + 2 m_W^2 - m_Z^2 \right)\,.
\eeq
The CP-odd states, however, mix thus their mass structure is less clear. It is, however, useful to expand the expressions for small $\theta$: at leading order in $\sin^2 \theta$, we obtain (the results are written in terms of the SM boson masses where possible):
\beq
m_{\eta_1}^2 \sim m_{N_0}^2 \sim m_{s}^2 (1-\delta) + \dots\,,  && m_{\eta_1^\pm}^2 \sim m_{N^\pm}^2 \sim m_{\eta_1}^2 + C_g \frac{m_Z^2 - m_W^2}{4 \sin^2 \theta} + \dots \\
 m_{\eta_2}^2 \sim m_{\eta_2^\pm}^2 \sim m_{h_2}^2 \sim m_{H^\pm}^2 &\sim & m_{s}^2 + C_g \frac{2 m_W^2 + m_Z^2}{16 \sin^2 \theta} + \dots \\
 m_{\eta_3}^2 \sim m_{\eta_3^\pm}^2 \sim m_{\Delta}^2 &\sim & m_{s}^2 (1+\delta) + C_g \frac{m_W^2}{2 \sin^2 \theta} + \dots
\eeq
The $\dots$ stand for higher order corrections in $v_{\rm SM}^2/f^2$.
We can clearly see that, for positive $\delta > 0$, the lightest states, $\eta_1$ and $\eta_1^\pm$, correspond approximately to the SU(2)$_R$ triplet $N$, and the splitting between the charged and neutral states is
\beq
\Delta m_{\eta_1}^2 = C_g \frac{m_Z^2 - m_W^2}{4 \sin^2 \theta} + \dots
\eeq
which is proportional to the hypercharge gauging (the only spurion that breaks SU(2)$_R$).
For negative $\delta < 0$, on the other hand, the doublet and the SU(2)$_L$ triplet may provide the lightest state.

In the following, we will focus on a case where $\delta \geq 0$, so that the lightest states always belongs to the $SU(2)_R$ triplet and contain a neutral singlet that may play the role of DM. 
The main reason behind this choice is to have a DM candidate with suppressed couplings to the EW gauge bosons, else Direct Detection bounds will strongly constraint the parameter space of the model. We leave the more complicated and constrained case of $\delta < 0$ to a future study.
Furthermore, we will choose a real $Y_{f0}$, so that the CP--even state that does not mix with other odd pNGBs is $\varphi \equiv A_0$, and chose $Y_{f0}$ aligned to the respective Yukawa matrix $Y_f$ to avoid flavour bounds. 
In the numerical results, for simplicity, we will also assume that $Y_{f0}/Y_f$ is a universal quantity, equal for all SM fermions. This assumption has the only remarkable consequence that it is the coupling of the top that is the most relevant for the DM phenomenology. Couplings to lighter quarks and leptons may also be relevant, but only if there is a large hierarchy between $Y_{f0}$ and $Y_f$, situation that can only be achieved by severe tuning between the Yukawas in Eq.~(\ref{eq:Yuk0}).

%%%%%%%%%%%%%%%%%%%%%%%%%%
\subsection{Phenomenology of the singlet $s$ }

We first focus on the DM--even scalar $s$ which  does not mix with other scalars, similarly to the singlet in the minimal case SU(4)/Sp(4). In terms of symmetries, $s$ is the only CP-odd singlet under the custodial SO(4) symmetry, like the $\eta$ in SU(4)/Sp(4): thus, together with the doublet aligned with the EW breaking direction of the vacuum, it can be associated with an effective SU(4)/Sp(4) coset inside the larger SU(4)$^2$/SU(4).  As already discussed in~\cite{Ma:2015gra}, a coupling to two gauge bosons is allowed via the WZW term, thus $s$ cannot be a stable particle.  In Appendix~\ref{appintA} we discuss in detail the origin of the single couplings of $s$ to SM particles, leading to its decays and single production at colliders.

The phenomenology of $s$ is very similar to the one in the minimal SU(4)/Sp(4) case, thus we refer the reader to Refs.~\cite{Galloway:2010bp,Arbey:2015exa} for more details.
In summary, the DM--even singlet $s$ is very challenging to see at the LHC due to feeble production rates~\cite{Arbey:2015exa}.

%%%%%%%%%%%%%%%%%%%%%%%%%%%%%%%%%
\subsection{Phenomenology of DM-odd  pNGBs}

In a model where the DM parity is exactly conserved, the odd pNGBs can only decay into each other.  In Appendix~\ref{appintB} we list the relevant couplings, which also enter in the production at colliders. The decays proceed as follows:

\begin{figure}[!htb]
\begin{center}
\includegraphics[width=0.49\textwidth]{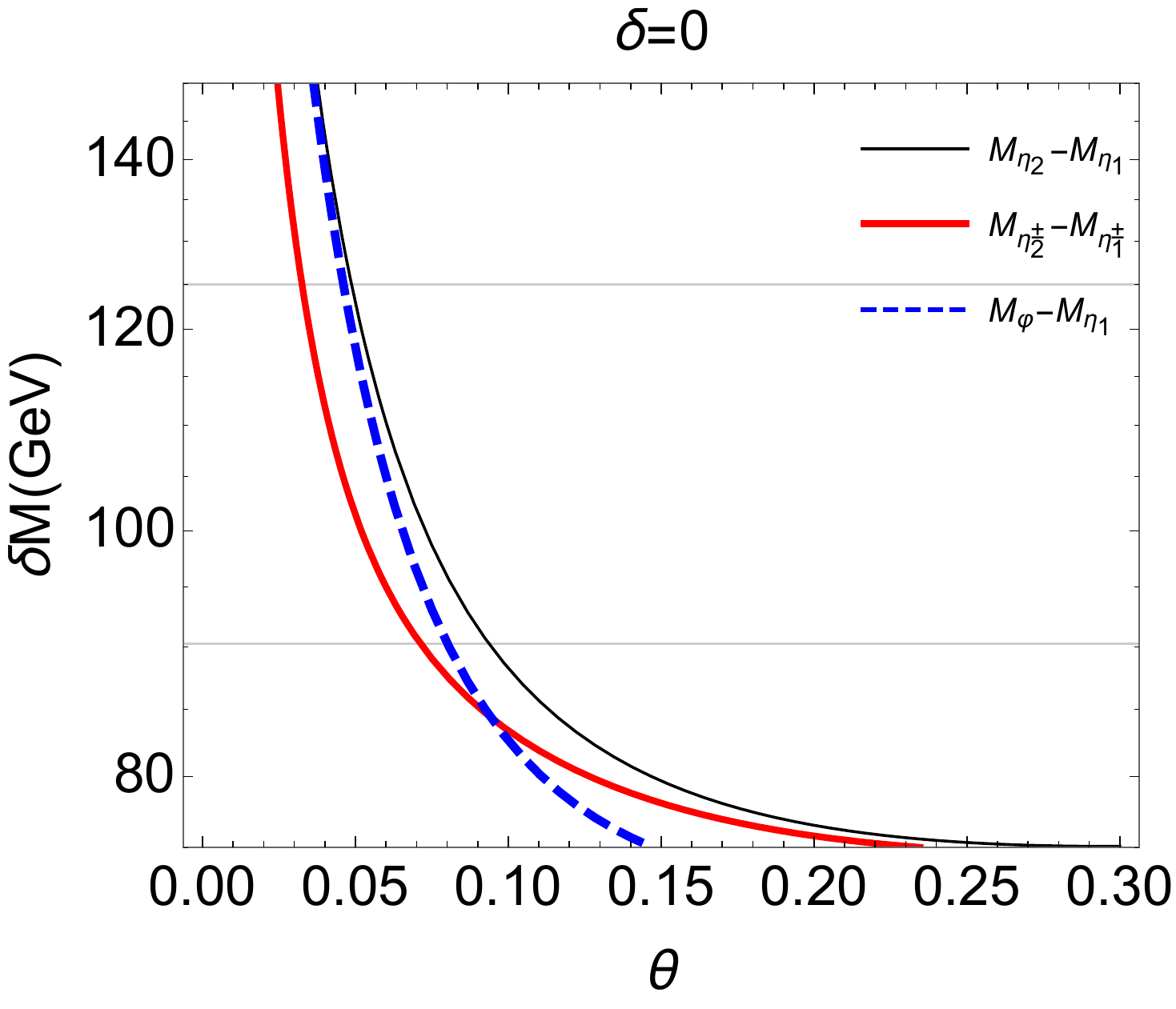}
\includegraphics[width=0.49\textwidth]{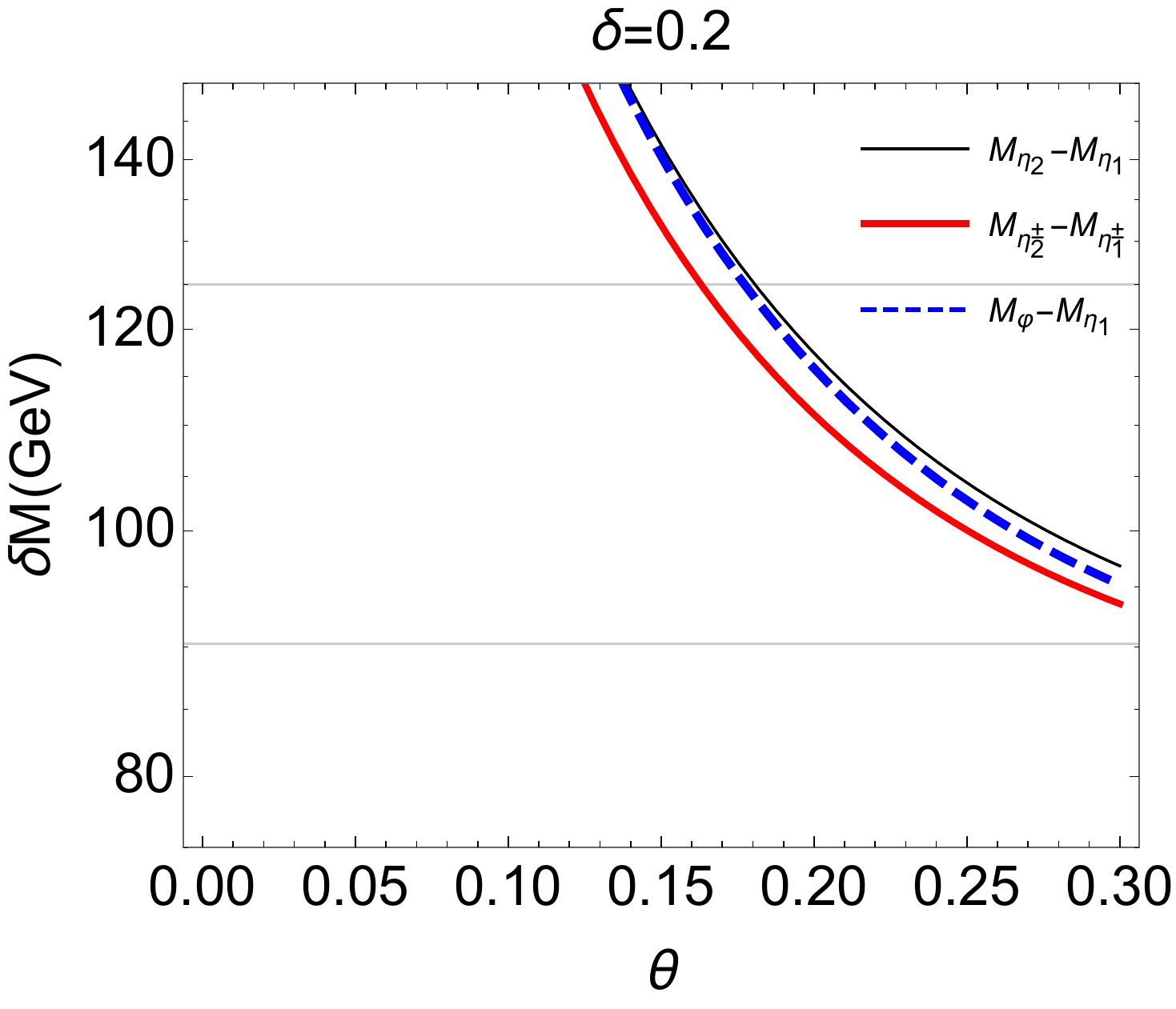}\\
\includegraphics[width=0.49\textwidth]{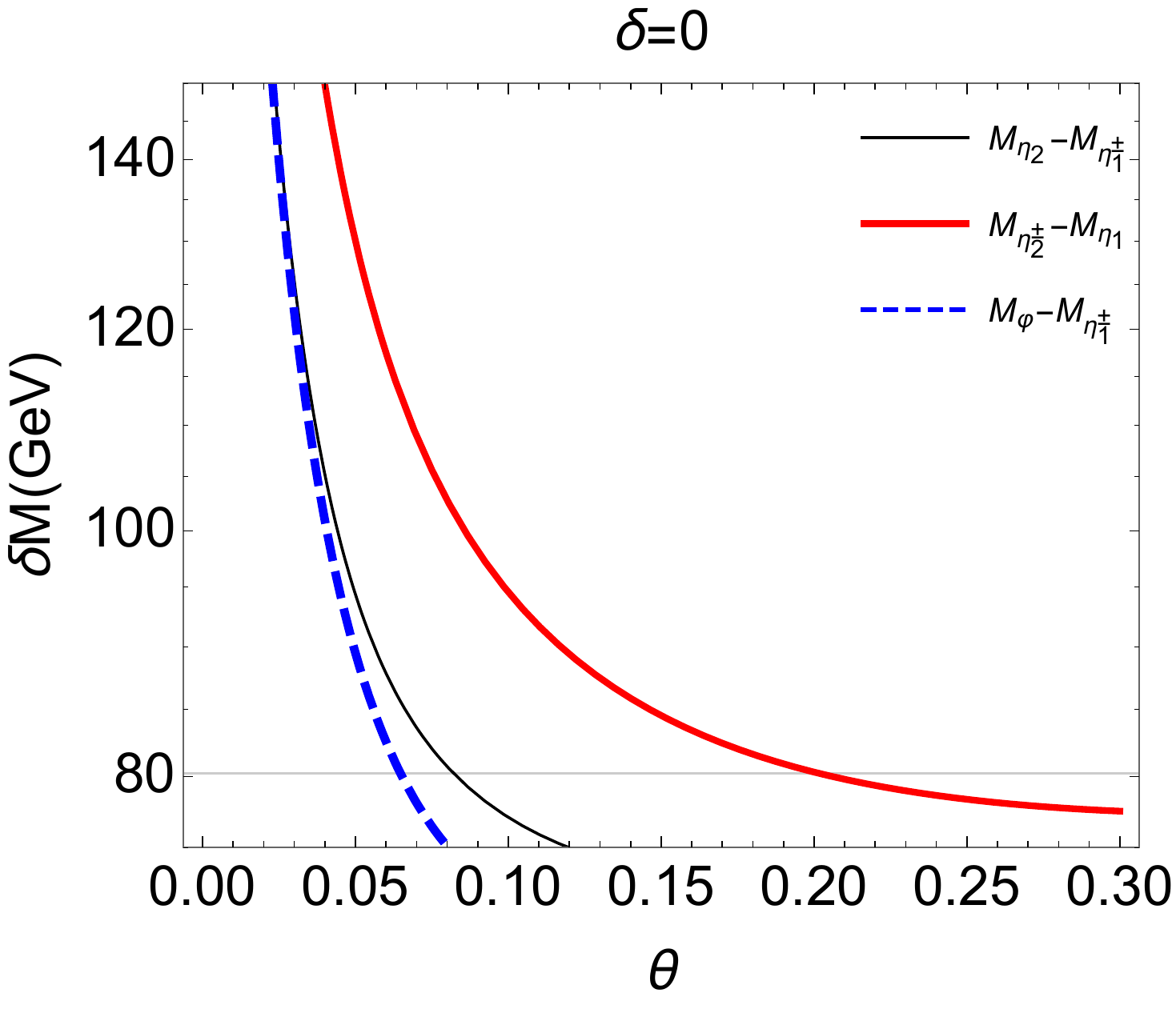} 
\includegraphics[width=0.49\textwidth]{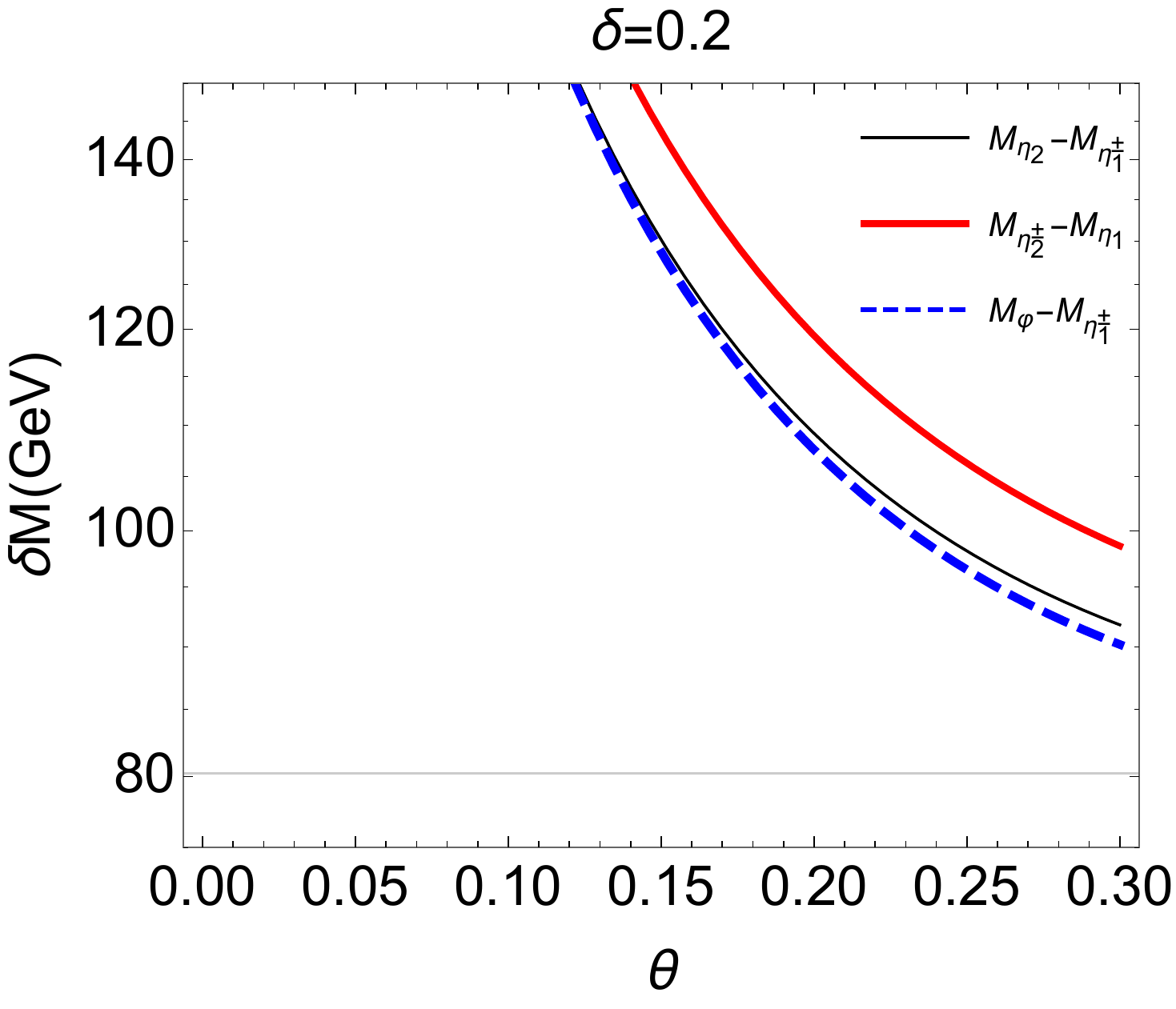}
 \end{center}
\caption{ Mass differences between the DM-odd pNGBs of the tier 2 and tier 1, for  $\delta=0$ (left column) and $\delta=0.2$ (right column). In the plots at the top row, the two grey lines correspond to the masses of Higgs and  $Z_\mu$ respectively. In the plots at the bottom row, the grey line corresponds to the mass of $W^\pm _\mu$ bosons. } \label{fig:mass_splitting}
\end{figure}

\begin{itemize}

\item[(A)] The two lightest states are $\eta_1$ and $\eta_1^\pm$, roughly corresponding to the $SU(2)_R$ triplet $N$ (for $\delta \geq 0$).  As the mass splitting is numerically very small, decays via a $W^\pm$ to the lightest state are kinematically forbidden, thus the only decays take place via a virtual gauge boson to a pair of light quarks or leptons. The branching ratios are thus independent on $\delta$:
\beq
\mbox{BR} (\eta_1^\pm \to \eta_1 j j) \simeq 65\%\,, \quad \mbox{BR} (\eta_1^\pm \to \eta_1 l^\pm \nu) \simeq 35\%\,.
\eeq
The width is very small, with values below $1$ keV.

\item[(B)] The second tier of states approximately form the second doublet: $\eta_2$, $\eta_2^\pm$ and the scalar $\varphi$. Due to the small mass splitting between them, they preferentially decay to a state of the lighter group plus a SM boson, $W$, $Z$ and Higgs.  The channels with a neutral boson are also constrained by CP invariance, so that the channels $\eta_2 \to Z\ \eta_1$ and $\varphi \to h\ \eta_1$ are forbidden. We also observe that the decays into the neutral bosons tend to be smaller than the decays into a $W$, and decrease until they vanish at increasing $\theta$. This effect can be understood in terms of the mass differences between states in this group and the lightest ones, shown in Figure~\ref{fig:mass_splitting}. For instance, we see that the channel $\eta_2 \to h\ \eta_1$ is kinematically close for $\theta \gtrsim 0.05$ for $\delta = 0$, because the mass splitting decreases below the $h$ mass.  From the right column of Fig.~\ref{fig:mass_splitting} we also see that the mass differences tend to increase for $\delta = 0.2$, thus pushing the kinematic closing of the channels to higher values of $\theta$. Interestingly, the mass differences never drop below the $W$ mass, so that decays via charged current are always open and dominate for large $\theta$ (i.e., smaller pNGB masses).

\begin{figure}[!htb]
\begin{center}
\includegraphics[width=0.41\textwidth]{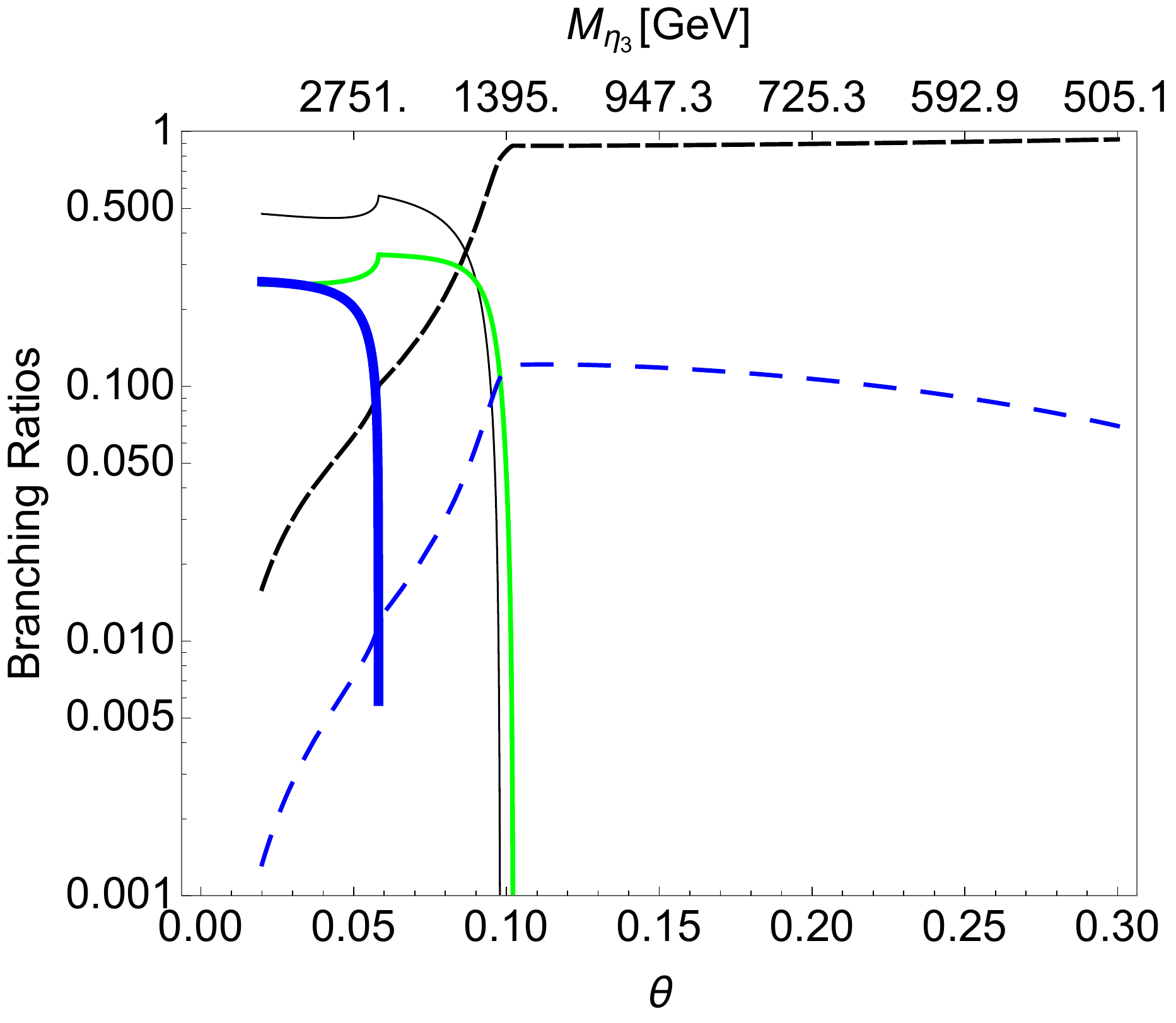}
\includegraphics[width=0.58\textwidth]{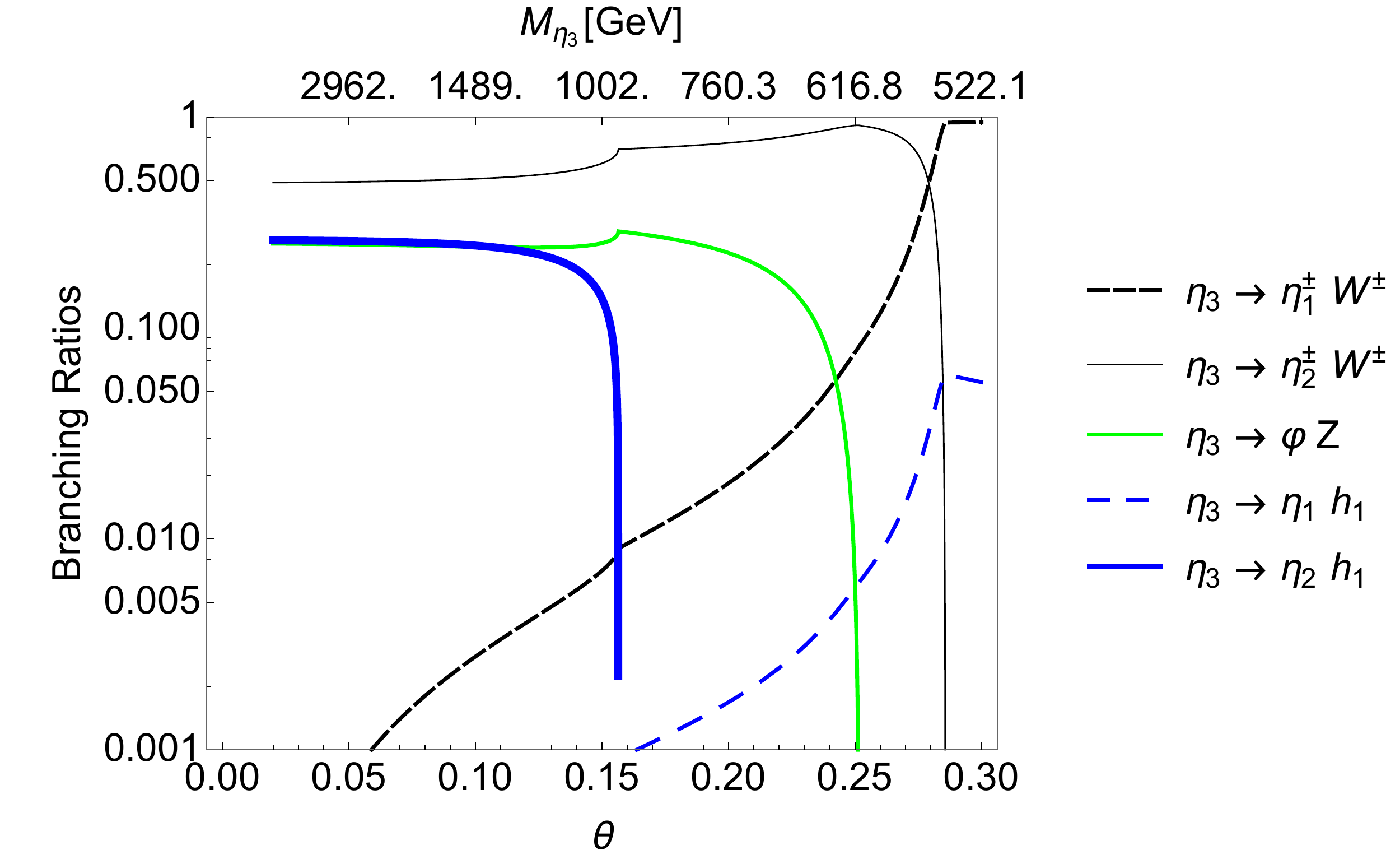}\\
\includegraphics[width=0.41\textwidth]{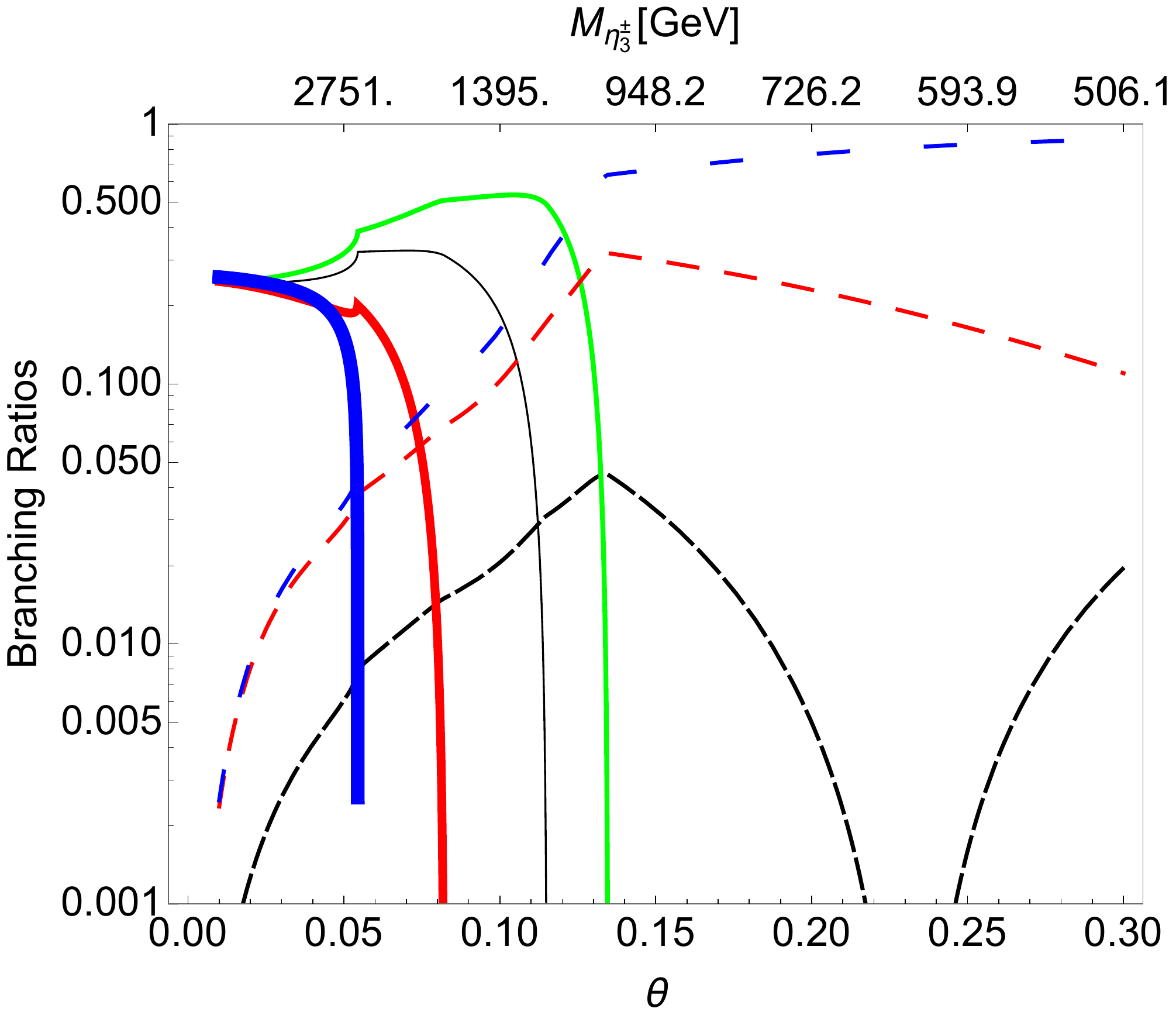}
\includegraphics[width=0.58\textwidth]{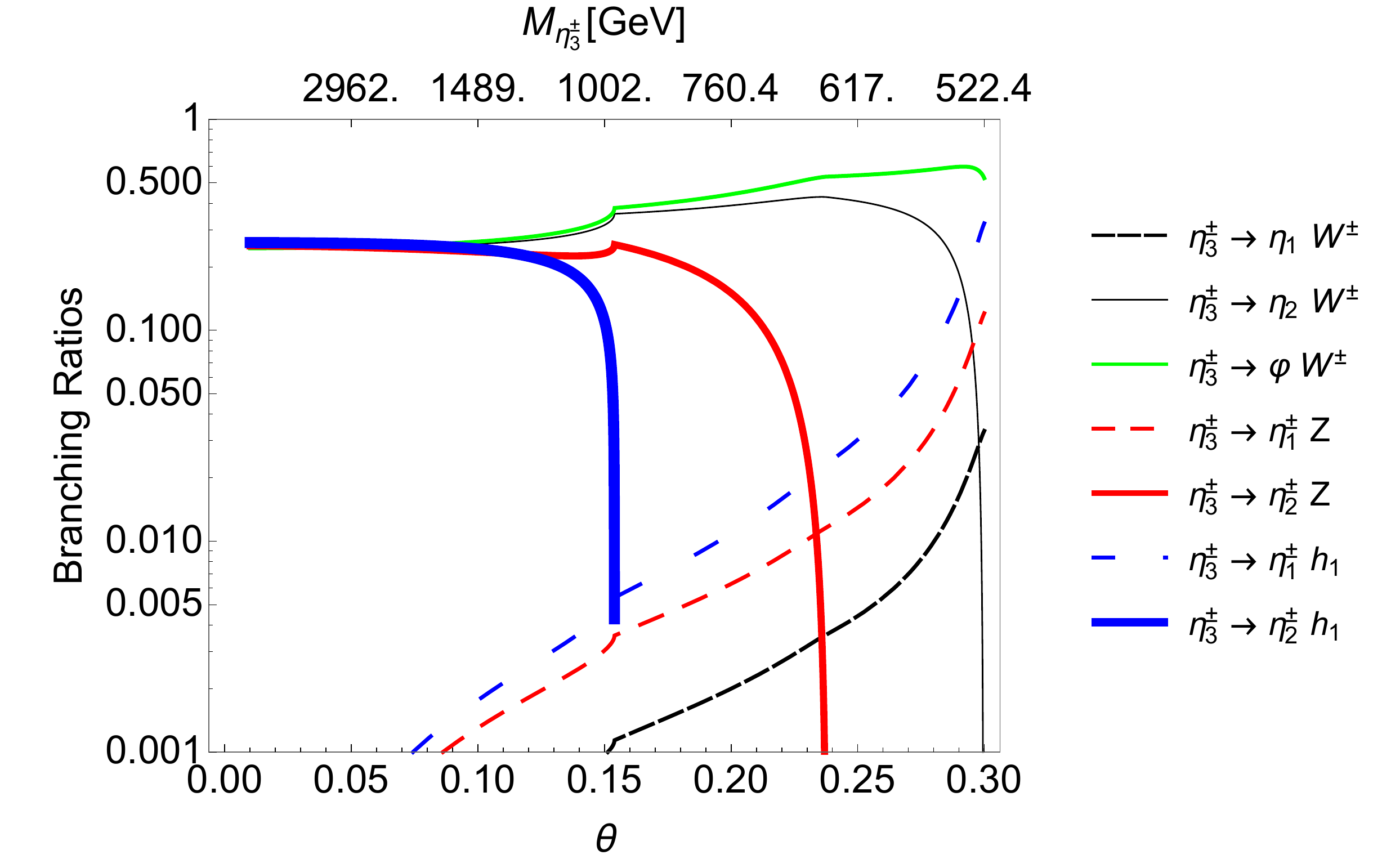}
 \end{center}
\caption{Branching Ratios of $\eta_3$ (top) and $\eta_3^\pm$ (bottom) for $\delta = 0$ (left) and $\delta = 0.2$ (right). } \label{fig:brDelta}
\end{figure}

\item[(C)] The most massive tier is mostly made of the $SU(2)_L$ triplet: $\eta_3$ and $\eta_3^\pm$. They can decay both to tier 2 and tier 1 states via appropriate EW bosons.
The Branching Ratios as a function of $\theta$ are shown in Figure~\ref{fig:brDelta}. 
The peculiar behaviour can, again, be understood in terms of mass differences. For small $\theta$, where the mass differences tend to be large, the preferred channels are to tier 2 states, due to the larger gauge couplings. For increasing $\theta$, the mass differences are reduced so that all channels into tier 2 states kinematically close, and the tier 2 states can only decay to the lightest tier 1 pNGBs. We recall that the patterns of decays to the neutral bosons $Z$ and $h$ depend on the CP properties of the scalars.

\end{itemize}

%%%%%%%%%%%%%%%%%%%%%%%%%%%%%
\subsubsection{Production rates at the LHC}

The odd pNGBs can only be pair produced at the LHC, and decay down to the lightest stable state which thus produces events with missing transverse energy ($E_T^{\rm miss}$). The main production modes are listed below:
\begin{itemize}
\item[] {\bf Vector Boson Fusion} (VBF): $q q' \to \pi_i \pi_j + 2 j$, via gauge interactions and s-channel Higgs exchange (singlet $s$ exchange provides subleading corrections).
\item[] {\bf Associated Production}:  $q\bar{q}' \to \pi_i \pi_j Z/W$, via gauge interactions.
\item[] {\bf Gluon Fusion}: $g g \to \pi_i \pi_j$, via top loops and Higgs s-channel exchange. This channel depends on $Y_{t0}$ (see Table~\ref{table:topyukawa} in Appendix~\ref{app:Yukawa}).
\item[] {\bf Drell-Yan}: $q \bar{q} \to \pi_i \pi_j$, via Gauge  interactions.
\item[] {\bf Associated top production}: $g g \to t \bar{t} \pi_i \pi_j$, via the top Yukawa. This channel is expected to be small because of the production of associated massive quarks.
\end{itemize}

\begin{figure}[!htb]
\begin{center}
\includegraphics[width=0.7\textwidth]{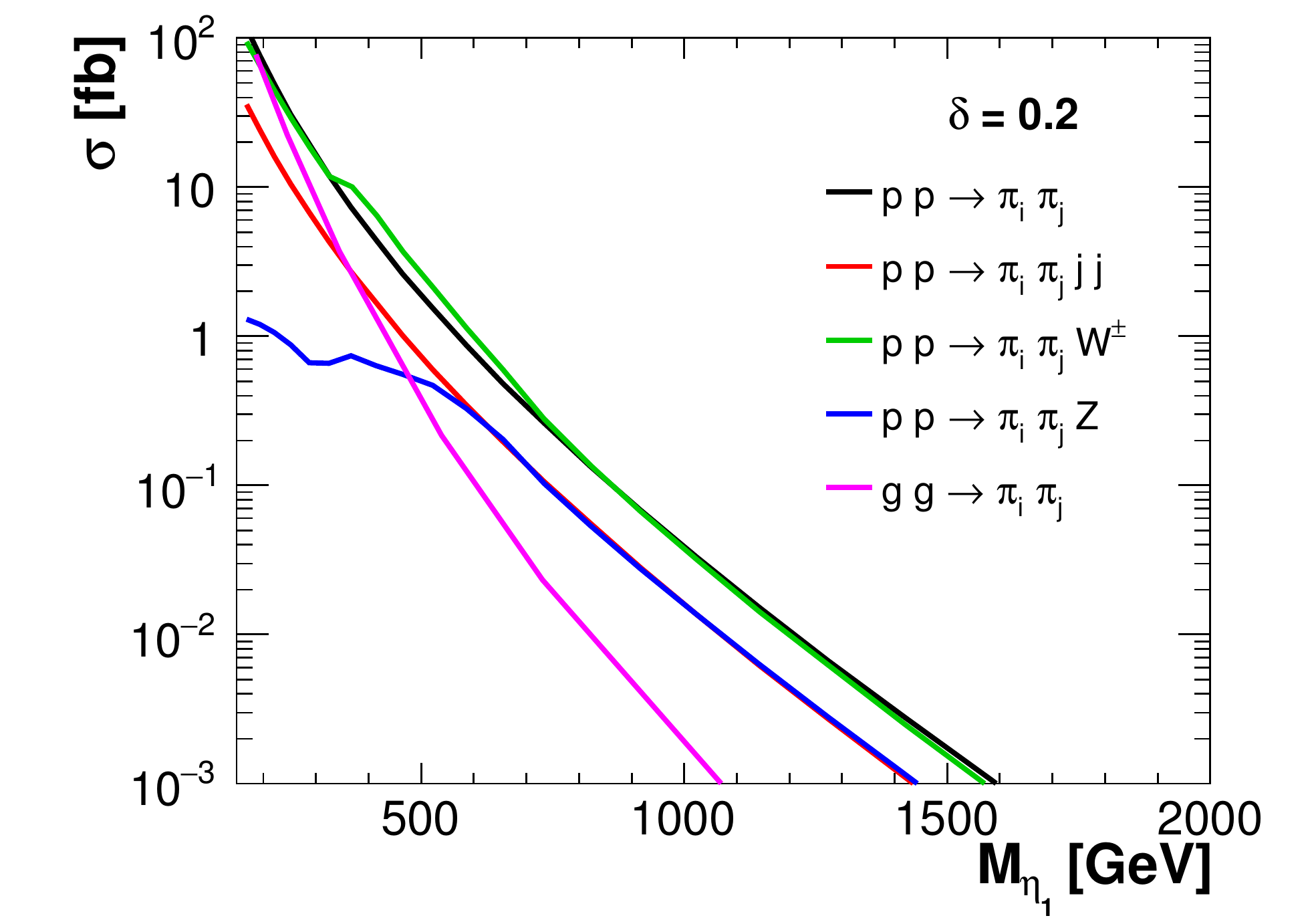}
\end{center}
\caption{Inclusive production cross sections  of  DM--odd pNGBs $\pi_i$ at the LHC with centre of mass energy of $14$ TeV,  as a function of  the  mass  for  five  different  channels. We applied a general cut on the transverse momentum of jets of  $p_T ^{jet} >20$ GeV. The dominant production channels are due to Drell-Yan and associated production with a $W$ boson.  } \label{fig:inclusive_cross_section}
\end{figure}

In Fig.~\ref{fig:inclusive_cross_section} we show the inclusive production channels for a pair of DM--odd pNGBs, in the approximation that the masses are degenerate. We can clearly see that the dominant production mode is always due to Drell-Yan and associated production via gauge interactions (in black and green respectively), while the cross section of all the other channels  is slightly below.
The cross sections are fairly large at low masses (corresponding to large $\theta$), providing rates between 100 fb and 10 fb for masses below $400$~GeV, nevertheless the sensitivity at the LHC crucially depends on the decay modes of the produced states. In most of the events, it is the heavier modes that are produced, as they have larger couplings to the SM gauge bosons, and only the decay products described in the previous sections will be observable at the LHC. The sensitive search channels will be similar to the ones employed in searches for supersymmetry, looking for the production of jets and leptons in association with large amounts of missing transverse energy.
As a complete analysis is beyond the scope of the present work, we decided to focus on two particular channels that provide more DM related signatures: production of one DM candidate $\eta_1$ in association of one other pNGB, and mono-jet signatures. The former class will generate events with large \met in association with the SM decay product of the heavier state, while the latter relies on radiation jets.

\subsection{The mono--$X$ + \met searches at the LHC}

 Having discussed the production rates of the DM-odd states, we can turn our attention to current searches at the LHC experiments: typical DM searches look for production of a single SM particle (mono--$X$) in association with large \met. In this model, there are 4 mono--$X$ signatures, where the $X$ can be a jet, a $W$, a $Z$ or a Higgs boson. We discuss each channel in detail below.

\begin{figure}[!htb]
\begin{center}
\includegraphics[width=0.49\textwidth]{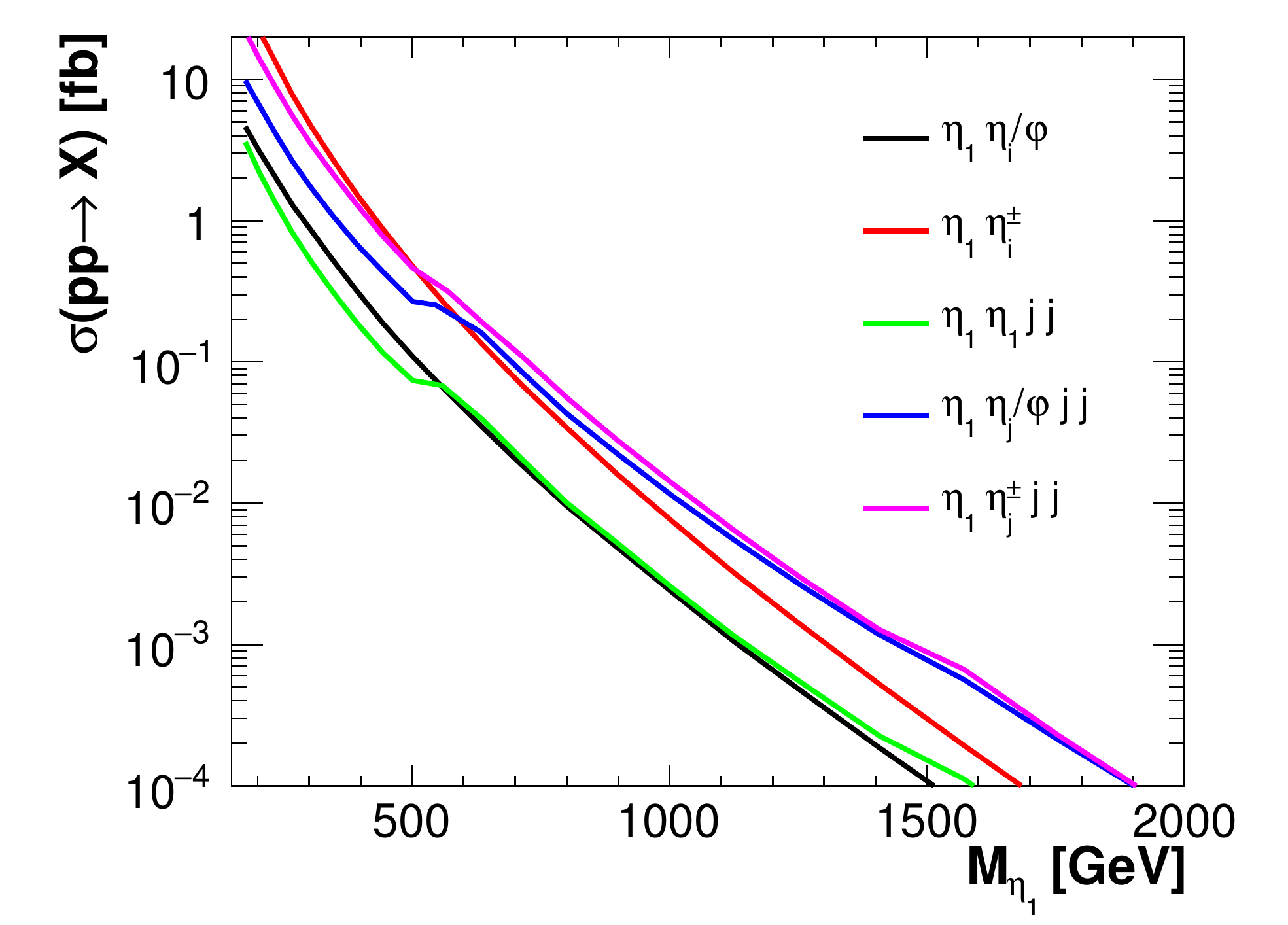}
\includegraphics[width=0.49\textwidth]{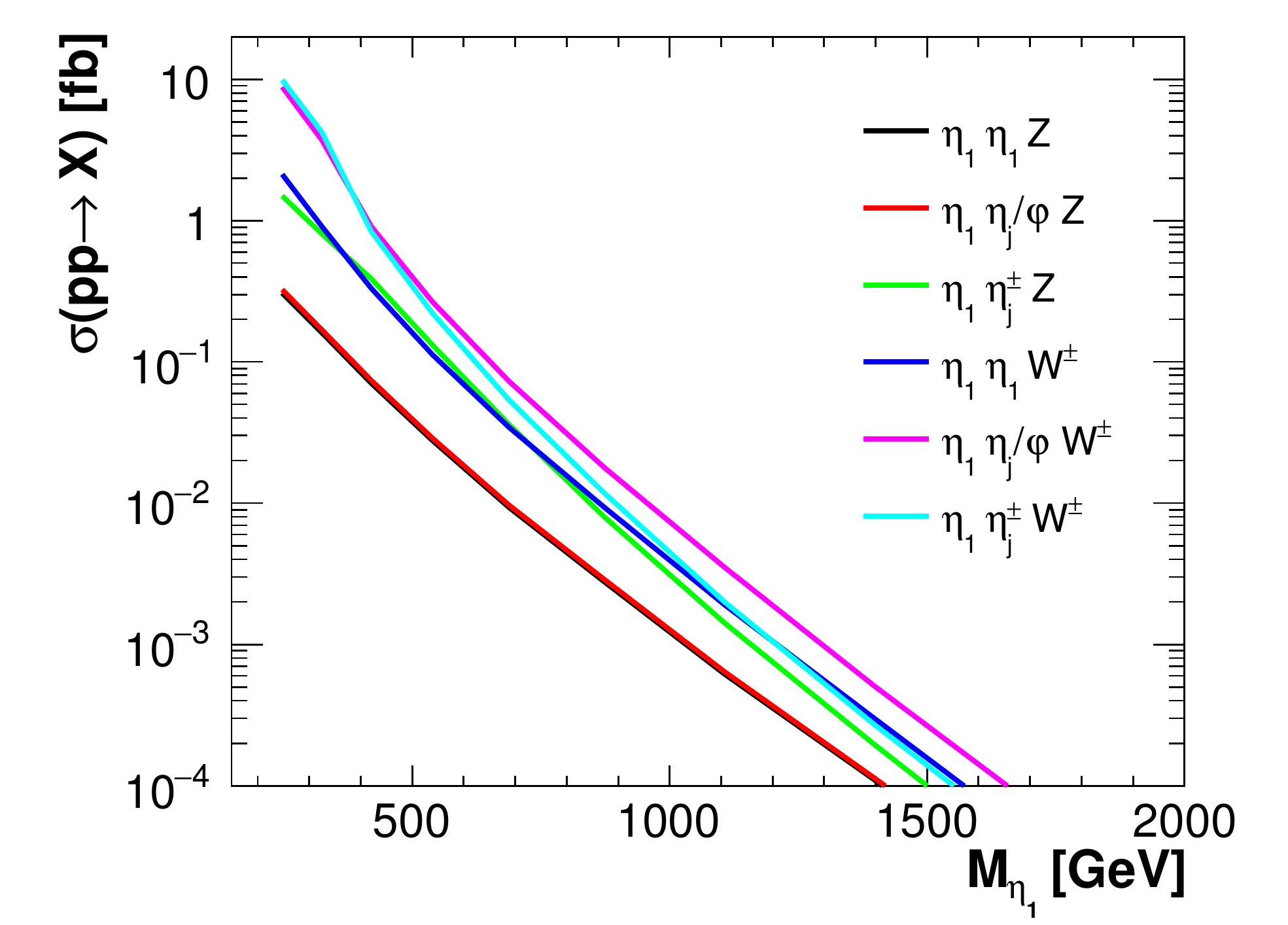}
\end{center}
\caption{ Production cross section  of  $\eta_1$ at  the LHC with  centre of mass energy of $14 $ TeV  as  a  function of  the  mass  for  different  channels.  We applied a general cut on the transverse momentum of jets of  $p_T ^{jet} >20$ GeV.} \label{fig:pr}
\end{figure}

\begin{itemize}

\item {\bf The mono-jet} signature has been widely used in the search for DM at both ATLAS~\cite{ATLAS-Monojet} and CMS~\cite{CMS-Monojet}: for the scalar DM case, the jet is radiated from the initial state in Drell-Yan and gluon fusion production.
In the model under study the production rates for $pp \to \eta_1 \eta_1 j$ are too small  compared to the experimental bounds: for instance, the $8$ TeV ATLAS search poses a bound of $3.4$ fb for $E_T ^{\rm miss}  \geq 700$ GeV~\cite{ATLAS-Monojet}. 
However, as it can be seen in Fig.~\ref{fig:pr}, the parton-level cross section for $pp \to \eta_1 \eta_1 jj$ with the requirement $p^{jet}_T > 20$ GeV is about 3 fb when the mass of $\eta_1$ is about 200 GeV (the cross section for $pp\to\eta_1\eta_1j$ is two orders of magnitude smaller than this one). This cross section is already smaller than the upper limit imposed by ATLAS, not to mention the stringent \met cuts they employ. We can thus conclude that mono-jet searches should be ineffective in this model set-up.

\item {\bf The mono-W/Z} signature can be obtained through Drell-Yan production of the pNGBs plus a gauge boson, or via decays of the heavier pNGB into the DM candidate (if the mass splitting is large enough).  ATLAS and CMS also have searched for dark matter in these channel~\cite{ATLAS-MonoWZ,CMS-MonoWZ}. For the parton level fiducial regions defined as $p_T^{W/Z}>250$ GeV, $|\eta^{W/Z}|<1.2$, $p_T^{\eta_1}>350\ (500)$ GeV, the upper limit on the fiducial cross section is $4.4\ (2.2)$ fb at 95\% C.L. However, we notice in Fig.~\ref{fig:pr} that the cross section of the Mono--W/Z processes without any cuts is below 4 fb when the mass of $\eta_1$ is around 200 GeV and even much smaller for larger mass region. Thus the Mono-W/Z still doesn't have any sensitivity in this model set-up.

\item  {\bf A mono-Higgs} signature can also be obtained when a Higgs boson $h$ is radiated by the DM-odd scalars
$\pi_i$.  However, the mono-H signal is very suppressed, giving cross sections smaller than $10^{-3}$ fb at 14 TeV LHC, thus beyond the  LHC capabilities. For reference, the searches at the LHC can be found in Refs.~\cite{ATLAS-MonoH-gaga-8T,ATLAS-MonoH-bb-8T,ATLAS-MonoH-bb-13T}.

\end{itemize}

\subsection{Associated production of DM}

 The DM candidate can also be produced in other channels that do not give rise to mono-X signatures: in this case, we will focus on the VBF and associated production channels with   final states containing two additional jets  as tags. 
A complete analysis is beyond the scope of this paper, due to the presence of a large number of possible final states: to give an example, we will focus on the production of a single DM particle, $\eta_1$, in association with a heavier pNGB.
The cross sections at 14 TeV are shown in Fig.~\ref{fig:pr}:  the VBF production and the associated production with a gauge boson, $W$ or $Z$,  are at the same order of magnitude. When $\eta_1$ is  heavier than about $400$ GeV, the cross section of all these channels is smaller than $1$ fb.

 To test the feasibility of the detection of these channels at the LHC, we selected 3 promising channels, organised in terms of the final state after decays, and compare each with the most important irreducible SM background.
In  Table~\ref{tab:LHC}, we show the cross section of the different  channels  and  their corresponding  SM  background   under different  \met  cuts and for  $\theta =0.2$.   The 3 channels are chosen as follows:

 \begin{table}[tt]
 \begin{center}
 \begin{tabular}{|c|c|c|c|c|}
\hline
\multicolumn{2}{|c|}{Channel} & \multicolumn{3}{c|}{Cross sections (fb)} \\ \hline
& $E_T ^{miss}  >$ & $50$ GeV & $100$ GeV &  $200$ GeV \\
\hline \hline
Channel $1$& $p p \to \eta_1  \eta_1 j j  $ &  $< 0.1$ & $ < 0.1 $ & $<0.1$  \\
\hline
SM BG & $j j \nu_l \bar{\nu}_l  $&  $4.29 \times 10^5 $ &$1.15 \times 10^5 $ & $1.41\times 10^4 $\\
\hline
\hline
Channel $2$ &$p p \to  \eta_1 \eta_1 j j l^{\pm} \nu_{l}   $ &  $<0.4$ & $ < 0.4$ & $<0.4$ \\
\hline
SM BG &$ j j l^{\pm} \nu_{l}  $ & $ 6.35 \times  10^5$ & $9.62\times 10^4 $  & $ 8.39 \times 10^3 $   \\
\hline
\hline
Channel $3$ & $  p p \to  \eta_1 \eta_1 j j l^\pm l^\mp $ &$<10^{-3}$ & $< 10^{-3} $ & $<10^{-3} $ \\
\hline
SM BG & $ j j  l^\pm l^\mp(Z) \nu_l \bar{\nu}_l $ &  $4.386 \times 10 $ & $2.10 \times10 $ & 5.14    \\
\hline
\end{tabular}
\caption{ Cross section for the three chosen channels with different \met cuts, for $\theta=0.2$. The main irreducible backgrounds (SM BG) are also reported.}\label{tab:LHC}
 \end{center}
\end{table}

\begin{itemize}

\item[1:] In this channel we consider production of the DM candidate in association with jets. There  are  three classes of processes:  jets from the decays of $W^\pm$/$Z$ bosons in $p p \to \eta_1 \eta_1 W/Z  \to  \eta_1 \eta_1 j j $ and $p p \to \eta_1 \eta_i/\eta_i ^\pm /\varphi  \to  \eta_1 \eta_1 j j $, and VBF production $p p  \to  \eta_1 \eta_1 j j $. The dominant background is $p p \to  j j Z  \to  j j  \nu \bar{\nu}$.  In Table~\ref{tab:LHC}, we  can see that  the  background  is many orders of magnitude above the signal, and that it is not effectively suppressed by the \met cuts. Additional cuts on the jets may be employed, like for instance the invariant mass reconstructing the mass of the $W$/$Z$ bosons to select the first process, or tagging forward jets to select the VBF production. A more dedicated study, including a more complete assessment of the background, would be needed to establish if the background can be effectively suppressed.

\item[2:]  This channel focuses on a leptonic $W$ produced with additional jets, and receives contributions from VBF production in $p p \to \eta_1 \eta^\pm _i   j j   \to \eta_1 \eta_1 j j l^\pm \nu$ and associated production in $p p \to \eta_1 \eta^\pm _i Z(\eta_1 \eta_i/\varphi W)   \to \eta_1 \eta_1 j j l^\pm \nu $. The irreducible background  of  this channel  mainly  comes from production of a single $W$ with jets, $p p \to  j j   W^\pm \to  j j    l^\pm  \nu$.   Like for Channel 1,  the  background is many orders of magnitude above  the signal, see Table~\ref{tab:LHC}, so that \met cut alone is not effective.

\item[3:] VBF and associated production can also produce leptons via the $Z$, in $p p \to  \eta_1 \eta_i/\varphi j j \to  \eta_1 \eta_1 j j l^+ l^- $ and $p p \to  \eta_1 \eta_i/\varphi  Z/W (\eta_1 \eta_i ^\pm Z) \to  \eta_1 \eta_1 j j l^+ l^- $ .  As the lepton pair comes  from a $Z$  boson,  the  main background  comes  from  $p p \to j j  Z Z   \to  j j l^+ l^- \nu \bar{\nu} $.  The  signal is  very  small because of the  leptonic  decay  of the $ Z$ and does not feature cross sections reachable at the LHC.

\end{itemize}

 We also checked production with additional jets, that may arise from hadronically decaying $W$/$Z$ bosons in
$p p \to  \eta_1 \eta_i/\varphi/\eta_i^\pm  j j  \to \eta_1 \eta_1  j j j j $ and $p p \to  \eta_1 \eta_i/\varphi/\eta_i^\pm  W/Z  \to \eta_1 \eta_1  j j j j $, where cross sections of several fb can be achieved thanks to the large hadronic branching ratios. However, the leading irreducible background $p p \to  j j j j Z  \to  j j j j \nu_l \bar{ \nu} _l$ is still overwhelming with cross sections of $10^5$ fb.

 The very simple analysis we performed here shows that the detection of \met signatures in this model from the direct production of the DM-odd pNGBs is very challenging, as \met cuts typically do not reduce the background enough to enhance the small signal cross sections.
 The main reason behind this is that the mass splitting between heavier scalars and the DM candidate is fairly small, so the decay products are typically soft, thus leading to small \met in the signal events.
 More dedicated searches in specific channels may have some hope, however, and we leave this exploration to further work.

%%%%%%%%%%%%%%%%%%%%%%%%%%%%%%%%%
\section{Relic density constraint on pNGB dark matter} \label{sec:relic}

As  discussed in  previous sections,  the lightest neutral composite scalar $\eta_1 $   is  stable under  the DM parity,  thus  it might be a candidate for annihilating thermal DM. In this section we will check if the correct relic density can be achieved within the available parameter space, and the constraints from Direct and Indirect DM detection experiments.
We will focus on positive values of the fermion mass splittings, $\delta\geq 0$, where the lightest state is mostly a gauge singlet. We expect, in fact, that direct detection bounds will be weaker in this region, furthermore it will be easier to characterize the general properties of this composite DM candidate. The more complex case $\delta < 0$ will be analyzed in a future work.

%%%%%%%%%%%%%%%%%%%%%%%
\subsection{Relic density}

The calculation of the relic density was carefully studied in~\cite{Wolfram:1978gp,Griest:1990kh}, and here we will simply recap the main ingredients of the calculation. We will stay within an approximation where analytical results can be obtained~\cite{Kolb:1990vq}, and consider the full co-annihilation processes between the odd pNGBs: this step is needed as the mass differences are small and they can be close or smaller than the typical freeze-out temperature.  The rate equation for annihilating DM is
\beq \label{eq:rate1}
\frac{dn}{dt} + 3Hn = - \langle \sigma ^{eff} v_{rel }  \rangle (n^2 - n^2_{eq})
\eeq
where $n = \sum_a  n (\pi_a)$ is the total number density of odd  scalar particles, i.e. $\pi_a = (\eta_i, \eta_i^\pm, \varphi)$ with $ i=1,2,3$, $n_{eq} =\sum_a n_{eq} (\pi_a) $ is the total number density that odd particles  would have in thermal equilibrium, and $H = \dot{R}/R$ is the Hubble constant (with $R$ being the scale factor). The assumption behind this equation is that all odd pNGBs freeze out at the same temperature, and the unstable ones decay into the DM particle promptly after freeze out. In this approximation, the DM relic density counts the total number density of all odd species at freeze--out. 

The averaged cross section, including co-annihilation effects, can be expressed as
\beq
\langle \sigma ^{eff} v_{rel }\rangle = \langle\sigma_{ab} v_{rel} ^{ab}\rangle\  \frac{n_{eq} (\pi_a) n_{eq} (\pi_b) }{n_{eq}^2 }\,,
\eeq
 where $\langle\sigma_{ab} v_{rel} ^{ab}\rangle$  is the velocity-averaged co-annihilation cross section for  $ \pi_a  \pi_b  \to  X X' $ where the final states includes any SM states and particles decaying into them, like the DM--even singlet $s$. In the fundamental composite 2HDM under consideration, the odd pNGBs  $\pi_a$ can annihilate into a pair of bosons (gauge vectors, Higgs and the singlet $s$)  or fermions:
\begin{multline}
 \sigma_{ab}  v_{rel} ^{ab}  =  \langle \sigma v \rangle_{\pi_a \pi_b \to V V } +   \langle \sigma v \rangle_{\pi_a \pi_b \to V h_1} +  \langle \sigma v \rangle_{\pi_a \pi_b \to V s}  +  \langle \sigma v \rangle_{\pi_a \pi_b \to f \bar{f}}  +
  \\
 \langle \sigma v \rangle_{\pi_a \pi_b \to h_1  h_1}  +  \langle \sigma v \rangle_{\pi_a \pi_b \to s s} + \langle \sigma v \rangle_{\pi_a \pi_b \to h_1  s}\,.
\end{multline}
The cross sections can be easily computed provided the relevant couplings (given in Appendix~\ref{appapp}).
  The cross  section of annihilation  with the singlet  $s$  is expected to be small as  the couplings  are small and  the  final states of  other annihilating  processes  are  lighter  compared to  the mass of the odd pNGBs.
 Furthermore,  as the colliding heavy particles are expected to be non-relativistic at the time of freeze-out, $\langle \sigma^{eff} v_{rel} \rangle$ can  be Taylor expanded  as $\langle \sigma^{eff} v_{rel} \rangle \simeq a^{eff} + b^{eff} \langle v^2 \rangle$ with  good  accuracy: in the following, we  only keep the s-wave term $a^{eff}$ for simplicity.

It is customary to rewrite the rate equation in terms of new variables, by dividing the number density by the entropy density of the Universe $\mathcal{S}$, and defining the variable $\mathcal{Y} = n/\mathcal{S}$.  Furthermore, the temperature can be introduced via a variable normalized by the mass of the DM candidate, $x = m_{\eta_1}/T$. Other relations can be obtained by use of the standard Friedmann-Robertson-Walker Cosmology, where the Hubble constant can be related to the energy density $\rho$ by $H=(\frac{8}{3} \pi G \rho)^{1/2}$, $G$ being the Newton constant. Combining all these ingredients, the rate equation (\ref{eq:rate1}) can be transformed to
\beq \label{eq:Density}
\frac{d\mathcal{Y}}{dx} = - \left( \frac{45}{\pi} G \right)^{-1/2} \frac{g^{1/2}_{\ast} m_{\eta_1}}{x^2} \langle \sigma^{eff} v_{rel} \rangle (\mathcal{Y}^2 - \mathcal{Y}^2_{eq})\,,
\eeq
where we have used the following relations between the entropy and energy densities and the temperature:
\beq
\mathcal{S} = h_{eff}(T) \frac{2 \pi^2}{45} T^3, \qquad \rho = g_{eff}(T) \frac{\pi^2}{30} T^4\,,
\eeq
with $h_{eff}$ and $g_{eff}$ being the effective degrees of freedom for entropy and energy densities. The parameter $g^{1/2}_{\ast}$, counting the degrees of freedom at temperature $T$, is defined as
\beq
g^{1/2}_{\ast} = \frac{h_{eff}}{g^{1/2}_{eff}} \left( 1 + \frac{1}{3} \frac{T}{h_{eff}} \frac{d h_{eff}}{d T} \right)\,.
\eeq
Before freeze-out, we work within the approximation $\Delta = \mathcal{Y} - \mathcal{Y}_{eq} = c \mathcal{Y}_{eq}$, with $c$ being a given constant, and neglect $d \Delta / dx \sim 0$ . 
Within these conditions, we get the identities
\beq
\mathcal{Y} = (1+c) \mathcal{Y}_{eq}\,,  \quad   \frac{d\mathcal{Y}}{dx} =  \frac{d\mathcal{Y}_{eq}}{dx}\,,
\eeq
which allow to simplify the rate equation.
 Substituting above identities into Eq.~(\ref{eq:Density}), we can find the freeze-out temperature $x_f = m_{\eta_1}/T_f$  and density $\mathcal{Y}_f$ by numerically solving the differential equation
\beq
\left( \frac{45}{\pi} G \right)^{-1/2} \frac{g^{1/2}_{\ast} m_{\eta_1}}{x^2} \langle \sigma^{eff} v_{rel} \rangle \mathcal{Y}_{eq}\ c (c + 2) = -\frac{d ln \mathcal{Y}_{eq}}{dx}\,,
\eeq
The numerical constant $c$ can be chosen by comparing the approximate solution to a full numerical solution of the differential equations and, in our numerical solutions, we fix the constant $c = 1.5$ following Ref.~\cite{Gondolo:1990dk}.
After freeze-out, we can neglect $\mathcal{Y}_{eq}$ and integrate the transformed rate equation~(\ref{eq:Density}) from the freeze-out temperature $T_f$ down to today's temperature $T_0 = 2.73K$:
\beq
\frac{1}{\mathcal{Y}_0} = \frac{1}{\mathcal{Y}_{f}} + \left( \frac{45}{\pi} G \right)^{-1/2} \int^{T_f}_{T_0} g^{1/2}_{\ast} \langle \sigma^{eff} v_{rel} \rangle d T\,.
\eeq
The odd particles are non-relativistic both at and after freeze-out, so they obey the Maxwell-Boltzmann statistics with $n_{eq} =\sum_a (\frac{m_{\pi_a} T}{2 \pi})^{3/2} e^{-m_{\pi_a}/T}$. We can then compute the relic density $\Omega h^2 = \rho_0 h^2/ \rho_c = m_{\eta_1} \mathcal{S}_0 \mathcal{Y}_0 h^2 / \rho_c$ knowing $\mathcal{Y}_0$ and the critical density $\rho_c = 3 H^2 / 8 \pi G$. The result is
\beq
\Omega h^2 = 2.83 \times 10^8\ \frac{m_{\eta_1}}{\rm GeV} \mathcal{Y}_0\,.
\eeq

\begin{figure}[!htb]
\begin{center}
\includegraphics[width=0.49\textwidth]{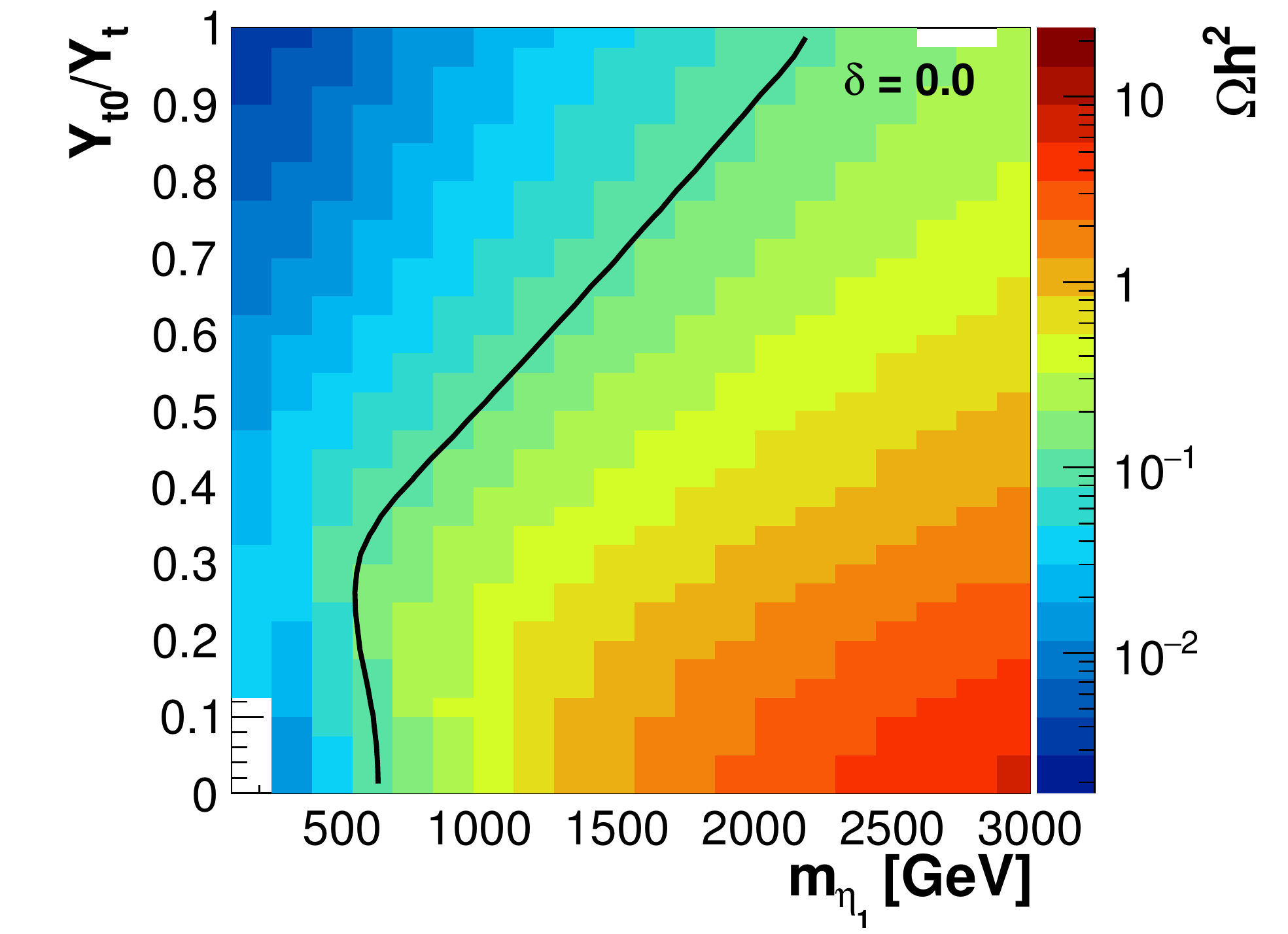}
\includegraphics[width=0.49\textwidth]{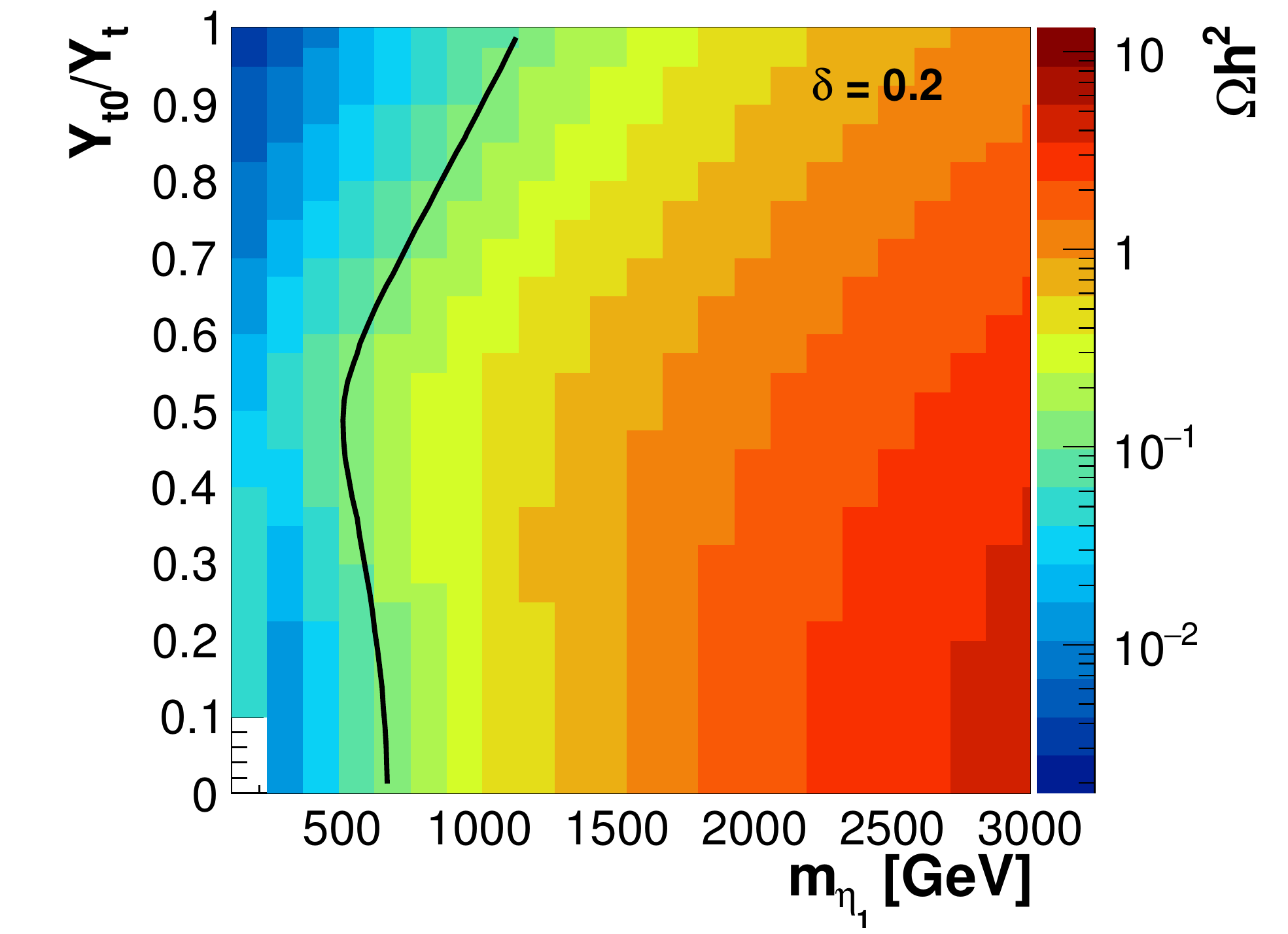}
 \end{center}
\caption{Plot of the thermal relic density $\Omega h^2$ of the DM candidate $\eta_1 $  in  the plane of its mass and $Y_{t0}/Y_t $ for $\delta =0$ (left) and $\delta =0.2$ (right). The  black  lines correspond to the observed value of DM relic density $\Omega h^2 = 0.1198$. The plots are symmetric under change of sign of $Y_{t0}$, so we only show positive values. } \label{fig: DMrelicdensity }
\end{figure}

\begin{figure}[!htb]
\begin{center}
\includegraphics[width=0.49\textwidth]{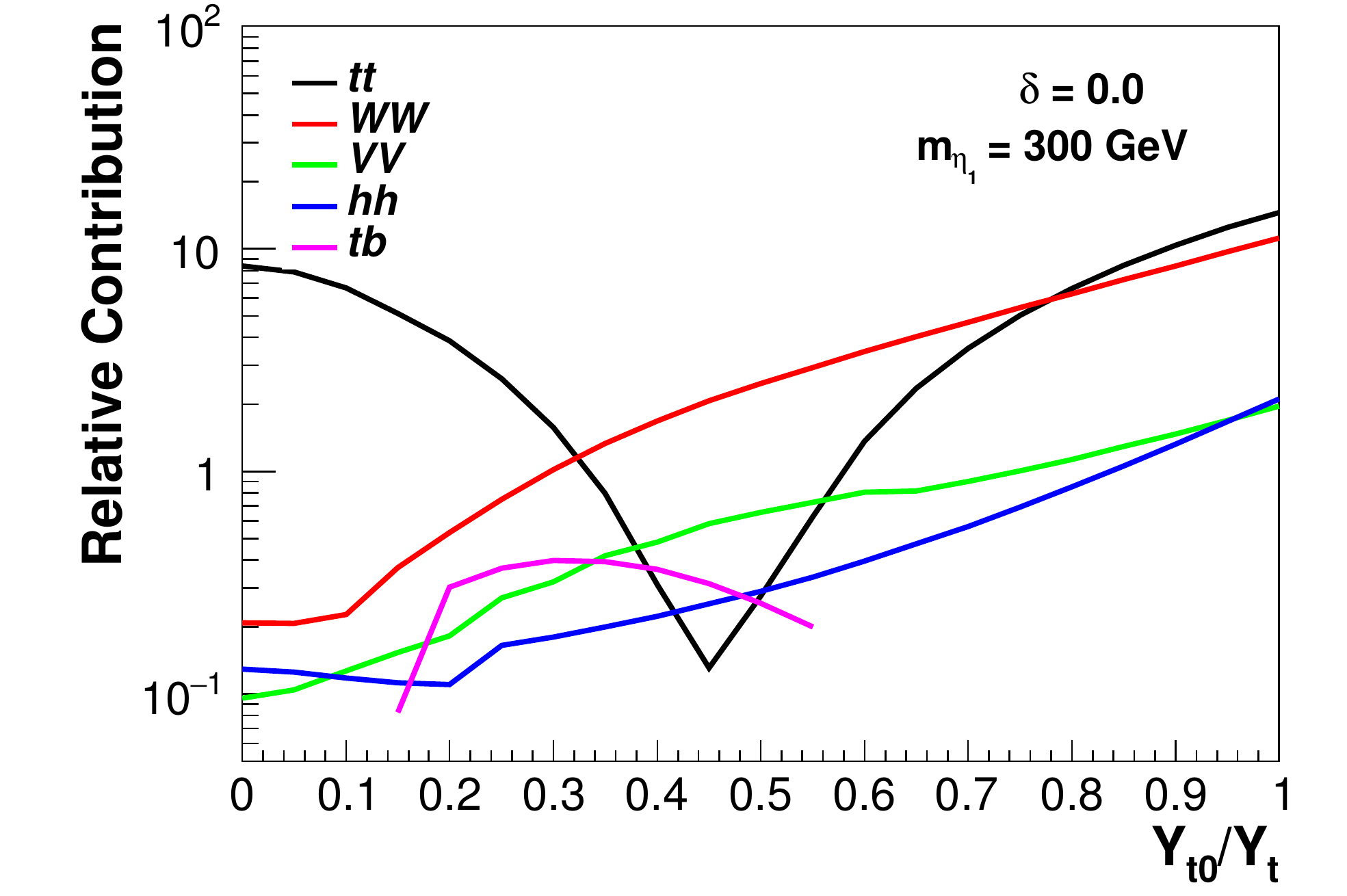}
\includegraphics[width=0.49\textwidth]{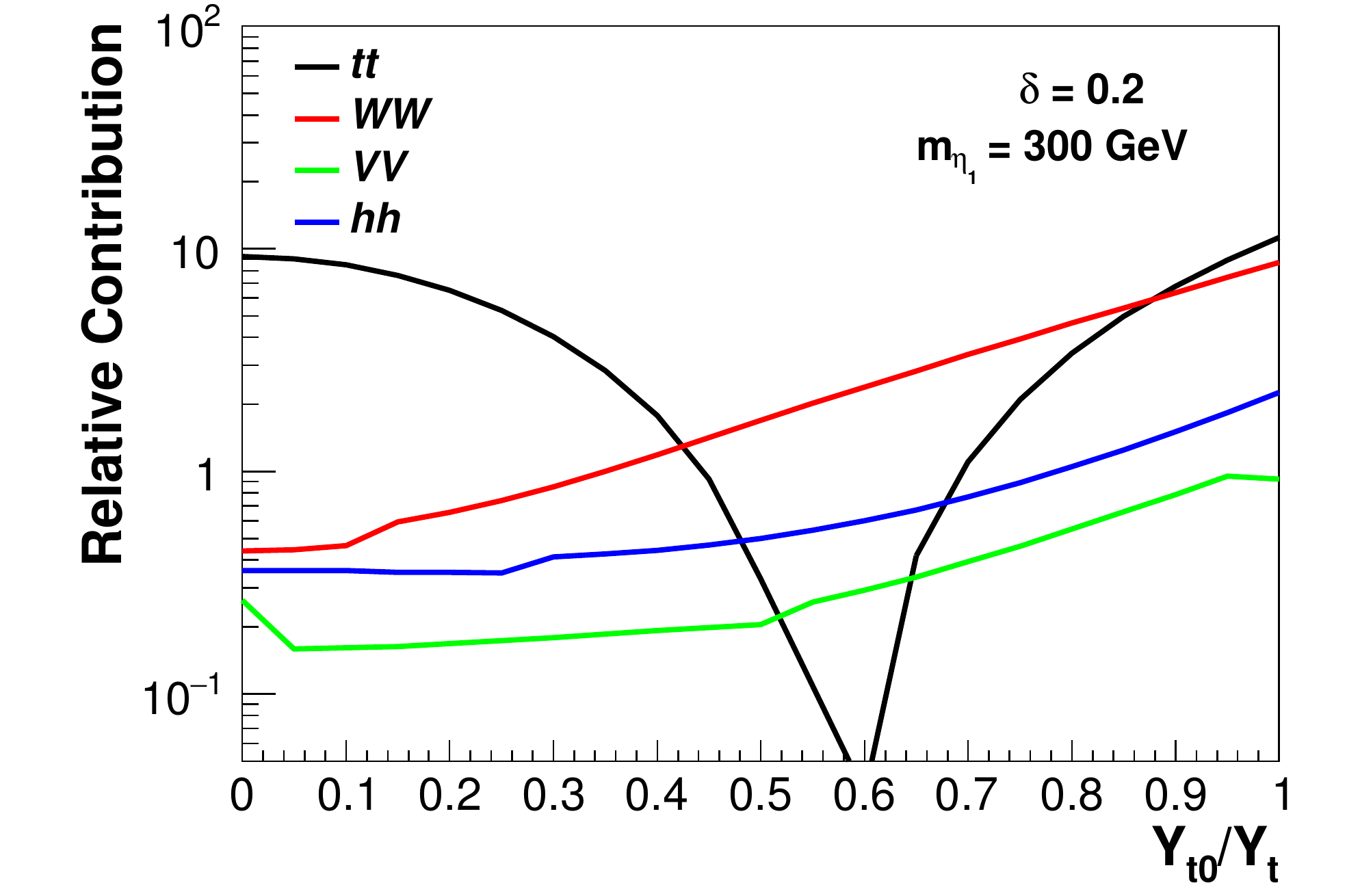}\\
\includegraphics[width=0.49\textwidth]{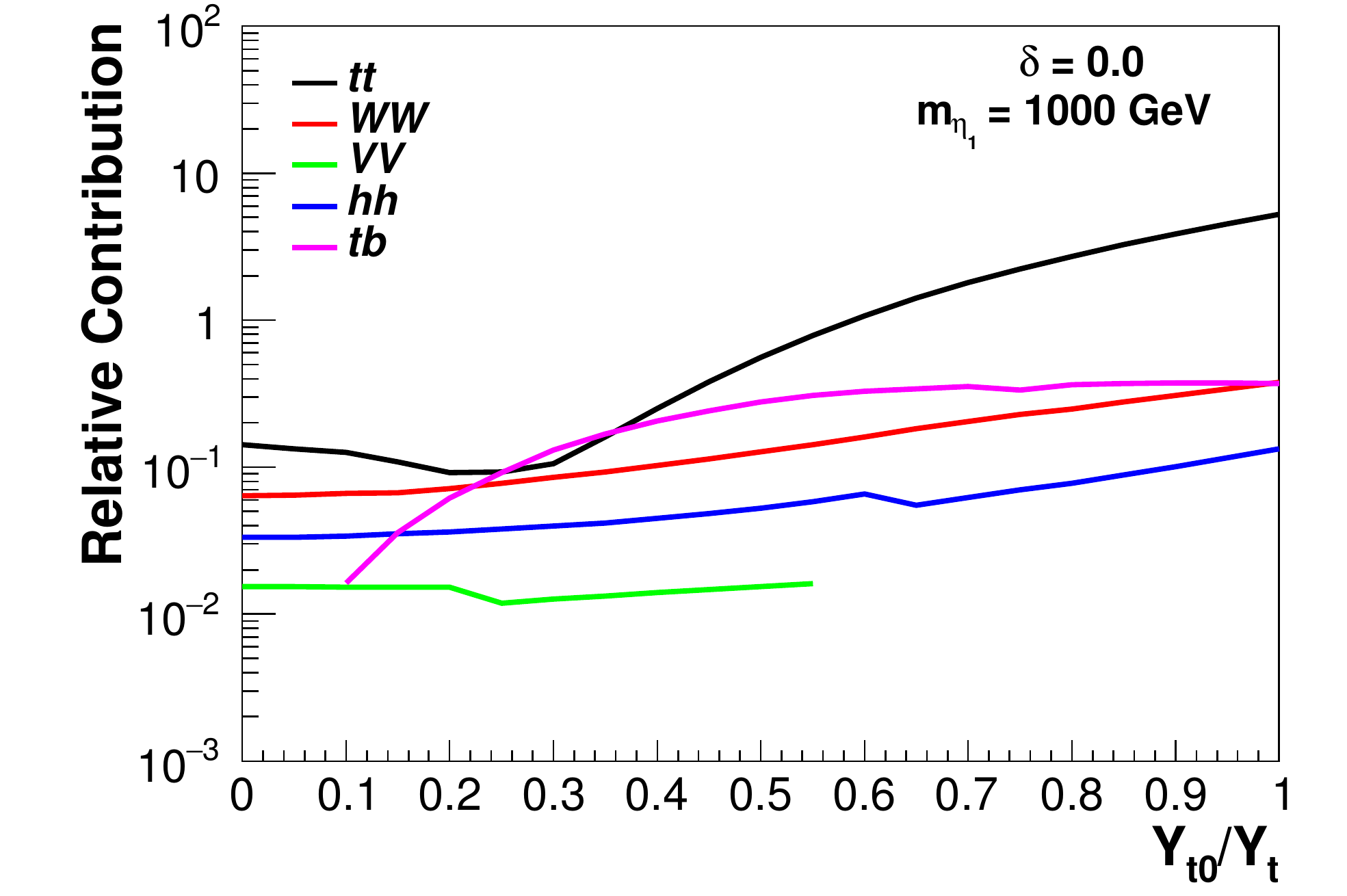}
\includegraphics[width=0.49\textwidth]{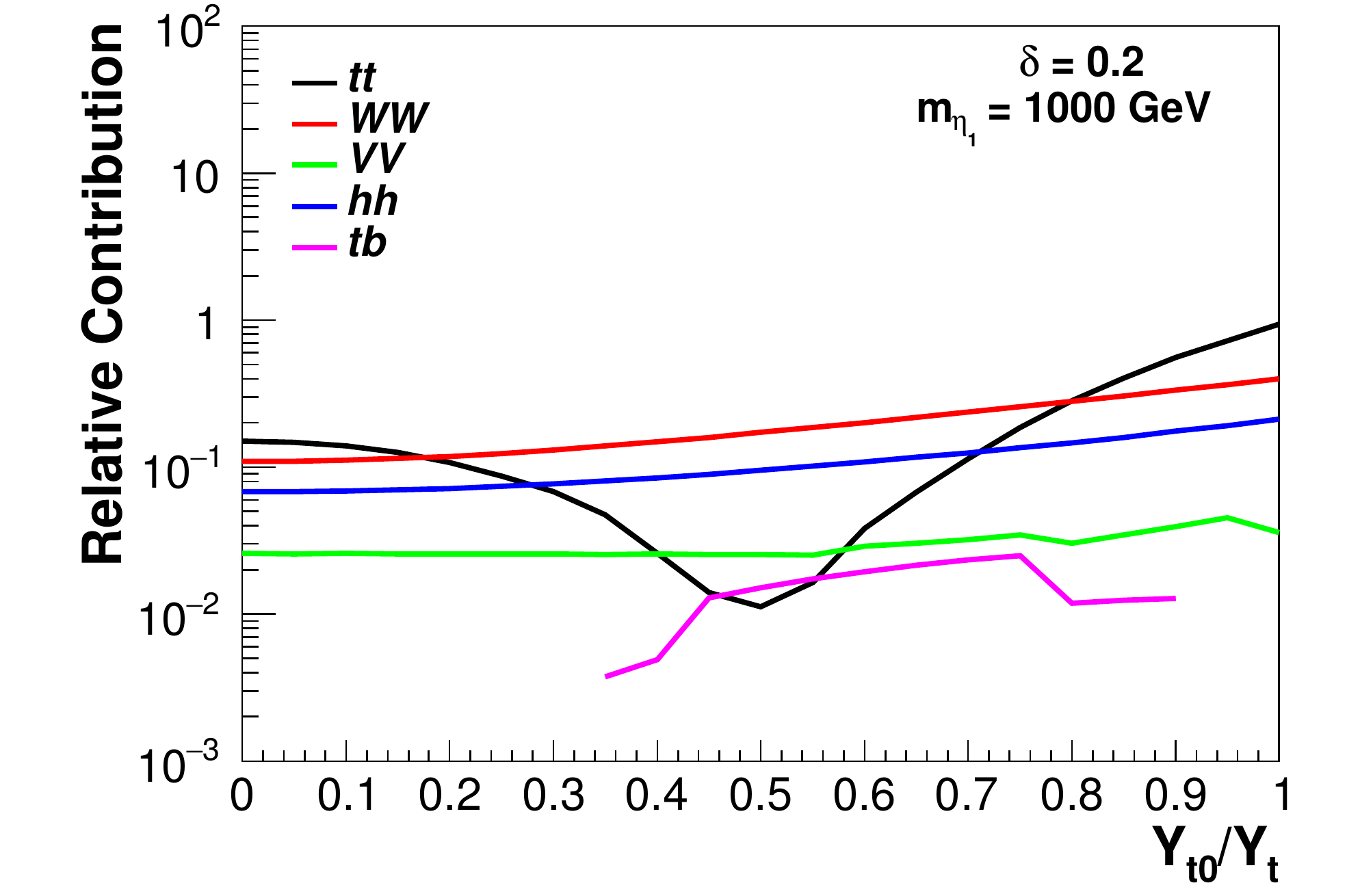}\\
\includegraphics[width=0.49\textwidth]{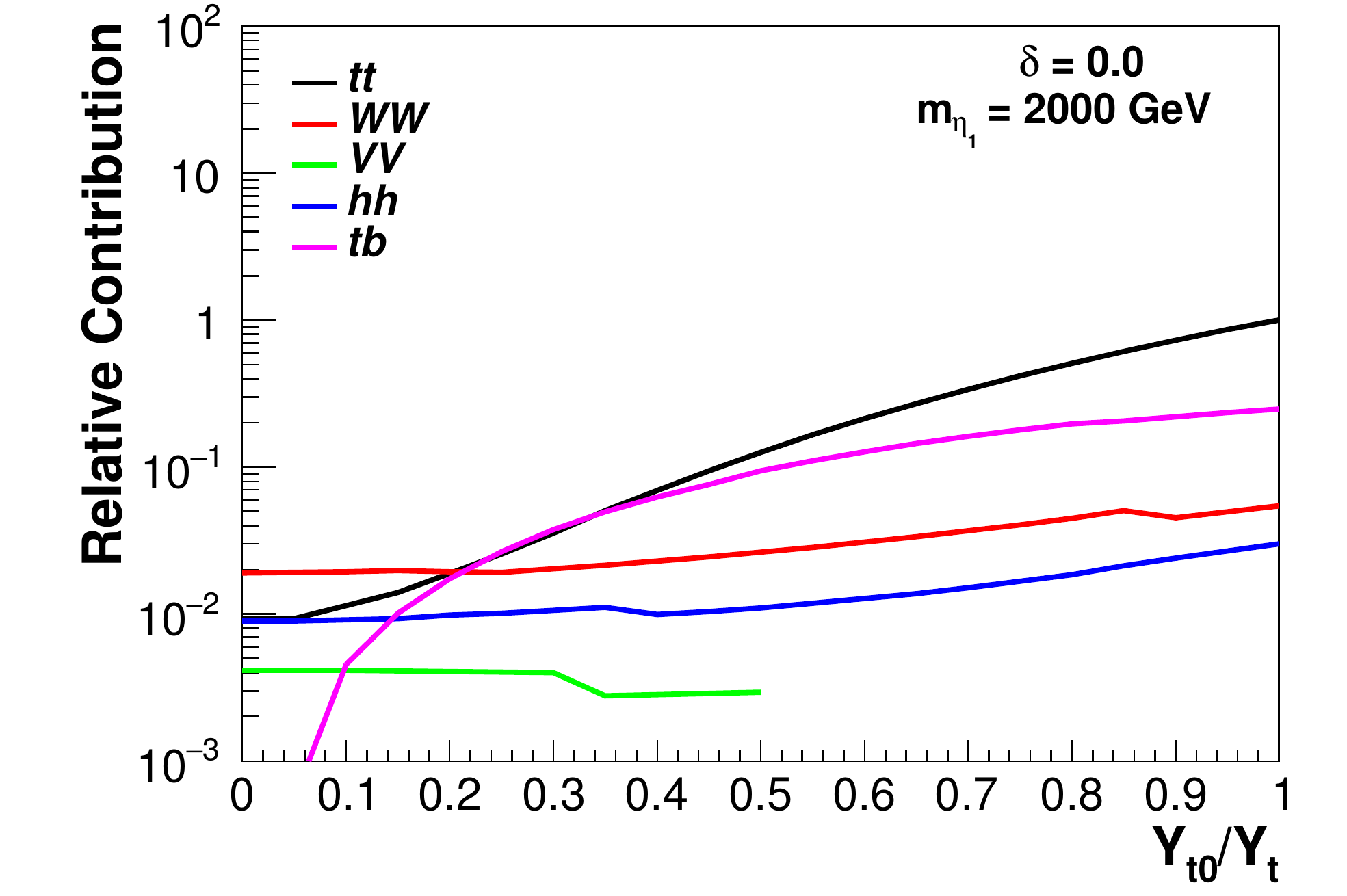}
\includegraphics[width=0.49\textwidth]{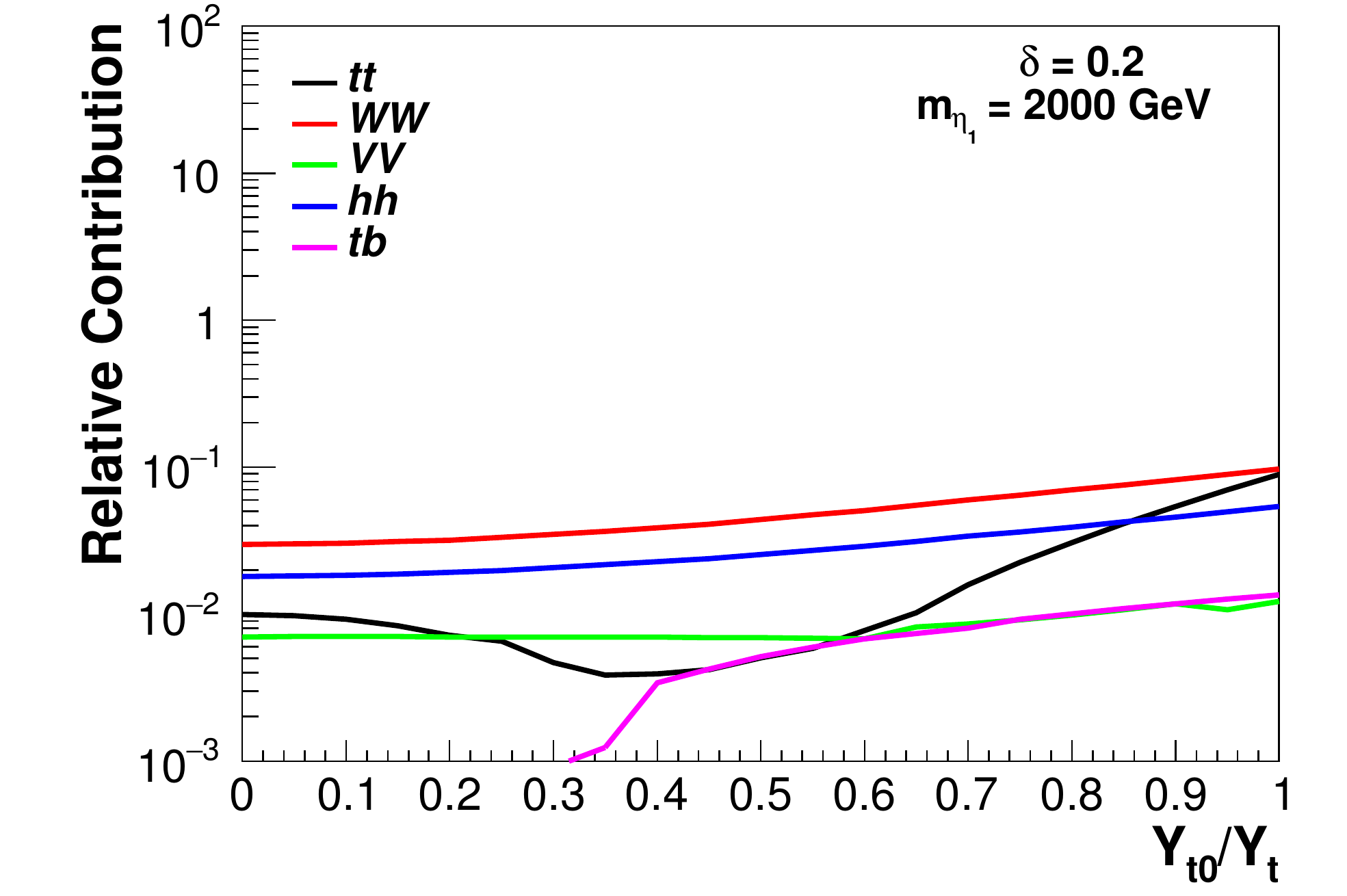}\\
 \end{center}
\caption{
Contribution of the main (co-)annihilation channels to the inverse relic abundance $1/\Omega_{\eta_1}$, which is roughly proportional to the average cross section,  as a function of $Y_{t0}/Y_{t}$ for different choice of the mass.  The numerical values are normalised to the observed value. All possible channels are categorised according to their final states ($VV$ collects all the di-vector channels except for $WW$).}
 \label{fig:anniSigma}
\end{figure}

In Figure~\ref{fig: DMrelicdensity } we show the values of the thermal relic density $\Omega_{\eta_1} h^2$ in the plane of $Y_{t0}/Y_t$ and the mass of the DM candidate $\eta_1$.   The  black  line  is  the   observed value  of  DM  relic density in the  universe as measured by Planck, $\Omega h^2 = 0.1198 \pm0.0015$~\cite{Ade:2015xua}. We recall that larger values of the relic density are excluded as they would lead to overclosure of the Universe, while smaller values are allowed if other DM candidates are present. 
From Figure~\ref{fig: DMrelicdensity } we see that  $\eta_1$ can saturate the needed relic abundance if its  mass is heavier than few hundreds GeV, with  values ranging from $500$ to a few  TeV in most of the parameter region. The pre-Yukawa $Y_{t0}$ parameterises the  mixing between the two triplets and the second doublet: if  $Y_{t0} =0 $,  the lightest state $\eta_1$ is the neutral component of the SU(2)$_R$ triplet, i.e $N_0$. We also observe that for large positive $\delta$, the mixing is suppressed, thus leading to larger relic density and lower DM masses. 
The dominant annihilation channels of $\eta_1$ are $WW$ ($hh$) and $t\bar t$ via mixing, while for small $\delta$ (i.e., small mass splitting) co-annihilation in $tb$ is also relevant.
% at low mass range $m_{\eta_1} <1$ TeV, while at high mass range $m_{\eta_1} \ge1$ TeV, the dominant channels are $tb$ and $t \bar t$. For $\delta =0.2$, the dominant channels are also $WW$ and $t\bar t$ at mass range $m_{\eta_1} <1$ TeV. 
The annihilation cross section into $t \bar t$ decreases at high mass faster than the $WW$ and $hh$ ones, hence these two channels are the dominant one at large $\eta_1$ mass. The behaviour of the main (co-)annihilation cross sections are illustrated in Figure~\ref{fig:anniSigma}, where we plot the inverse relic abundance normalized to the measured DM relic abundance, i.e. $\Omega_{DM}/\Omega_{\eta_1}$, without considering all the other channels. This quantity is roughly proportional to the average cross section in each channel, and we plot it as a function of $Y_{t0}/Y_t$ for various values of $m_{\eta_1}$. We find that the annihilation cross section into $WW$ (and other di-boson channels) increases with growing $Y_{t0}$, while the $t\bar{t}$ one decreases before growing again for large couplings.
Therefore, for small $Y_{t0}$ the top channel dominates over the $WW$ one, then becomes subdominant near the minimum value before dominating again for large $Y_{t0}$. From this figure, we also see that the cross sections of the dominant annihilation channels decrease monotonically with $m_{\eta_1}$. 
The behaviour of the dominant cross sections explains the observed dependency of the relic abundance on $Y_{t0}$ and $m_{\eta_1}$: we observe, therefore, that for fixed $Y_{t0}$ the relic abundance increases with the mass of the DM candidate,  while for fixed mass it increases with $Y_{t0}$ first and then decreases following the $t\bar{t}$ channel. A maximum is reached for $Y_{t0} \sim 0.5\; \cdot Y_t$ for $\delta =0.2$ and $Y_{t0} \sim 0.3 \; \cdot Y_t$ for $\delta =0$, which provides the smallest value of the DM state saturating the relic abundance at $m_{\eta_1} \sim 500$~GeV. 

We stop the plots at $Y_{t0} = Y_t$ because, as we will see in the next section, larger values of $Y_{t0}$ are excluded by Direct DM detection experiments. Furthermore, the cross sections are not sensitive to the sign of $Y_{t0}$, so that the results are symmetric under sign change. 
Remarkably, the mass range for the DM particle to saturate the relic abundance are in the few hundred GeV to 2 TeV, thus corresponding to moderate values of $\theta$ that do not require very high fine tuning in the pNGB potential.

Very large values of the DM mass are excluded by the overclosure of the Universe, however lighter states (which correspond to less fine tuning in the potential) are allowed once the pNGBs does not fulfill the DM relic density: it  is interesting that in our  model there  exists another kind  of DM,  i.e. the  lightest  Techni-baryon  which  is  protected by a conserved TB number  $ U(1)_{TB} $ and might be a second DM candidate.

 %%%%%%%%%%%%%%%%%%%%%%%%%%%%%%%%%
\subsection{Direct  detection  constraints} \label{sec:DD}

According  to the previous  discussion, we find that the scalar DM  should  not be  too heavy  in this model as  its  mass  upper  bound is  around  few  TeV  in  most   parameter space.   We will consider now bounds from Direct Detection experiments, which are sensitive to the scattering of the DM off nuclei. The lightest odd pNGB couples to a pair  of  quarks  via operators of the form:
\beq
\mathcal{L}_{\eta_1qq} =  \frac{a_q}{f} \eta_1 ^2 \bar{q} q   +  \frac{d_q m_q }{m_{h_1} ^2} \eta_1 ^2 \bar{q} q\,,  \label{eq:DMfermions}
\eeq
where   $a_q $ and $ d_q$ are constants  and  $m_q $ is the mass of the quark $q$. The first  term comes from  high  order  terms  in the expansion of the effective quark Yukawas and the coefficients $a_q$ are related to the couplings in Table~\ref{table:topyukawa} via the mass diagonalization.
The second term is generated  by  exchanging the Higgs (the singlet $s$ gives subleading corrections).
In both cases, the dominant contribution is proportional to the mass of the quark via its Yukawa coupling.
 These  couplings give rise to spin  independent  elastic cross section  $\sigma_{SI} $, which  may  be  potentially  within the reach of  present and future Direct DM search experiments.   Note  that, on this case,  the  spin-dependent cross section $\sigma_{SD}$ is  always much  smaller than  $\sigma_{SI}$,  so  in the following discussion  we  only  consider  $ \sigma_{SI} $.  The spin-independent elastic scattering  cross section  of  $\eta_1  $   off  a nucleus can  be  parameterised as
 \beq
 \sigma_{SI} =\frac{4}{\pi m_{\eta_1} ^2} (\frac{m_{\eta_1} m_n   }{m_{\eta_1}  +m_n  })^2 \frac{[Z f_p +(A-Z)f_n   ]^2}{A^2}
 \eeq
where  $m_n $ is  the neutron mass, $Z $ and $A-Z $ are  the number of  protons and neutrons in  the nucleus and  $f_p  $ ($f_n$ )  describes the coupling between $\eta_1 $  and  protons (neutrons):
\beq
f_{n,p} =\sum_{q=u,d,s} f_{T_q} ^{n,p} c_q \frac{m_{n,p}}{m_q} + \frac{2}{27} f_{T_G}  ^{n,p} \sum_{q=c,b,t} c_q\frac{m_{n,p}}{m_q}  \,,
\eeq
where $c_q = a_q/f + d_q m_q/m_{h_1}^2$ from Eq.~(\ref{eq:DMfermions}).
The  hadron matrix elements  $f_{T_q} ^{n,p}  $  parameterize the quark content of the nucleons, and we take their  values from~\cite{Cheng:2012qr}:
\beq
f_{T_u} ^p &=& 0.017, \quad  f_{T_d} ^p =0.022, \quad  f_{T_s} ^p =0.053,\nonumber \\
f_{T_u} ^n &=& 0.011, \quad  f_{T_d} ^n =0.034, \quad  f_{T_s} ^n =0.053, \nonumber \\
f^{p,n} _{T_G} &=& 1- \sum_{q=u,d,s} f^{p,n} _{T_q}.
\eeq
For an alternative numerical evaluation of the above parameters, we refer the reader to Refs.~\cite{Alarcon:2011zs,Alarcon:2012nr}.

\begin{figure}[!htb]
\begin{center}
\includegraphics[width=0.49\textwidth]{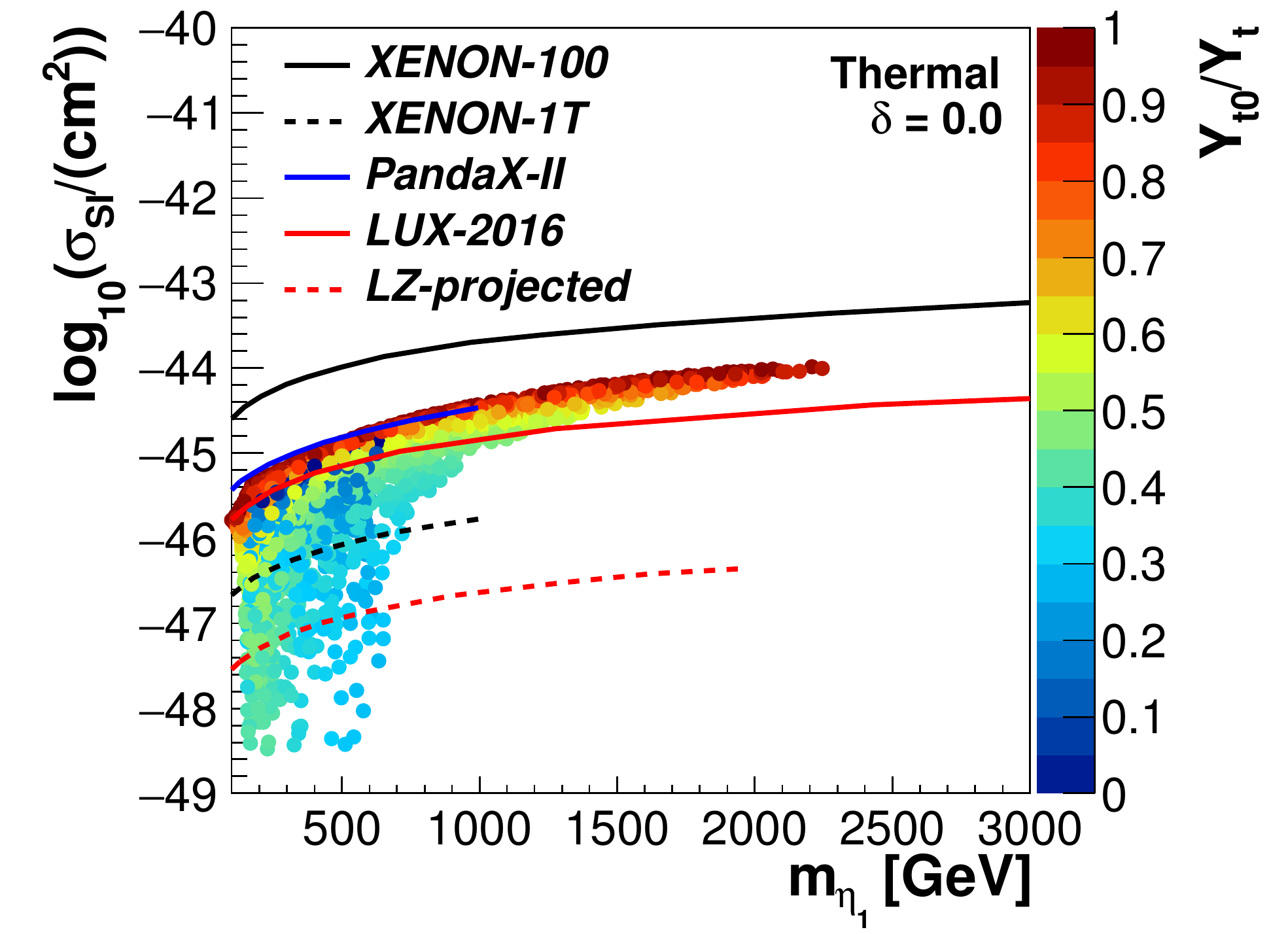}
\includegraphics[width=0.49\textwidth]{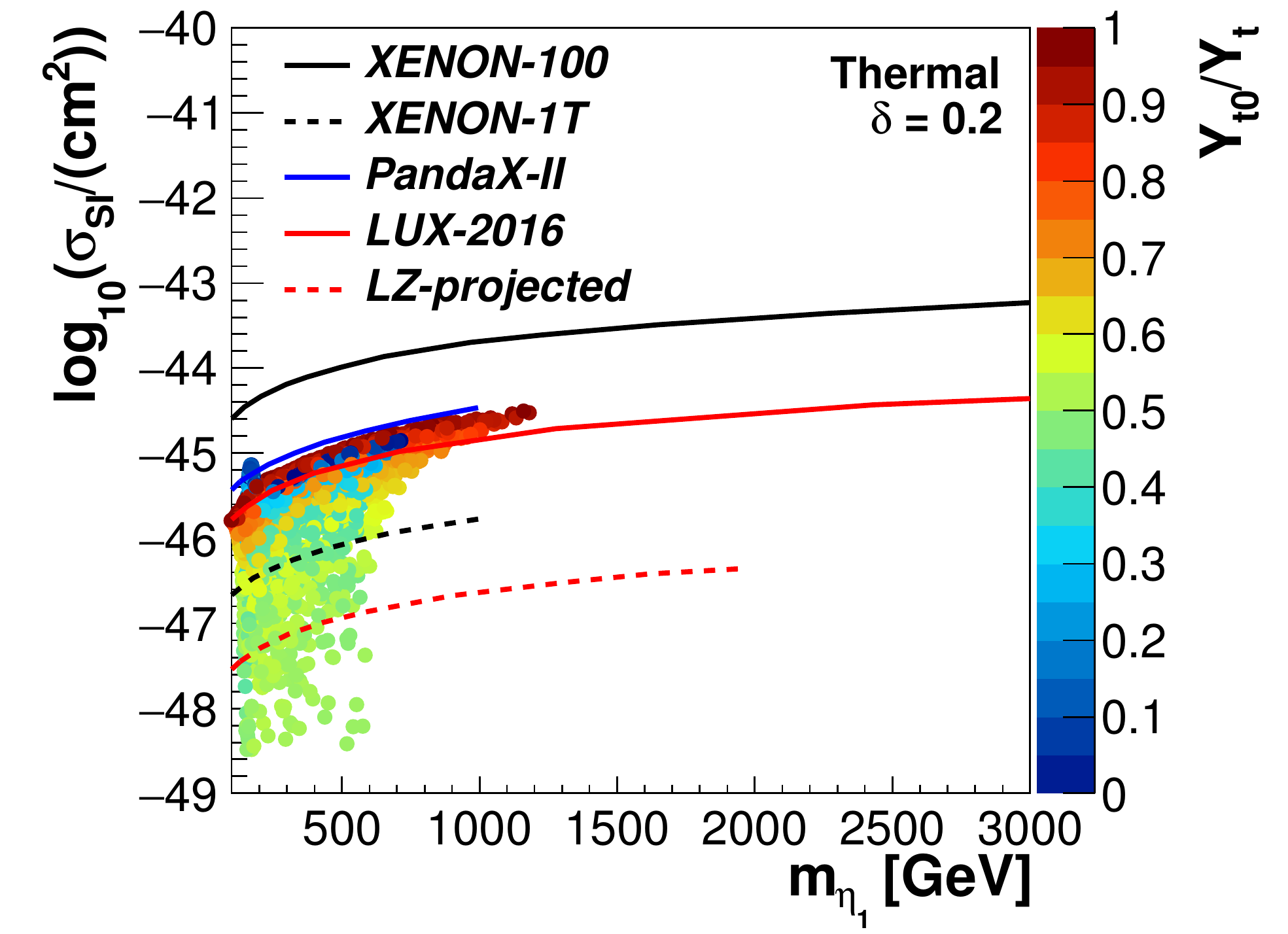}\\

\includegraphics[width=0.49\textwidth]{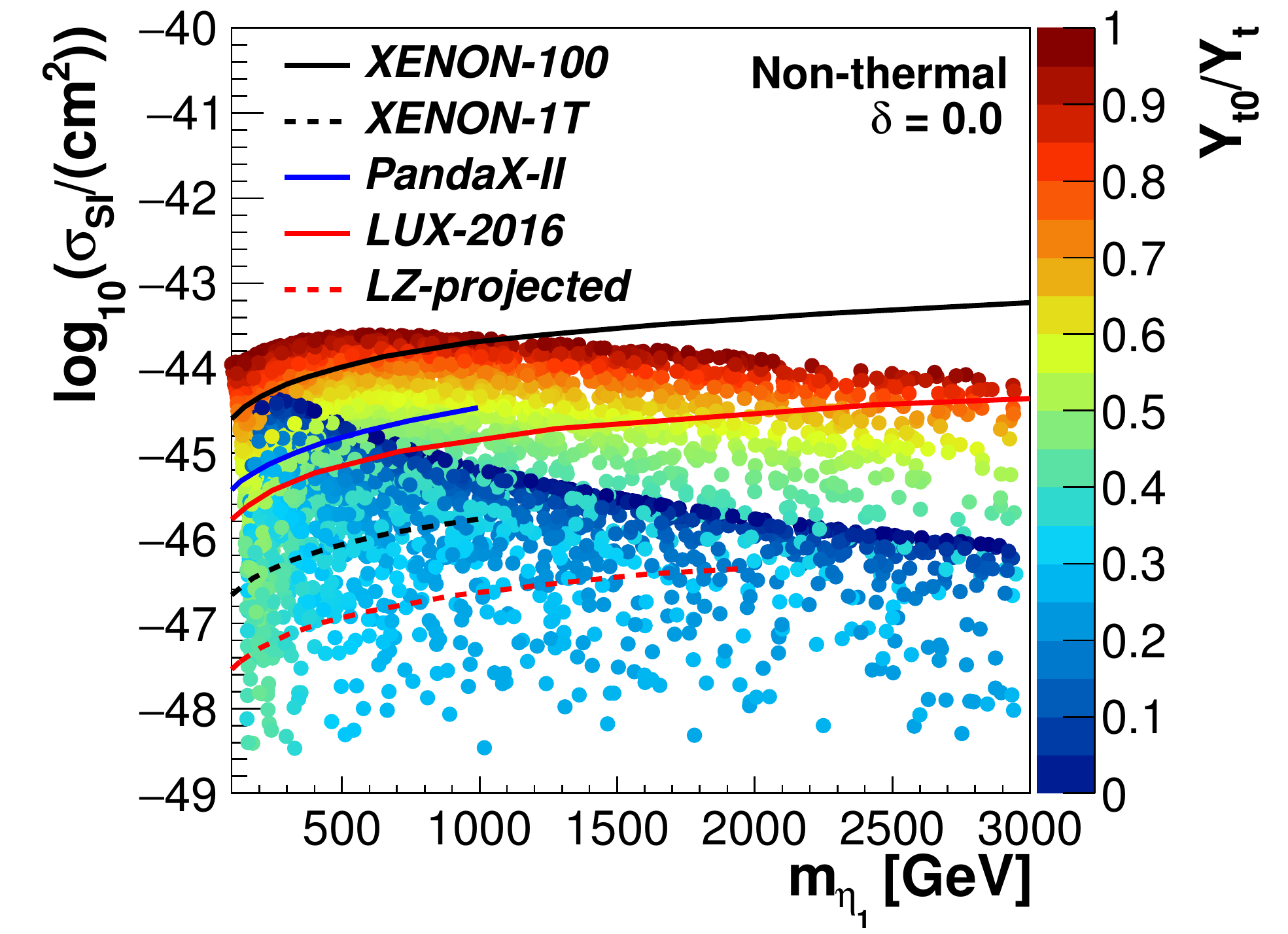}
\includegraphics[width=0.49\textwidth]{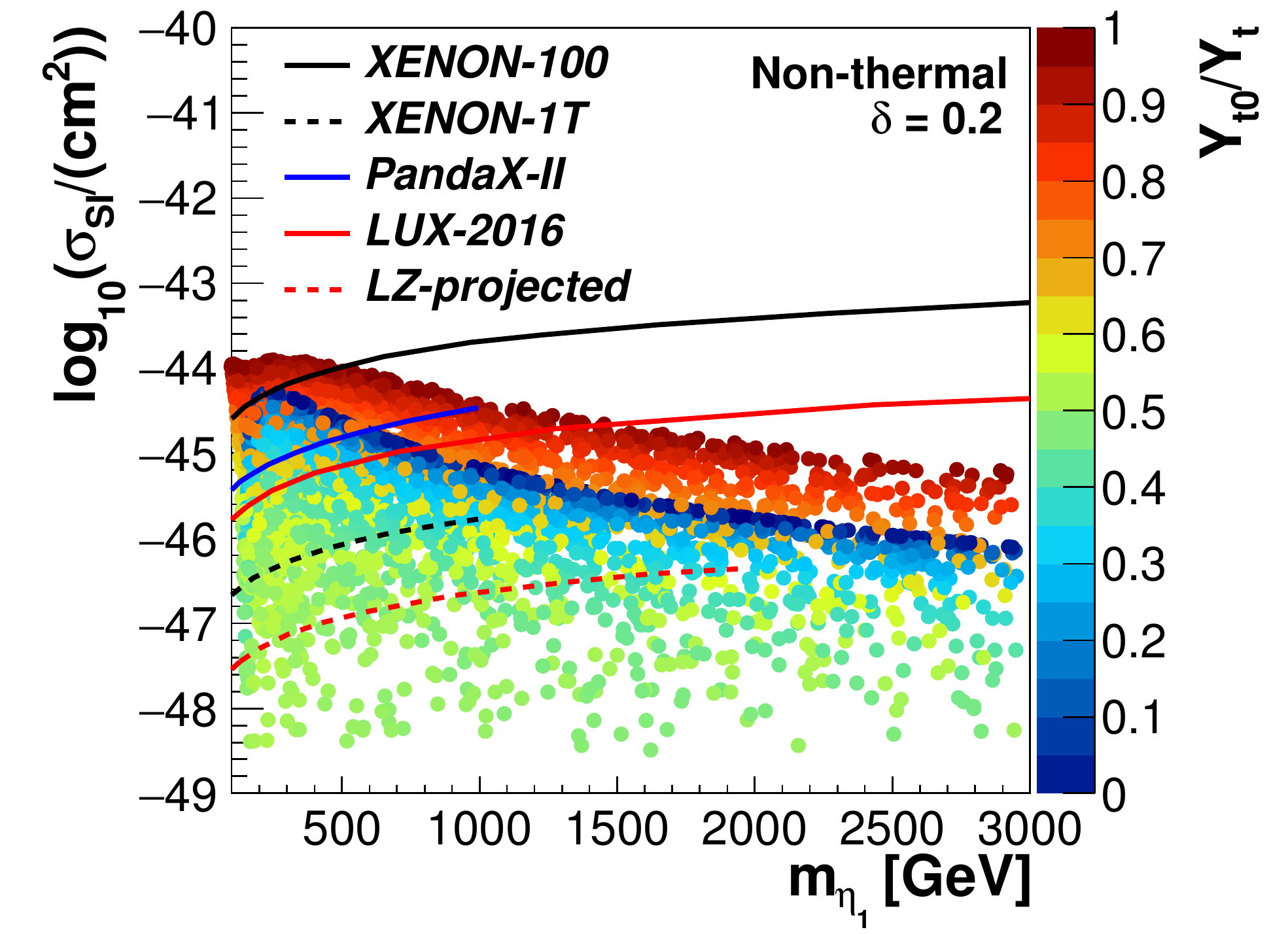}
 \end{center}
\caption{Spin-independent cross section for  the elastic scattering of  the DM candidate  $\eta_1$ off nuclei compared to current and future experimental sensitivities. The upper plots show cross sections rescaled to the thermal relic abundance (overabundant points are not shown), while the lower ones assume relic abundance equal to the measured one.} \label{fig:directdetection}
\end{figure}

The effective couplings in Eq.~(\ref{eq:DMfermions}) depend on $Y_{f0}$ via the mass mixing terms and the presence of off-diagonal couplings in Table~\ref{table:topyukawa}, we therefore decided to compute the elastic cross sections for fixed values of $Y_{f0}/Y_f$ to be compared to the relic abundance calculation in Figure~\ref{fig: DMrelicdensity }.
Since the thermal relic density  of $\eta_1$ is a function  of  its  mass, to compare the theoretical cross section to the experimental constraints we need to take into account the difference between the local DM density and the actual density of $\eta_1$: therefore, in Figure~\ref{fig:directdetection} we rescale the inelastic cross section according to
    \beq  \label{eq:sigmaSIEXP}
  \sigma_{SI} ^{EXP} = \frac{\Omega_{\eta_1}}{\Omega_{DM}}  \sigma_{SI}\,, 
    \eeq
to compare with the experimental bounds, which assume the standard density of DM around the Earth,  $\rho_0   = 0.3$GeV/cm$^3$, assuming that the DM density is saturated by the particle under consideration (and standard halo profile).

The effective cross section $\sigma_{SI} ^{EXP}$ is thus compared to the current most constraining bounds, presently coming from LUX experiment~\cite{Akerib:2016vxi} on $\sigma_{SI}  $:  the results are shown on the top row plots in Figure~\ref{fig:directdetection} for $\delta=0$ and $0.2$. We  find that  the region where $Y_{f0} <0.1 Y_{f}$ and $Y_{f0}  >0.8 Y_{f}$ are almost excluded.  For intermediate values of $Y_{f0}$, $0.1Y_{f} <Y_{f0}  < 0.8 Y_{f}$,  the upper limit on the $\eta_1$ mass is around $1$ TeV for both cases, $\delta=0$ and $0.2$. The future experiment XENON1T~\cite{XENON1T}  and  LZ-projected~\cite{LZ-CDR}  can stringently limit the model parameter space and eventually exclude most of the parameter space in case of thermally produced DM. We remark that the right edge of the points corresponds to the parameter space saturating the measured relic abundance. 
If non-thermal production mechanisms for the DM are allowed, it may be possible that the correct relic density is obtained in the whole parameter space: under this pragmatic assumption, we compared the cross section with the experimental bounds in the bottom row of Fig.~\ref{fig:directdetection}. In this case, the cross section becomes nontrivially dependent on the value of $Y_{f0}$ and LUX limits can exclude DM masses below $800$ GeV for $Y_{f0}< 0.1 Y_{f}$ and $3000$ GeV for $Y_{f0}>0.9 Y_f$ in the case of $\delta=0$. For $\delta =0.2$, the lower limits on DM mass are $800$ GeV and $1500$ GeV for this two region of $Y_{f0}$ respectively. The projected reach of future experiments will be able to extend the exclusion to a few TeV.

%%%%%%%%%%%%%%%%%%%%%%%
\subsection{Indirect detection constraints } \label{sec:ID}

Indirect DM detection relies on astronomical observations of fluxes of SM particles reaching Earth to detect the products of annihilation or decay of DM in our galaxy and throughout the cosmos.
The differential flux of DM annihilation products can be written as
\beq \label{eq:flux}
\Phi(\psi,E) = (\sigma v)\  \frac{dN_i}{dE}\ \frac{1}{4\pi m_{DM} ^2}  \int_{\rm line\ of\ sight} ds\ \rho^2(r(s,\psi))\,,
\eeq
where $E$ is the energy of the particle $i$ (either direct product of the annihilation, or generated as a secondary particle during the diffusion in the Galaxy) and $\psi$ is the angle from the direction of the sky pointing to the centre of our Galaxy.
The cross section $(\sigma v)$ is the annihilation in a specific final state that produces the particle $i$ as primary or secondary product with a differential spectrum $dN_i/dE$, while $\rho$ is the DM profile distribution in the Galaxy at a distance $r$  from the centre of the Galaxy, and $s$ is the distance from Earth running along the line of sight defined by $\psi$.
 Once the DM distribution profile in the Galaxy  $\rho$ is determined, the flux of particle $i$ can constrain  the DM annihilation cross section.
 From each annihilation channel, the expected flux for each detectable particle specie $i$ can be computed~\cite{Cirelli:2010xx}.
 We will use these results to constrain our model by demanding that the annihilation cross section of $\eta_1$ does not exceed the observed value of the various fluxes.

 Similarly to the case of Direct detection, the DM distribution profile $\rho_{\eta_1} $ need to be rescaled to take into account the actual thermal relic abundance where $\eta_1$ only represent part of the total DM density (assuming its profile follows the standard lore):
 \beq
\rho_{\eta_1} =\frac{\Omega_{\eta_1 }}{\Omega_{DM}} \rho.
 \eeq
Thus, the physical annihilation cross section should have the following relation with the experimental
value according to Eq.~(\ref{eq:flux}):
\beq \label{eq:sigmavEXP}
 (\sigma v)_{ EXP}  = \left(\frac{\Omega_{\eta_1 }}{\Omega_{DM}}\right)^2 (\sigma v)_{\eta_1}\,.
\eeq

In this model, the Sommerfeld enhancement~\cite{Feng:2010zp} is very small because the couplings of one Higgs/$W$/$Z$  to $\eta_1$ pairs are not large enough. Therefore, we will neglect this effect. There are the following annihilation modes:
\begin{itemize}
\item  $\eta_1$ can annihilate into  leptons  and light quarks pairs via direct couplings: however, under the assumption that $Y_{f0} \sim Y_f$ so that the coupling to light fermions is roughly proportional to their mass, these annihilation modes are very small and no significant constraints emerge.

\item Annihilation into $b \bar{b}$ can be larger than that into light fermions because of  the larger Yukawa coupling of the bottom quark. In the top row of  Fig.~\ref{fig:IDthermal} we show the theoretical prediction (coloured region) for the rescaled velocity-averaged cross section $(\sigma_{b\bar{b}} v)_{EXP } $ for the annihilation channel $b\bar{b}$.  The solid black and red curves are the  limits based on Fermi-LAT gamma-ray observation and HESS respectively. We found that, varying $0 \leq Y_{b0} \leq Y_b$, the rescaled  cross section ranges between $(\sigma_{b\bar{b}} v)_{EXP } = 10^{-27} \div 10^{-29}$ cm$^3$/s in the allowed DM mass range. We can see that  the limits from Fermi-LAT and HESS are about two to three orders of magnitude too weak to impose any useful bounds  for the whole mass region. Releasing the assumption that DM is thermally produced, i.e. assuming that it is always saturating the DM abundance, the cross section are enhanced for low masses, still remaining below the Fermi-LAT current bound, as shown in the first row of Fig.~\ref{fig:IDnonthermal}.

\begin{figure}[!htb]
\begin{center}
\includegraphics[width=0.49\textwidth]{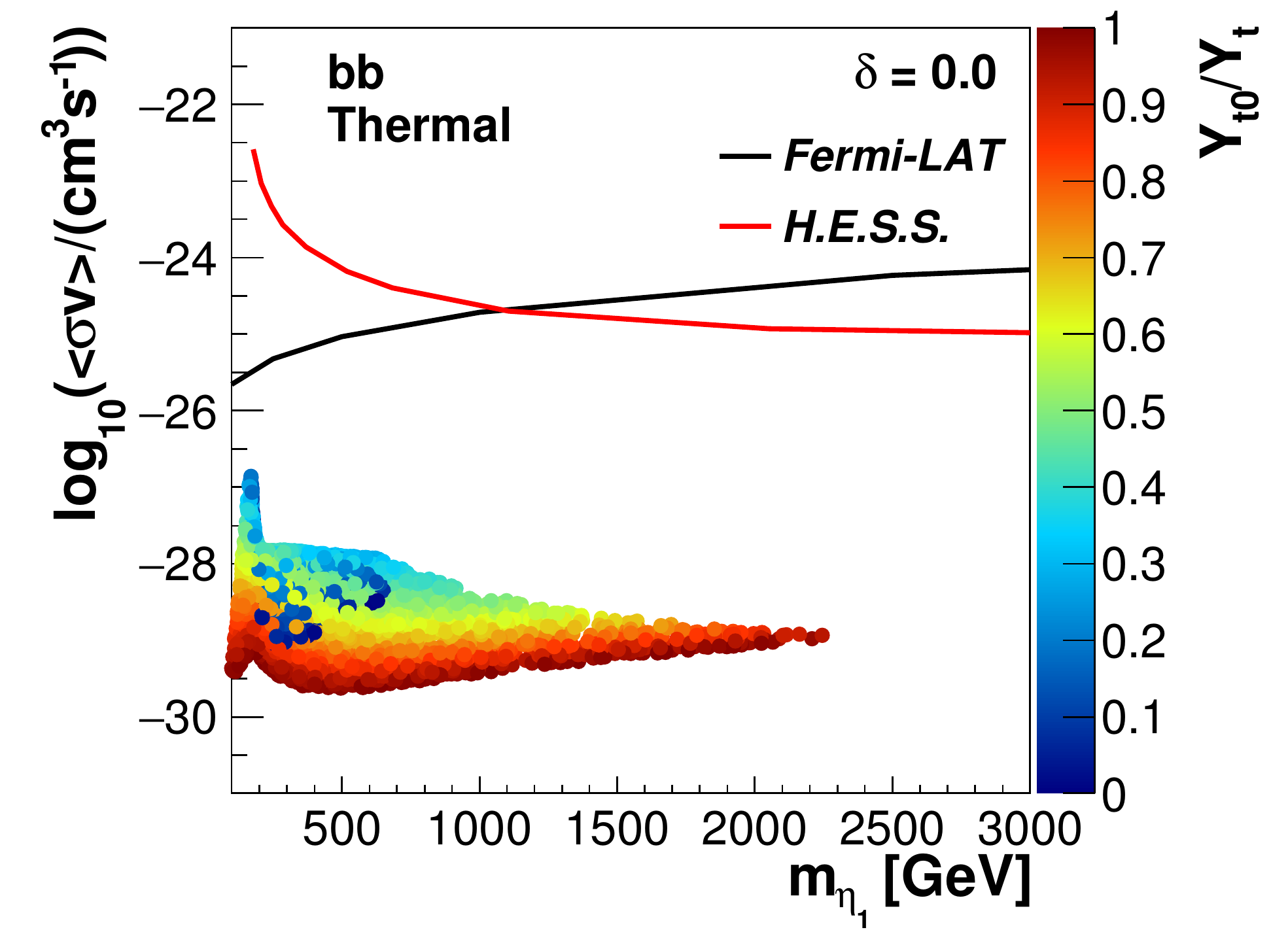}
\includegraphics[width=0.49\textwidth]{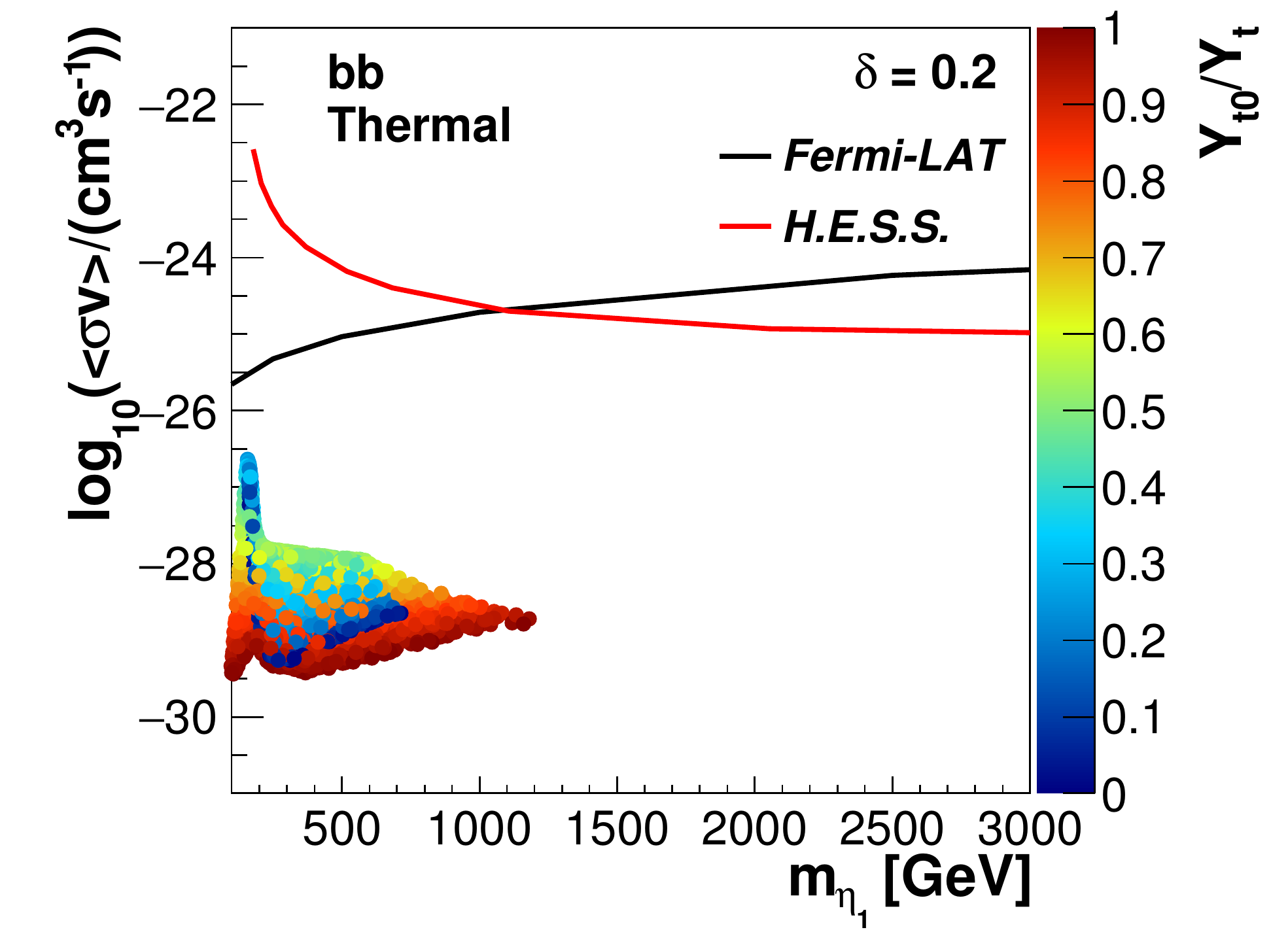}\\
\includegraphics[width=0.49\textwidth]{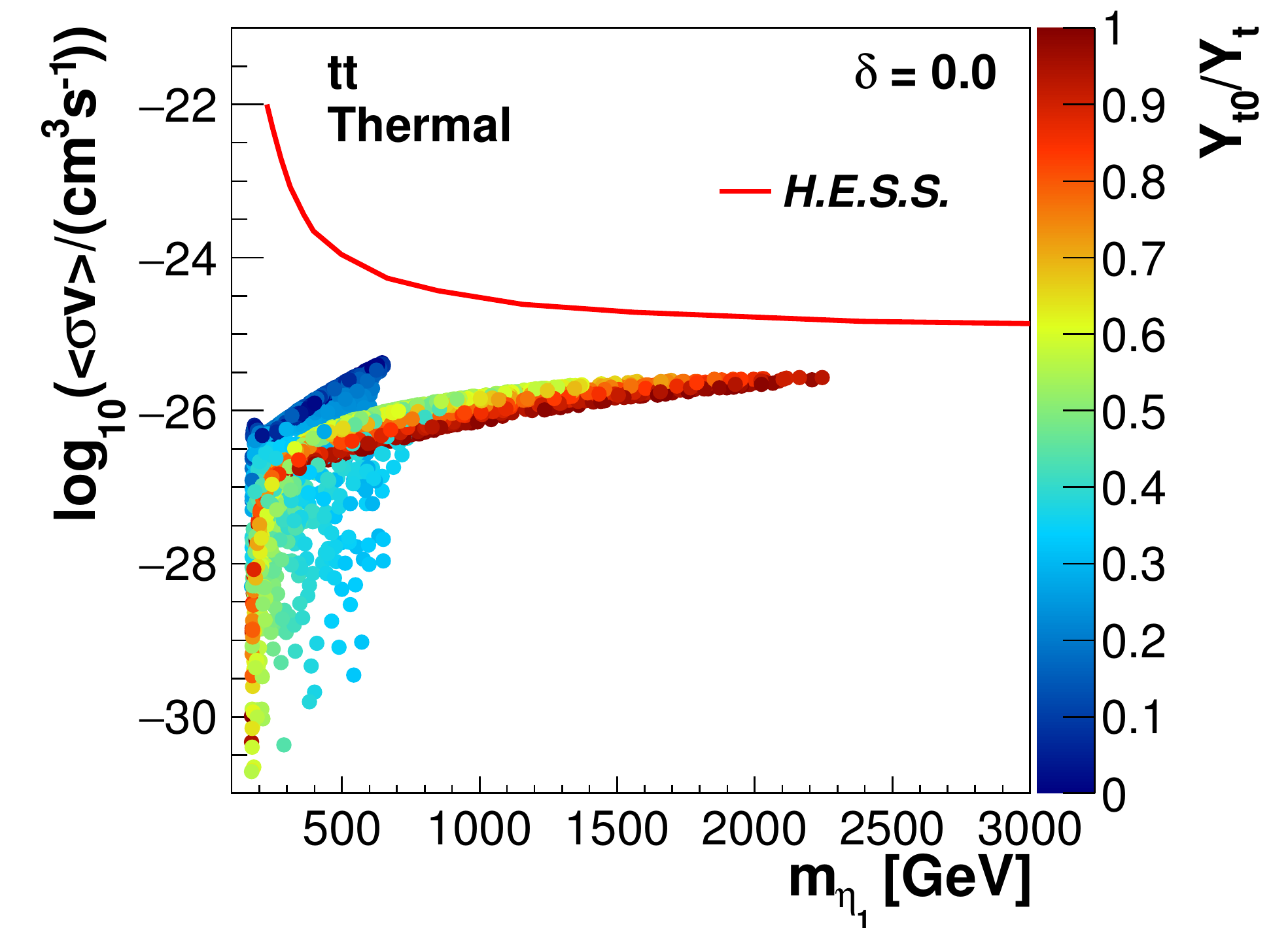}
\includegraphics[width=0.49\textwidth]{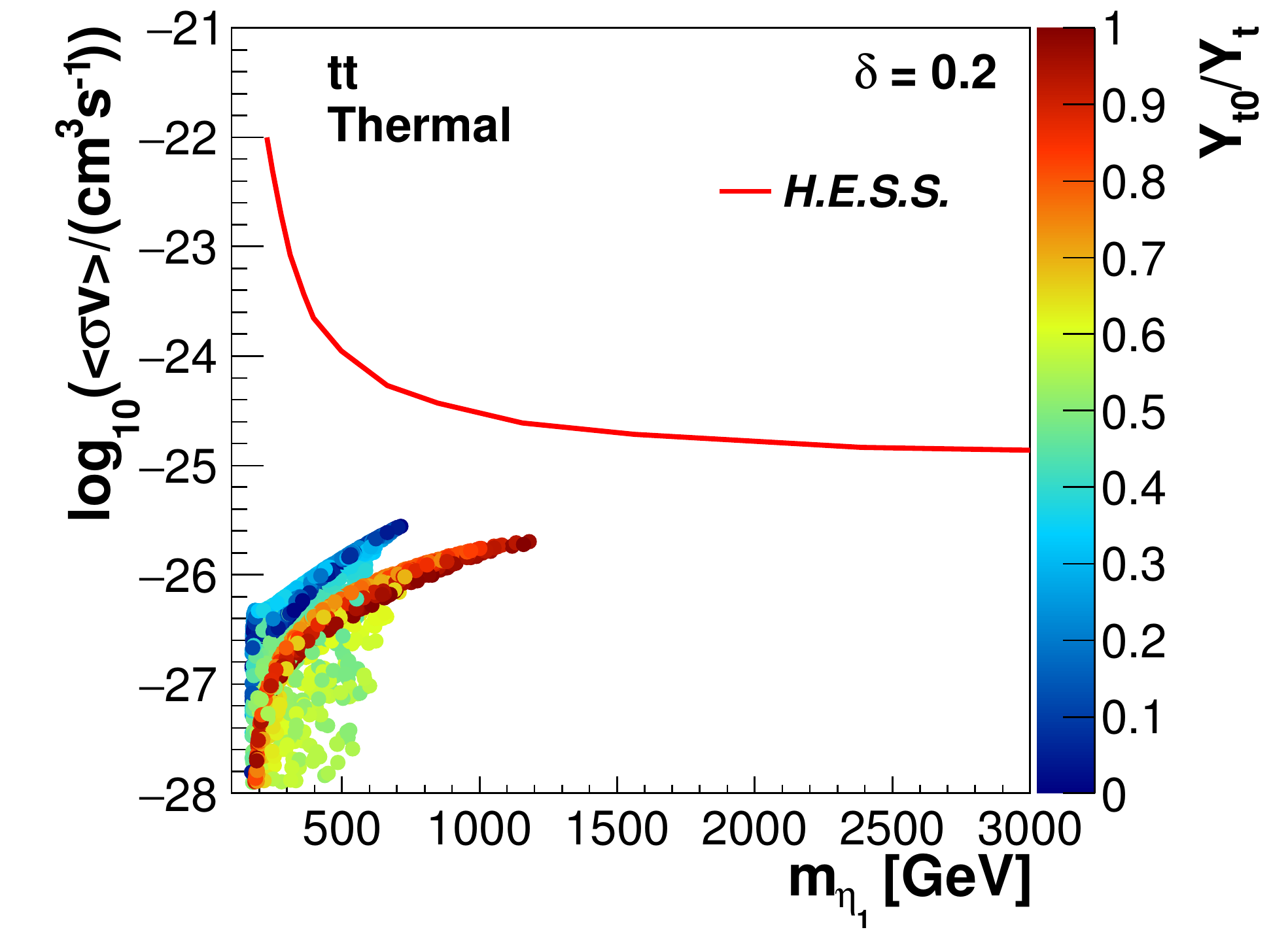}\\
\includegraphics[width=0.49\textwidth]{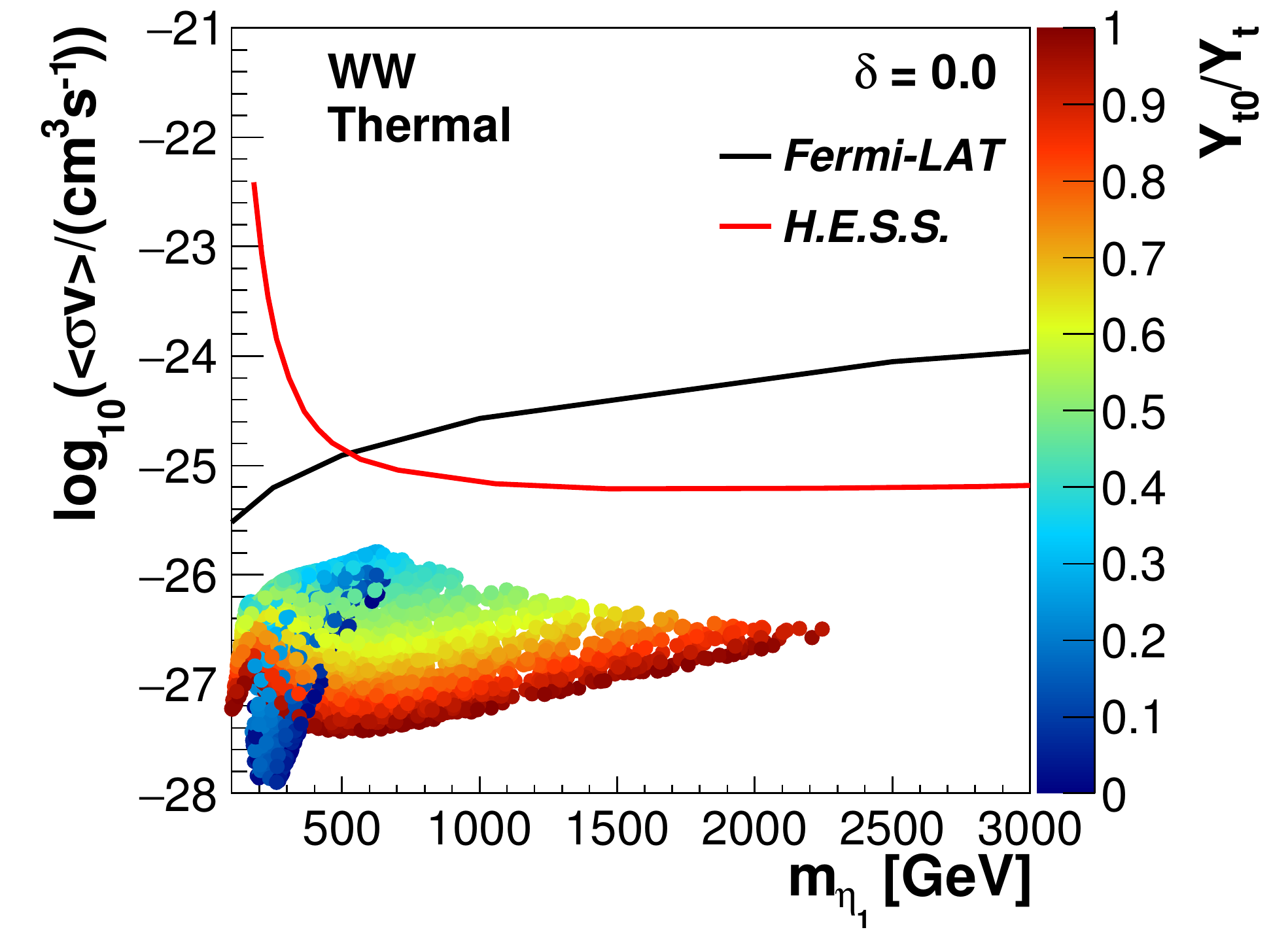}
\includegraphics[width=0.49\textwidth]{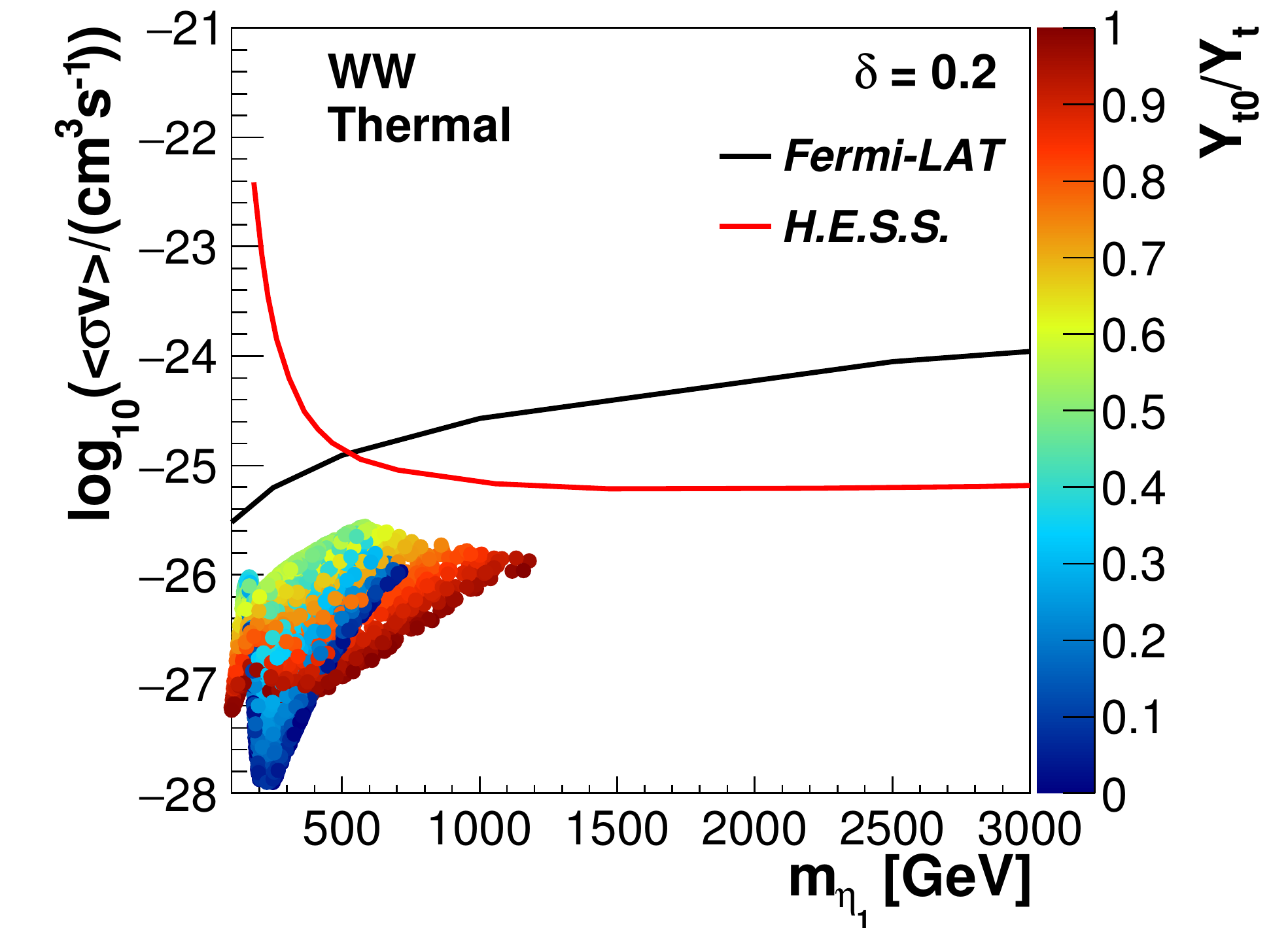}
 \end{center}
\caption{Theoretical  prediction (coloured region) for the velocity-averaged cross section for  $\eta_1$  annihilation into bottom pair (top row), top pair (middle row) and $WW$ (bottom row) for $\delta =0$ (left)  and $0.2$ (right), compared to the upper limits from Fermi-LAT (black) and HESS (red) gamma-ray observation. The colour code indicates the variation of the Yukawa coupling $Y_{f0}$ from $0$ to $Y_{f}$. The cross sections are rescaled to the thermal relic abundance.} \label{fig:IDthermal}
\end{figure}

\begin{figure}[!htb]
\begin{center}
\includegraphics[width=0.49\textwidth]{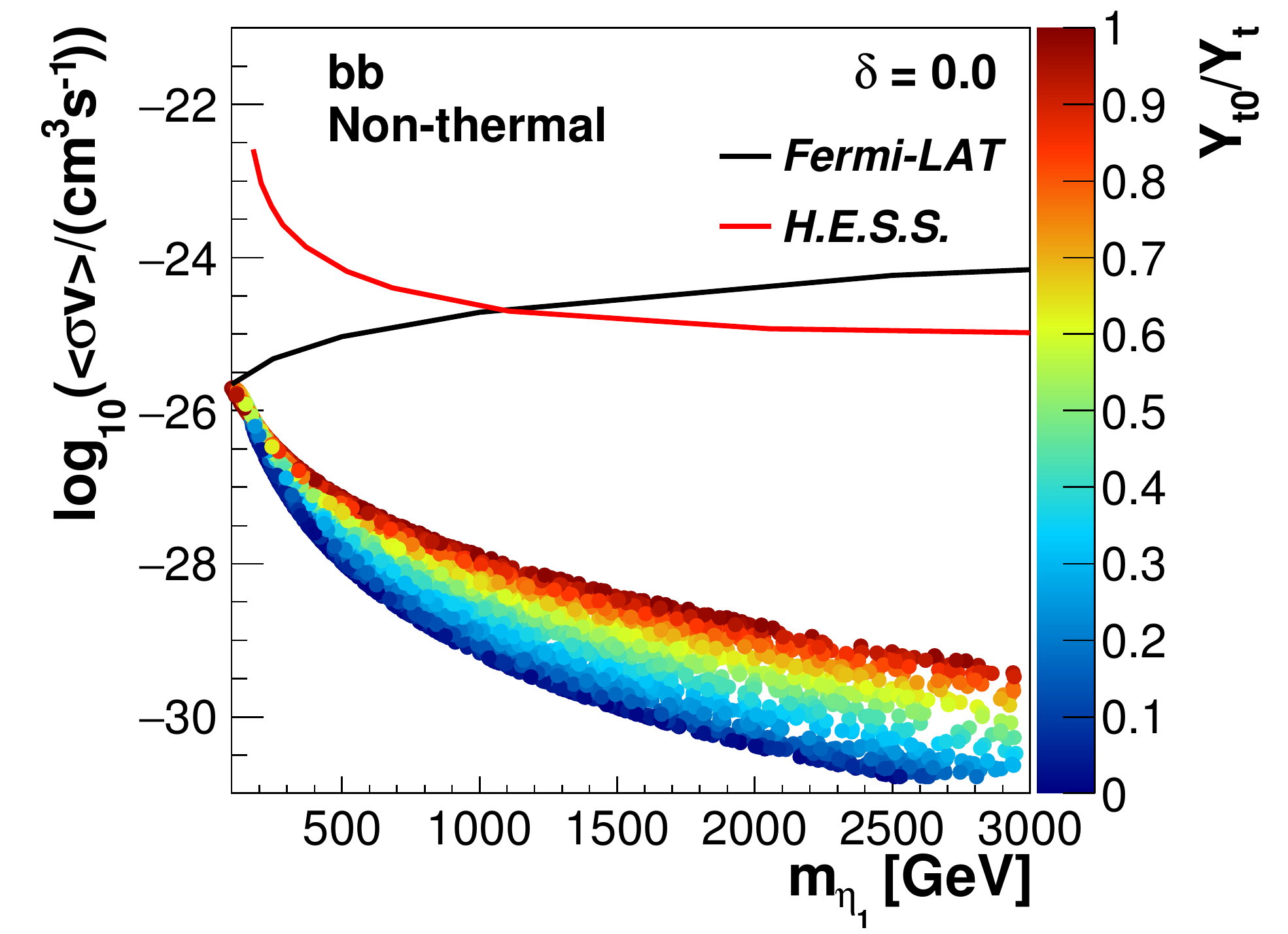}
\includegraphics[width=0.49\textwidth]{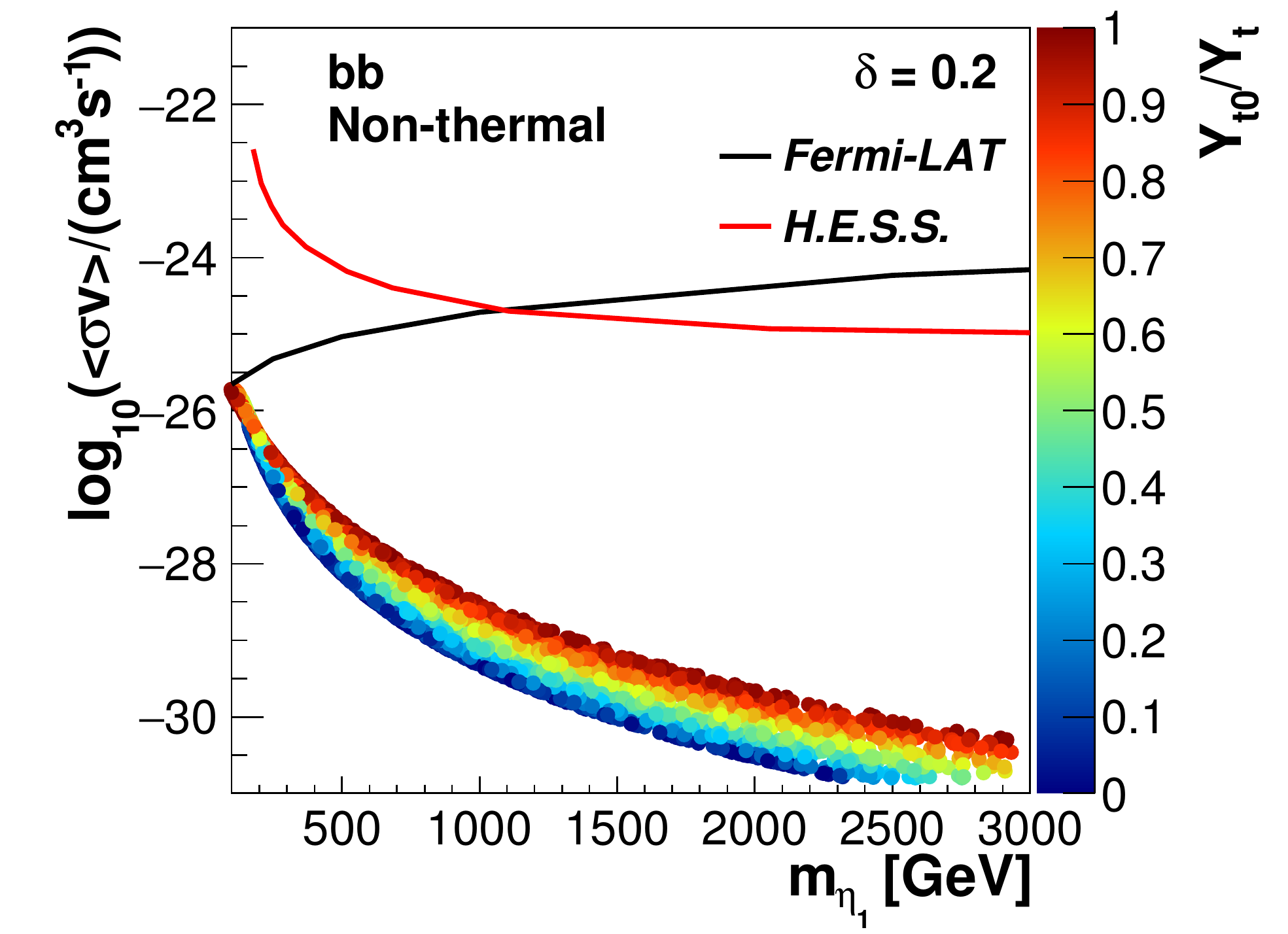}\\
\includegraphics[width=0.49\textwidth]{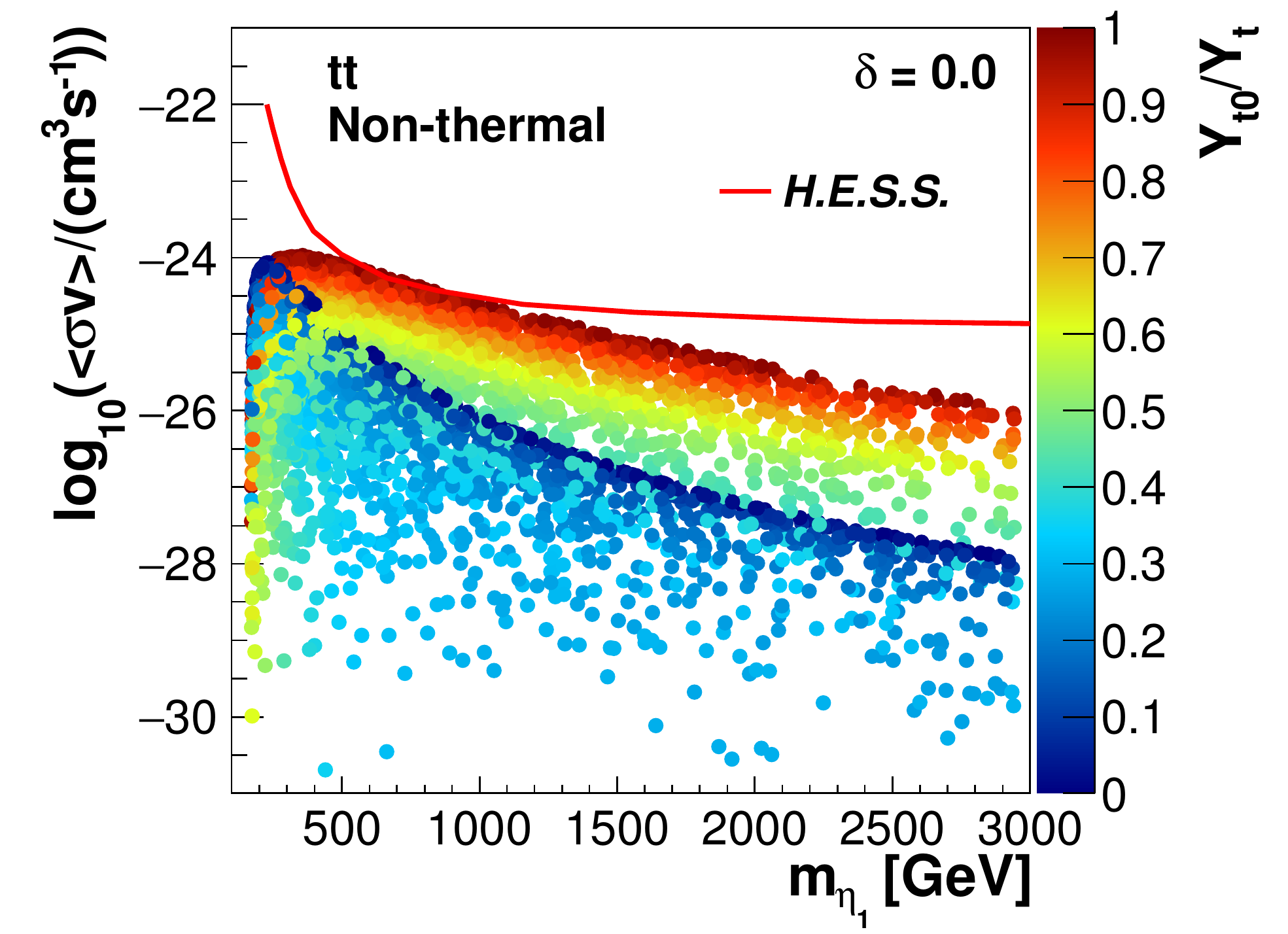}
\includegraphics[width=0.49\textwidth]{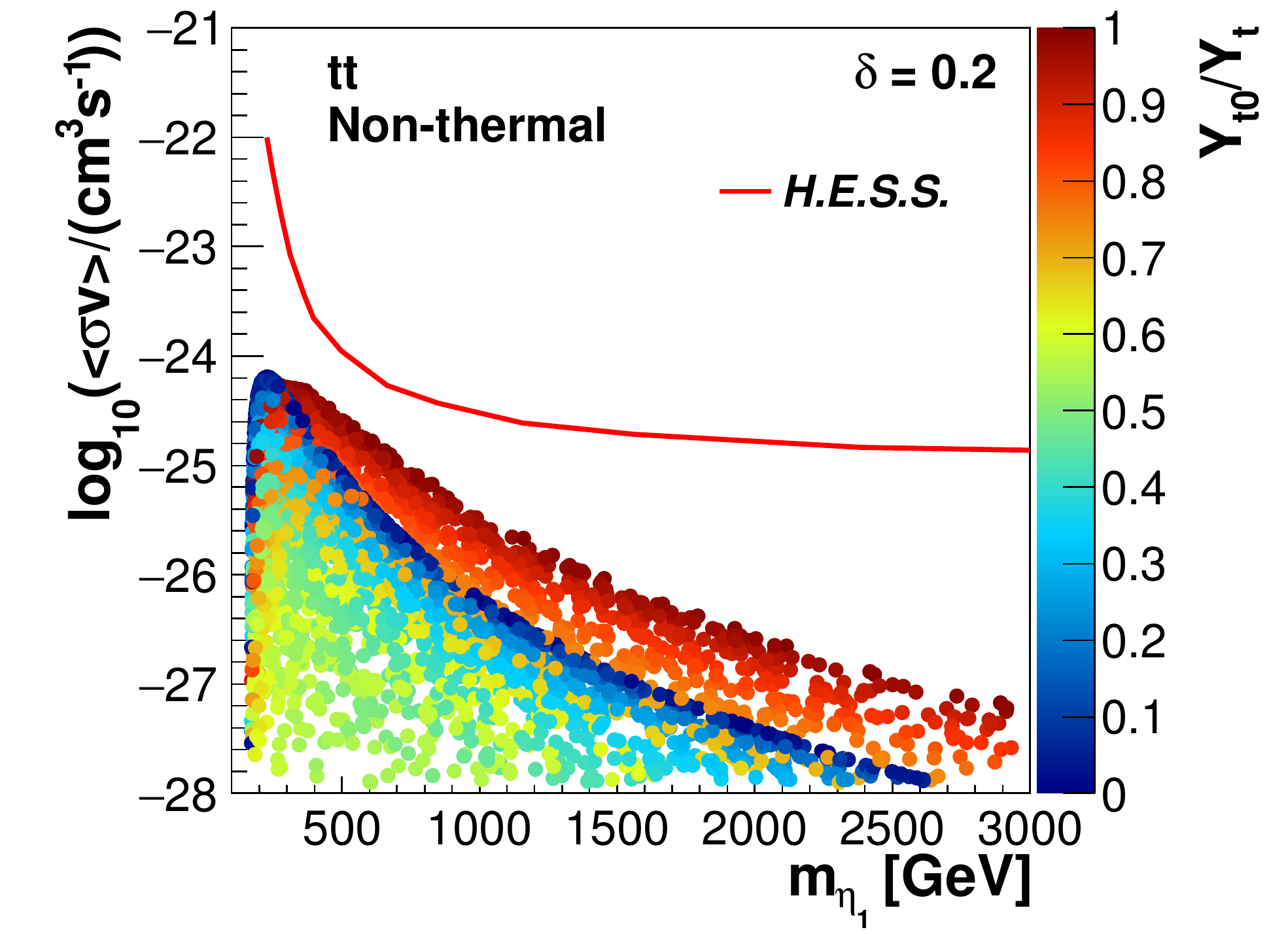}\\
\includegraphics[width=0.49\textwidth]{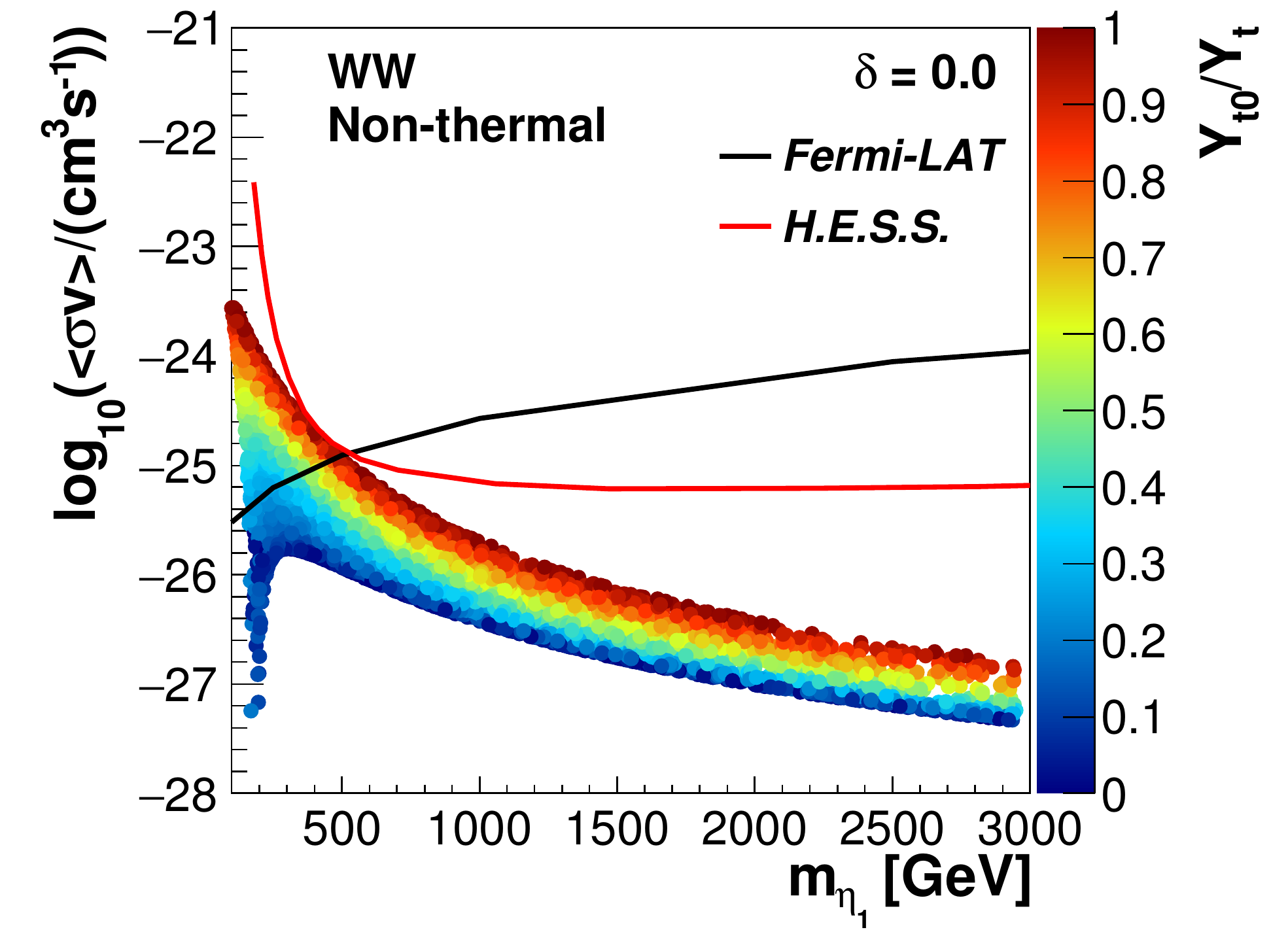}
\includegraphics[width=0.49\textwidth]{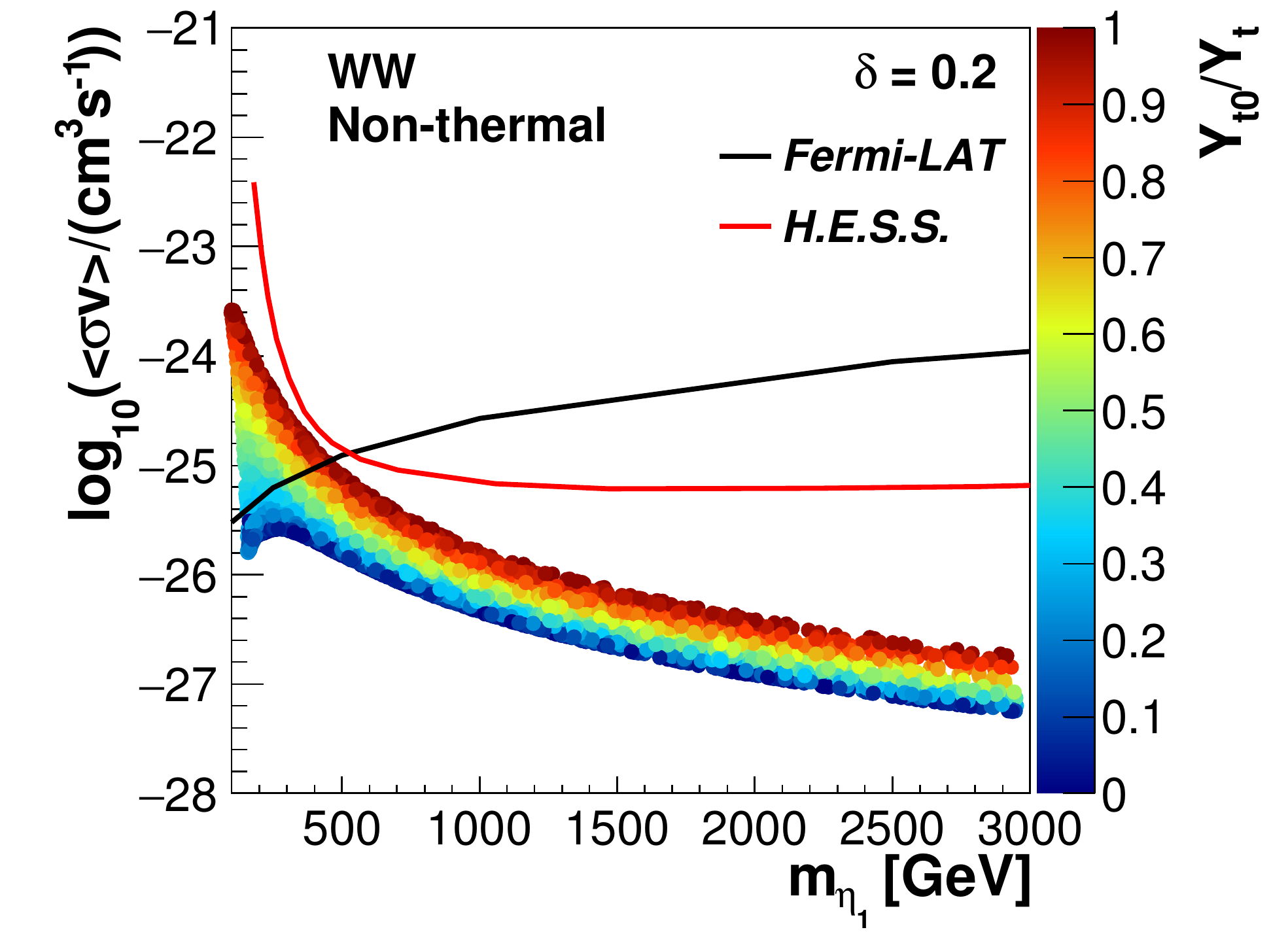}
 \end{center}
\caption{Same as Fig.~\ref{fig:IDthermal}, assuming saturated relic abundance in the whole parameter space.} \label{fig:IDnonthermal}
\end{figure}

\item  Annihilation into top pairs $t \bar{t} $ is further enhanced by the top Yukawa coupling. We find that, in the same range as for the bottom, the rescaled cross section shown in the middle row of Fig.~\ref{fig:IDthermal} varies between $(\sigma_{t\bar{t}} v)_{EXP} = 10^{-25} \div 10^{-31}$ cm$^3$/s for $\delta=0$ and between $(\sigma_{t\bar{t}} v)_{EXP} = 10^{-25} \div 10^{-28}$ cm$^3$/s for $\delta=0.2$ .  The cross section is always below the experimental limit coming from the $\gamma$-ray  observed by experiments like HESS: the cross section into tops is bound to be smaller than $10^{-23}$ cm$^3$/s for $m_{\eta_1} < 1$~TeV, and $5 \times 10^{-24}$ cm$^3$/s  for $1 < m_{\eta_1} < 10$~TeV~\cite{Abdallah:2016ygi}. Even releasing the assumption on thermal relic abundance (middle row of Fig.~\ref{fig:IDnonthermal}), the cross section remains below the experimental sensitivity.

\item  Annihilation into $W^+ W^-$ occurs through Higgs mediated s-channel, charged scalar mediated u- and t-channel, and four particles vertices.  As the case for annihilation into fermion pairs, in the bottom row of Fig.~\ref{fig:IDthermal}  we also show theoretical predictions (coloured regions) for the rescaled velocity-averaged cross section $(\sigma_{WW} v)_{EXP } $: we see that this channel has larger cross sections than the fermionic channels and $(\sigma_{WW} v)_{EXP }$ increases with mass. However, no bound is imposed on this channel by Fermi-LAT nor HESS. Releasing the assumption of thermally produced relic abundance (bottom row of Fig.~\ref{fig:IDnonthermal}), the cross section increases in the low mass region,  and exclusions  around $500$ GeV are observed, depending on the value of $Y_{f0}$. \\

\end{itemize}

  As already mentioned above, Indirect detection reach on DM models can be significantly enhanced if large Sommerfeld enhancement factors are possible. In our model, $\eta_1$ is almost a $SU(2)_L$ singlet except for small mixing with the $SU(2)_L$ doublet and
  triplet,  thus it has  small couplings to the EW gauge boson. We also checked the possible Sommerfeld enhancement induced by Higgs exchange and find this effect to also be quite small: in fact, the ``Yukawa'' potential $V(r) =-\frac{\alpha_h}{r} e^{-m_h r}$ is suppressed by $\theta ^2$. In conclusion, we find that DM indirect detection can not impose significant constraints on this model if $\eta_1$ is a thermally produced DM component.

%%%%%%%%%%%%%%%%%%%%%%%%%%%%%%%%%%%%
\subsection{Summary of DM constraints}
\begin{figure}[!htb]
\centering
\includegraphics[width=0.49\textwidth,trim=50 0 150 0]{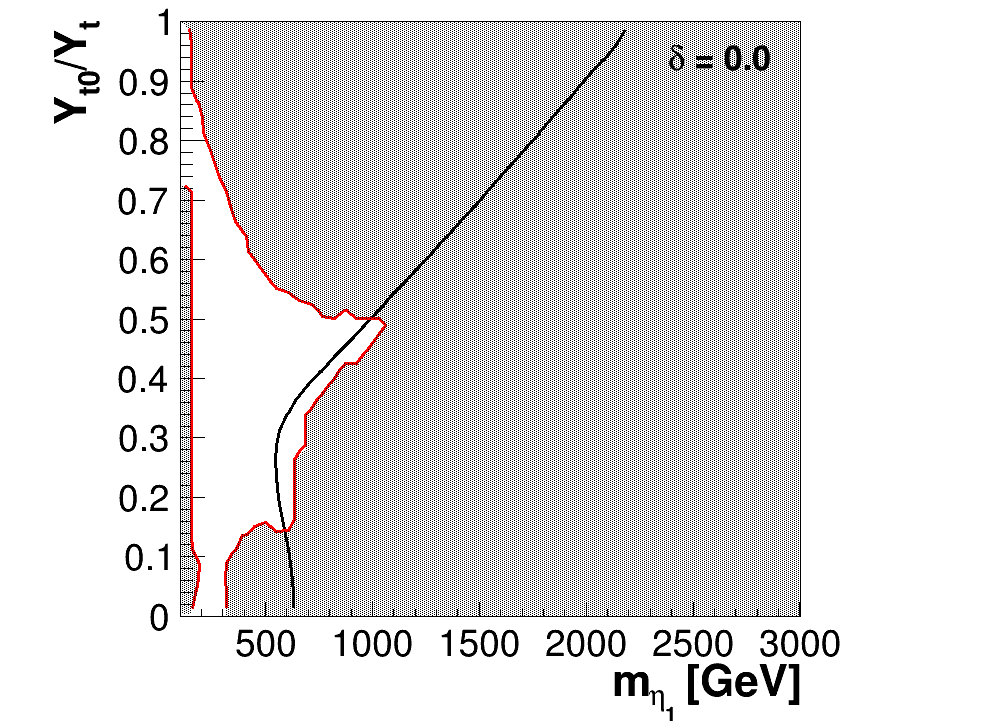}
\includegraphics[width=0.49\textwidth,trim=50 0 150 0]{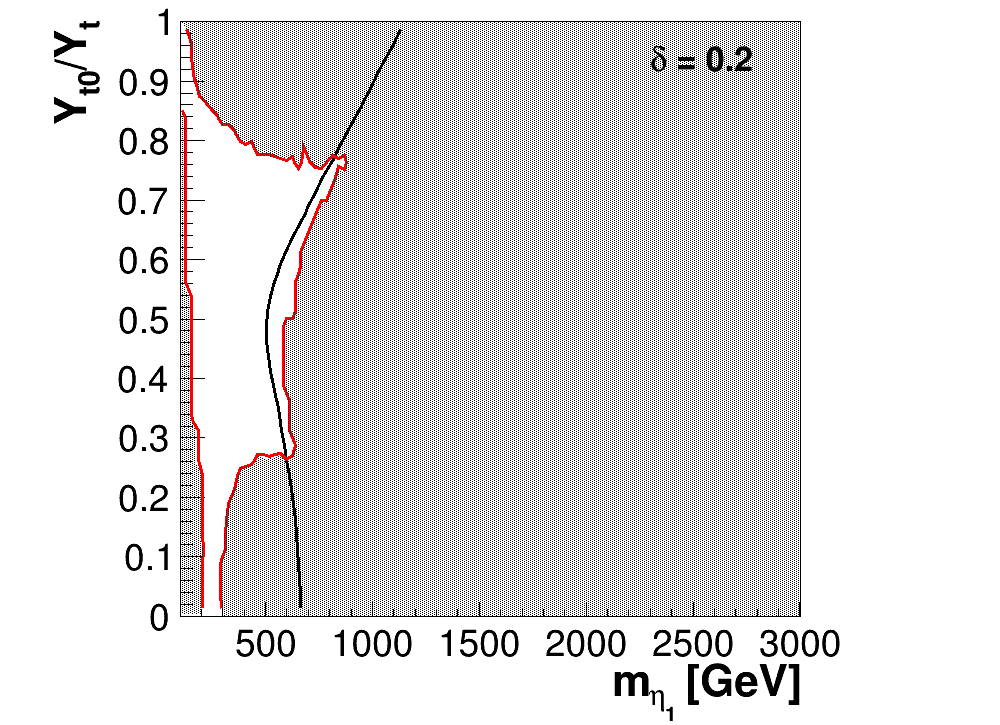}
\caption{Combination of constraints from the relic density measurement and direct detection experiments. The region to the right of the black line is excluded by overclosure of the Universe, while the region within the red line is still allowed by the direct detection.}
\label{fig:combine_results}
\end{figure}
Previously we separately discusses the constraints on the DM candidate $\eta_1$ from different DM experimental tests. We now combine all these constraints to extract the allowed parameter space in order to guide future research projects. The final summary is shown in Figure~\ref{fig:combine_results}, where we show the constraints from DM relic density and direct detection measurements (indirect detection does not impose any stronger bounds on $\eta_1$). The region allowed by direct detection is within the red line (the grey area being excluded), while the black line corresponds to the parameters fulfilling the thermal relic density. We recall the reader that the region on the right of the black line is excluded by overclosure, while on the left $\eta_1$ can only partly accommodate for the full DM density, as shown in Fig.~\ref{fig: DMrelicdensity }.
The surviving region where $\eta_1$ can be the unique DM candidate consists of $m_{\eta_1} \in [500, 1000]$ GeV and $Y_{t0}/Y_t \in [0.12 ,0.5]$ for $\delta =0$ and $m_{\eta_1} \in [500, 800]$ GeV and $Y_{t0}/Y_t \in [0.28 ,0.75]$ for $\delta =0.2$.
Larger (and smaller) values of $Y_{t0}$ are only allowed for small $m_{\eta_1}$ (and low DM thermal relic density), which are however in tension with electroweak precision tests: $\theta >0.2$ roughly corresponds to $m_{\eta_1} >500$ GeV. The DM candidate $\eta_1$ still has a wide parameter spaces to provide enough relic density. 
%However, future DM direct detection will impose an upper limit on $\eta_1$ mass around $500$ GeV if it is thermally produced, see Fig.\ref{fig:directdetection}, which can almost exclude the $\delta >0$ case.    

%%%%%%%%%%%%%%%%%%%%%%%%%%%%%%%%%%%%%%%%%%%%%%%%%%%%%%%%%
\section{ Technicolor Interacting Massive Particles as  Dark  Matter } \label{sec:LTB}

In this section, we would like to briefly discuss the possibility that the Techni-baryons, protected by a global TB number, may play the role of DM. This possibility was first discussed in~\cite{Nussinov:1985xr} in the context of Technicolour theories.
As the Techni-baryons are strongly interacting, it is difficult to calculate the annihilation rate: however, approximating it by the geometrical cross section allows to estimate the value of the required masses  to be, generically, around $100$ TeV. This value would point toward values of the condensation scale of at least a few TeV, thus in a more fine tuned region of the parameter space in terms of generating the pNGB Higgs boson.

Another intriguing possibility is that an asymmetry is generated via the electroweak baryogenesis~\cite{Petraki:2013wwa,Zurek:2013wia}, in the same way as a lepton and baryon asymmetry is generated. This mechanism does, however, require that the TB number is violated by some anomalies that preserve a combination of it with the SM baryon and/or lepton numbers~\cite{Barr:1990ca}. In our case, the strong sector being non-chiral with respect to the SM gauging, such anomaly is absent and the electroweak sphaleron would be ineffective in balancing the asymmetries.

We are therefore left with two mechanisms to generate a Techni-baryon asymmetry:

\begin{itemize}
\item break $U(1)_{TB}$ by a gauge sphaleron  process at a higher scale than the electroweak phase transition;
\item explicitly break $U(1)_{TB}$.
\end{itemize}

For the former case, we need to assume that the theory is UV completed by an extended gauge sector (in a similar fashion to Extended Technicolor models~\cite{Dimopoulos:1979es}) which is chiral. At the scale where the new gauge theory is spontaneously broken (by condensation or a Higgs mechanism), we need a strong first order phase transition and sufficient CP violation in order to generate an asymmetry in the TB number in a similar way as baryogenesis~\cite{Sakharov:1967dj}. The net asymmetry is then left unchanged at the electroweak phase transition. This mechanism can be very attractive if the model is UV completed by a chiral theory that gives mass to the top at a scale not far from the condensation scale~\cite{Cacciapaglia:2015yra}.

In the latter case, we assume that the TB number is violated explicitly by new interactions at a very high scale that couple the Techni-fermions to SM particles.  Depending on the number of FCD-colours, the low energy effect may be a linear mixing to SM fermions (odd $N_{\rm FCD}$) or couplings to fermion bilinears (even $N_{\rm TC}$). For instance, in the case of $N_{\rm TC} = 3$, a 4-Fermion interaction may be induced in the form:
\beq
\frac{1}{\Lambda^2}\ \bar{l}\ \epsilon^{abc} \psi^a \psi^b \psi^c \to \lambda_l \bar{l} L\,,
\eeq
where $L = \langle \epsilon^{abc} \psi^a \psi^b \psi^c \rangle$ is one Techni-baryon and $a,b,c$ stand for the FCD--colour indices. The field $l$ is a generic lepton (including neutrinos), so that this interaction preserves L + TB. The mediator generating such interaction can thus be invoked as generator of TB asymmetry in the same way as in Grand Unification baryogenesis models~\cite{Sakharov:1967dj}: at the UV scale, out of equilibrium decays of the mediator will produce an asymmetry if CP is also violated. However, the same operator will induce decays of the composite Techni-baryons at low energy. This drawback can be avoided if the theory is conformal in the UV, as it can be achieved by adding additional flavours~\cite{Lombardo:2014pda}. A mass gap is only generated at the low scale where conformality is lost, and the running of the operator will generate a large suppression if the anomalous dimensions of the Techni-baryon operator is small~\cite{Pica:2016rmv}. In fact, $\lambda_l$ is related to the UV scale by
\beq
\lambda_l \sim \left(\frac{\Lambda_{\rm TC}}{\Lambda} \right)^{2-\gamma}\ \Lambda_{\rm TC}\,,
\eeq
where $\gamma$ is the anomalous dimension of the $\langle \epsilon^{abc} \psi^a \psi^b \psi^c \rangle$ composite operator. A very small $\lambda_l$, i.e. a large $\Lambda$, can thus guarantee long lived Techni-baryons at Cosmological scales. 
Assuming that the mass of the baryon is $\Lambda_{TC}$, the mixing angle between lepton $l$ and Techni-baryon is $\sin \theta_l  = \frac{\lambda_l}{\Lambda_{TC}}$  for $\lambda_l \ll \Lambda_{TC} $.  The Techni-baryon main decay channels are $L \to l \, W/Z$ through gauge interaction. By simple calculation and assuming $\gamma \sim 0$, we find that the UV scale should be above $\Lambda \sim 10^{13}$ GeV to ensure Cosmologically stable Techni-baryons. 
Mixing with quarks can also be written if additional coloured Techni-fermions are added, as in Ref.~\cite{Vecchi:2015fma}.

Direct detection experiments typically impose very stringent bounds on a composite fermionic DM candidate because of its magnetic moment. However, it is possible that the lightest Techni-baryon is made of electromagnetically neutral Techni-fermions (and potentially mixes to the right-handed neutrino) and thus has vanishing magnetic moment, thus avoiding the constraints.

%%%%%%%%%%%%%%%%%%%%%%%%%%%%%%%%%%%%%%%%%%%%%%%
\section{Conclusions}

It's tantalising to think that the Higgs boson may be a composite state of a more fundamental confining dynamics. In this work, we explore the possibility that, along with the Higgs, the dynamics produces also a light scalar DM candidate, arising from the same symmetry breaking responsible for the breaking of the EW symmetry. The minimal model featuring this property is based on the symmetry breaking SU(4)$\times$SU(4)/SU(4). It is also the only consistent model where a DM pNGB can be shown to exist, also featuring a simple underlying description in terms of a gauge-fermion theory.

The main features of the model have been described in Ref.~\cite{Ma:2015gra}: in this work we focus on the parameter space where the DM candidate is stable, and numerically explore the constraints coming from thermal relic abundance and Direct/Indirect DM detection. The neutral DM candidate is accompanied by several neutral and charged companions, whose mass is close to the DM one. One needs, therefore, to consider co-annihilation. The main free parameters determining the properties of the DM candidate are an additional Yukawa coupling $Y_{f0}$,  which does not enter the mass of the light SM fermions, and a mass splitting between the underlying Techni-fermions $\delta$. We also focus on the case $\delta \geq 0$, where the lightest odd states are mostly made of gauge singlets, in order to avoid strong constraints from Direct detection experiments.

We show that, under reasonable simplifying assumptions, the thermal relic abundance can saturate the measured value for masses of the lightest neutral pNGB between 500 GeV to 2 TeV, depending on the value of two free parameters. Remarkably, larger values of the masses are excluded by overabundance, thus providing an interesting upper limit on the condensation scale and on the fine-tuning parameter $\sin \theta$.
We also study the constraints coming from Direct and Indirect detection. Under the assumption of thermally produced DM, we find that only Direct Detection is sensitive to the available parameter space. In fact, regions with small and large values of the Yukawa coupling $Y_{f0}$ are almost excluded by the LUX bound, while intermediate regions still survive with roughly $0.1\ Y_f < Y_{f0} < 0.8\ Y_f$. In this window, the largest allowed mass is about 1 TeV for both choices of the mass splitting $\delta$ that we consider. Future projections of Direct detection experiments show that more parameter space will be probed in the near future. 
On the other hand, the cross sections for Indirect Detection always fall short of the current bounds by 1 to 2 orders of magnitude.
The fact that low mass DM candidates are preferred gives a change to the LHC for complementary reach on this model, even though we showed via very preliminary estimation that it is very challenging to produce and detect the additional scalar pNGBs.

We also tested the case where the DM is not thermally produced, thus matching the relic abundance to the observed one in the whole parameter space.  Direct detection, thus, excludes large values of the Yukawa up to large masses (3 TeV for $\delta = 0$ and 1.5 TeV for $\delta = 0.2$), while small values of $Y_{f0}$ are probed up to smaller masses (roughly 800 GeV in both cases). For indirect experiments, we find some sensitivity in the $t \bar{t}$ channel around 500 GeV, while the $WW$ channel allows to exclude masses up to 600 GeV for large $Y_{t0}$ (in the range $0.3 \div 1 \ Y_t$).

This class of models also has an additional candidate in the form of Techni-baryons made of $N$ Techni-fermions transforming as the fundamental of the SU($N$) confining group. While the thermal relic abundance is typically too low, an asymmetry may be generated in a similar fashion as baryogenesis in the SM, giving rise to an asymmetric DM candidate. However, in the simplest realization of this model, the TB number is a conserved quantity, and thus the sphaleron mechanism is not active, unless the UV physics generates additional interactions.

%%%%%%%%%%%%%%%%%%%%%%%%%%
\section*{Acknowledgments}

We thank the France China Particle Physics Laboratory (FCPPL) for supporting this project. T.M. is supported in part by project Y6Y2581B11 supported by 2016 National Postdoctoral Program for Innovative Talents. 
G.C. acknowledges partial support from the Labex-LIO (Lyon Institute of Origins) under grant ANR-10-LABX-66 and FRAMA (FR3127, F\'ed\'eration de Recherche ``Andr\'e Marie Amp\`ere").

%%%%%%%%%%%%%%%%%%%%%%%%%%%%%%%%%%%%%%%%%%%%%%%%%%%%%
\appendix

\section{Appendix: Interactions of the exotic pNGBs} \label{appint}

\subsection{Singlet $s$ } \label{appintA}

Below, we will review and discuss all the possible coupling leading to decays (and single production) for the DM-even singlet.

\begin{itemize}
\item[] {\bf Gauge and Chiral Lagrangian Interactions:} we first consider couplings generated by the gauged chiral Lagrangian. As we work in a basis where we expand around the true vacuum of the theory (which includes the misalignment), this term is fully invariant under a shift symmetry of all the pNGBs except the Higgs candidate. This is enough to ensure that no couplings linear in $s$ appear.

 \item[] {\bf WZW anomaly:} direct couplings of  $s$ to two gauge bosons are generated by the WZW anomaly, giving rise to the couplings
\begin{multline}
\mathcal{L}_{WZW}  = \frac{5iCg^2}{4c_{\theta_W} ^2} s \  \epsilon^{\mu \nu \alpha \beta  } (2\sqrt{2}c_{\theta_W} ^2 \cos \theta \partial_\beta W^+ _\mu  \partial_\nu W^- _\alpha   + \\
\sqrt{2}s_{ 2\theta_W }\cos \theta \partial_\beta Z _\mu  \partial_\nu A _\alpha   + \sqrt{2}c_{2\theta_W } \cos \theta \partial_\beta Z _\mu  \partial_\nu Z _\alpha     )
\end{multline}
with  $C=\frac{iN_{TC}}{240\pi ^2}  $.  The two-body  decay  width for  each channel  is
\beq
\Gamma( s \to W_\mu ^+ W_\mu  ^-)& =& \frac{g^4 \cos^2 \theta(M_{s}^2 -4m_W ^2  )^{\frac{3}{2} }    }{16384 \pi ^5  f^2} \,;  \nonumber \\
\Gamma(s \to Z_\mu  Z_\mu  )& =&   \frac{g^4 c_{2\theta_W} ^2 \cos^2 \theta(M_{s}^2 -4m_Z ^2  )^{\frac{3}{2} }    }{131072 \pi ^5  f^2 c_{\theta_W} ^4}\,;  \nonumber \\
\Gamma( s \to Z_\mu  A_\mu  ) &=& \frac{g^4s_{\theta_W}^2 \cos^2 \theta }{16384\pi ^5 M_{s} ^3  c_{\theta_W}^2 f^2} (M_{s} ^2- m_Z ^2   )^3 \,.
\eeq
Note that, like in SU(4)/Sp(4), no coupling to two photons appears~\cite{Arbey:2015exa}.

\item[] {\bf Yukawa couplings:}  the Yukawa coupling can potentially generate linear couplings of $s$ to a pair of fermions, however according to  Eq.~(\ref{eq:genralyukawa}) such coupling is  absent~\footnote{Additional couplings may be generated depending on the representation of the composite fermions in the partial compositeness scenario, see~\cite{Gripaios:2009pe,Serra:2015xfa} for examples.}.  This is  because, although  the general Yukawas break the parities $A$ and $B$, they are invariant under a global $SU(4) $ symmetry under which $s$ is odd:
\beq
  \Sigma \to  P^ \dagger \Sigma P\,,   \quad  P =\left(\begin{array}{cc}
0 & 1 \\
-1 & 0
\end{array} \right).
\eeq
Under the above parity:
\beq
  s \to -s \quad  H_2 \to -H_2 \quad  \Delta \leftrightarrow N.
\eeq
 The operator $P$ is an element of $SU(4)$ which exchanges the $SU(2)_L  $ and $SU(2)_R $ subgroups of $SU(4)$, and it is thus broken by the EW gauging.
According to this symmetry, the allowed couplings need to involve at least another pNGB:
\beq
s (\Delta - N)\ \bar{f} f'\,, \quad s H_2\ \bar{f} f\,, \quad s ^2\ \bar{f} f\,.
\eeq
From Table~\ref{table:topyukawa} in Appendix~\ref{app:Yukawa}, however, we see that only the coupling with the triplets is generated besides the one with $s ^2$, furthermore such coupling vanishes in the DM-preserving Yukawas (as it is proportional to $Y_{fD}$). Thus, we can assume that no decay of $s$ is generated by the Yukawa couplings.

\item[] {\bf Higher order couplings to fermions:}  as the symmetries forbidding the couplings of $s$ to two fermions are finally broken, higher order Yukawa-like operators will generate such couplings, in particular via breaking of the parity described in the above paragraph.
Such breaking is due to the gauging of the EW group, and to different techniquark masses for the two doublets (i.e. the parameter $\delta$). Following~\cite{Galloway:2010bp} we will focus here on the latter as an illustration~\footnote{See~\cite{Arbey:2015exa} for an example of the effect of gauge couplings plus Yukawas.}. The higher order operator we consider can be written as (where we take the top as an example):
\beq \label{eq:YukM}
\mathcal{L}_{\rm Yuk-2} = - (\bar{Q}_{L}^\alpha t_R) \, \Phi_\alpha + h.c.\,,
\eeq
with
\begin{multline}
\Phi =  \left[ \mbox{Tr} [M_{TQ} \Sigma   P_{1,\alpha} (y''_{t1} \Sigma + y''_{t2} \Sigma^\dagger)] + (i\sigma_2)_{\alpha \beta} \mbox{Tr} [M_{TQ} \Sigma P_2^\beta (y''_{t3} \Sigma + y''_{t4} \Sigma^\dagger)] \right]  \\
+ \left[ \mbox{Tr} [M_{TQ} \Sigma^\dagger   P_{1,\alpha} (y'_{t1} \Sigma + y'_{t2} \Sigma^\dagger)] + (i\sigma_2)_{\alpha \beta} \mbox{Tr} [M_{TQ} \Sigma^\dagger  P_2^\beta (y'_{t3} \Sigma + y'_{t4} \Sigma^\dagger)] \right] \,,
\end{multline}
where $M_{TQ}$ is the Techni--fermion mass matrix.
Even though, in principle, 8 new Yukawas appear in this operator, we can follow the simplifying assumptions that they are proportional to the lowest order Yukawas up to form factors: $y''_{ti} = a'' y_{ti}$ and $y'_{ti} = a' y_{ti}$.
We   find single  $s$  Yukawa coupling:
\beq
i \frac{(m_L +m_R)}{f} (a'' -a') \delta\ Y_t \sin \theta\; s\ \bar{t}_L t_R  + h.c. \label{eq:scoupling}
\eeq
which is proportional to the masses of the fermions, and also to the symmetry breaking parameter $\delta$. Note also the proportionality to $\delta_a = a'' - a'$ that breaks the A-parity.
To estimate the size of this effect, following~\cite{Arbey:2015exa} we compare it to the shift in the top mass generated by the same operators in Eq.~(\ref{eq:YukM}):
\beq \label{eq:deltamtop}
\delta_{ m_t} = \frac{1}{\sqrt{2}}(m_L +m_R) \sin 2\theta\  ((a'' + a') Y_t + \delta_a (Y_{t0} + (Y_{tD} + Y_{tT}) \delta)\,.
\eeq
Assuming that the combination of Yukawas appearing in Eqs.~(\ref{eq:scoupling}) and (\ref{eq:deltamtop}) are of the same order, the coupling of $s$ can be estimated as
\beq
g_{s\bar{t}_L t_R} \sim i \frac{2 \delta m_{t}}{v_{SM}} \delta \tan \theta \sim i\ 0.14 \ \delta \tan \theta\,,
\eeq
where we have allowed for a maximum 10\% correction to the top mass.

 Thus, the  partial decay  width  of the channel  $s \to \bar{f} f$ is
 \beq
 \Gamma(s \to \bar{f} f  ) =  \frac{6 |g_{s\bar{f}_L f_R} |  ^2}{16 \pi M_{s} ^2}  \sqrt{(M_{s} ^2  -4m_f ^2 )}  (M_{s} ^2 -2m_f ^2 ). \nonumber \\
 \eeq
As  $s$ couples to  quarks,  so  it  can couple to both gluon and photon  pairs  by quark loop (dominated by the top). The decay  widths of  these two  channels are~\cite{Arbey:2015exa}:
\beq
\Gamma(s \to g g ) &=& \frac{\alpha \alpha_s ^2 M_s ^3}{8\pi^2 m_W ^2 s^2 _W} \frac{|g_{s\bar{t}_L t_R}|^2 v^2  }{m_t ^2} F_1 ^2 ( x_t ).
\eeq
\beq
\Gamma(s \to  \gamma \gamma  ) = 1/2 N_c ^2 \frac{\alpha^2}{\alpha_s ^2}( \frac{2}{3}) ^4 \Gamma(s \to g g ),
\eeq
where  $F_1(x_t)$ is  the form factor of  top loop  and  $x_t =\frac{4m_t ^2}{M_s ^2} $,
\beq
F_1 (x_t) = 1/2 x_t (1+ ( 1-x_t)\sin ^2 (x_t ^{-1/2} )    ).
\eeq

\item[] {\bf Scalar potential:} decays of $s$ to a pair of pNGBs can also be generated via the potential that aligns the vacuum. Following the leading potential we used in this paper, the couplings read:
\begin{multline} \label{eq:pipis}
\mathcal{L}_{\rm pot} \supset \frac{C_t f}{\sqrt{2}} \sin^2 \theta Y_f \mbox{Im}(Y_{f0})\, s\left( \frac{1}{2} (N_0^2 - \Delta_0^2) + (N^+ N^- - \Delta^+ \Delta^-) \right) + \\
\frac{C_g f}{32} \sin \theta\ s\ \left\{(4 g^2 \pm (g^2-{g'}^2) \cos \theta)\, A_0 \Delta_0/N_0 \right. \\
\left. + i (4 g^2 - 2 {g'}^2 \pm (g^2 + {g'}^2) \cos \theta)\ (H^+ \Delta^-/N^- - \Delta^+/N^+ H^-) \right\}\,;
\end{multline}
where the couplings are evaluated at the minimum of the potential. We thus see that $s$ can only decay into a pair of DM-odd pNGBs, however those decays will always be kinematically unaccessible~\footnote{In principle it is possible to have different mass spectra, however the couplings will also be modified.}.

\end{itemize}

\subsection{DM-odd scalars} \label{appintB}

The interactions relevant for production and decays are the following, where we label by $\pi_i$ a generic odd scalar.

\begin{itemize}

\item[]  {\bf Gauge interactions} are generated by the lowest order chiral Lagrangian, and contain couplings of the form  $g_{\pi_1 \pi _2 V } V_\mu \pi_1 \overleftrightarrow{ \partial^\mu } \pi_2$ (see Ref.~\cite{Ma:2015gra} for the explicit expressions). The corresponding   decay  width  is
 \beq
 \Gamma(\pi_2  \to  \pi_1  V_\mu  )  = |g_{\pi_1 \pi _2 V }|^2   \frac{\{(M_{\pi_2} ^2 -(M_{\pi_1 } +m_V)^2 )(M_{\pi_2} ^2 -(M_{\pi_1 } -m_V)^2 )   \} ^{ \frac{3}{2}}  }{16\pi m_V ^2 M_{\pi_2} ^3}\,.
 \eeq

\item[]  {\bf Yukawa couplings} generate interactions of two pNGBs with two fermions, as in Table~\ref{table:topyukawa}, thus decays of the form $\pi_2 \to \pi_1 \bar{f} f$ are generated. However,  this process has usually very small rates, so we will not consider it further.

\item[] {\bf The scalar potential} generates couplings between 3 pNGBs, thus giving rise to decays $\pi_2 \to \pi_1 \phi$, with $\phi = s, h_1$.  This  partial  decay  width  is
\beq
 \Gamma(\pi_2  \to  \pi_1   \phi ) = \frac{|g_{\pi_2 \pi_1 \phi}|^2}{16\pi M_{\pi_2} ^3} \sqrt{  \lambda(M_{\pi_2} ^2, M_{\pi_1} ^2, M_{\phi} ^2) }
\eeq
with  $\lambda( x,y,z) =  x^2 +y^2 +z^2 -2xy -2yz -2xz   $.
The couplings with the singlet $s $ are in Eq.~(\ref{eq:pipis}): within our choice of potential, the masses of the 3 pNGBs will always make the decay kinematically inaccessible.  On the other hand, decays with the lighter Higgs boson $h_1$ are allowed: keeping only the terms proportional to the top Yukawas, the couplings can be written as
\begin{multline} \label{eq:pipih}
\mathcal{L}_{\rm pot} \supset \frac{C_t f}{2 \sqrt{2}} Y_f^2 \sin (2 \theta)\, h_1 \left( \frac{1}{2} (h_2^2 + A_0^2 + \Delta_0^2 + N_0^2) + H^+ H^- + \Delta^+ \Delta^- + N^+ N^-\right) \\
+ \frac{C_t f}{2} Y_f\, h_1 \left( \mbox{Re} (Y_{f0}) \cos \theta\ h_2 + \mbox{Im} (Y_{f0}) \cos (2\theta)\ A_0 \right) (\Delta_0 + N_0 ) \\
+ \frac{C_t f}{2} Y_f\, h_1 \left( \mbox{Re} (Y_{f0}) \cos \theta\ H^+ (N^- - \Delta^-) + i \mbox{Im} (Y_{f0}) \cos (2 \theta) \ H^+ (N^- + \Delta^-) \right) + h.c,
\end{multline}
plusing additional corrections from the gauge interactions.

\end{itemize}

%%%%%%%%%%%%%%%%%%%%%%%%%%%%%%%%
\section{Appendix: Relevant couplings} \label{appapp}

In this section we list all the couplings of the pNGBs, relevant for the calculation of the relic abundance, and other properties of the DM candidate.

\subsection{Potential}

\subsubsection*{Trilinear couplings}

Defining:
\beq
\mathcal{R}_Y &=& \mbox{Re} [Y_u\cdot Y_{u0}] - \mbox{Re} [Y_d\cdot Y_{d0}] - \mbox{Re} [Y_e\cdot Y_{e0}]\,, \\
\mathcal{I}_Y &=& \mbox{Im} [Y_u\cdot Y_{u0}] + \mbox{Im} [Y_d\cdot Y_{d0}] + \mbox{Im} [Y_e\cdot Y_{e0}]\,.
\eeq

\beq
\mathcal{L}_{\rm tri} = \frac{m_h^2}{2 v_{\rm SM}} \cos \theta \left( h_1^3 + h_1 s^2 \right) + g_{s\pi\pi}\ s \pi_i \pi_j + g_{h\pi\pi}\ h_1 \pi_i\pi_j\,,
\eeq

\beq
g_{sN_0N_0}  = - g_{s\Delta_0\Delta_0} &=& \frac{C_t f}{2 \sqrt{2}} \sin^2 \theta\, \mathcal{I}_Y\,,\\
g_{sN^+N^-}  = - g_{s\Delta^+\Delta^-} &=& \frac{C_t f}{\sqrt{2}} \sin^2 \theta\, \mathcal{I}_Y\,,\\
g_{sA_0N_0/\Delta_0} &=& \frac{C_g}{16 \sqrt{2}} \frac{4 m_W^2 \mp (2 m_W^2 - m_Z^2)\cos \theta}{v_{\rm SM}}\,,\\
g_{sH^+N^-/\Delta^-} = g^\ast_{sH^-N^+/\Delta^+} &=& i \frac{C_g}{16 \sqrt{2}} \frac{6 m_W^2 - 2 m_Z^2 \mp m_Z^2 \cos \theta}{v_{\rm SM}}\,.
\eeq

\beq
g_{hh_2h_2} &=& \frac{m_h^2}{2 v_{\rm SM}}\cos \theta\,,\\
g_{hA_0A_0} &=& \frac{\cos \theta}{2 v_{\rm SM}} \left( m_h^2 + \frac{C_g}{8} (2 m_W^2 - m_Z^2) \right)\,,\\
g_{hH^+H^-} &=& \frac{\cos \theta}{v_{\rm SM}} \left( m_h^2 + \frac{C_g}{8} m_Z^2 \right)\,,\\
g_{hN_0N_0/\Delta_0\Delta_0} &=& \frac{\cos \theta}{2 v_{\rm SM}} \left( m_h^2 - \frac{C_g}{16} (2 m_W^2 - m_Z^2) \pm \frac{C_g}{4} \frac{m_W^2}{\cos \theta} \right)\,,\\
g_{hN_0\Delta_0} &=& - \frac{C_g}{16 v_{\rm SM}} (2 m_W^2 - m_Z^2) \cos \theta\,, \\
g_{hN^+N^-/\Delta^+\Delta^-} &=& \frac{\cos \theta}{v_{\rm SM}} \left(m_h^2 - \frac{C_g}{16} m_Z^2 \pm \frac{C_g}{8} \frac{3 m_W^2 - m_Z^2}{\cos \theta} \right)\,, \\
g_{hN^+\Delta^-} = g_{h\Delta^+N^-} &=& - \frac{\cos \theta}{16 v_{\rm SM}} C_g m_Z^2\,,\\
g_{hh_2N_0} = g_{hh_2\Delta_0} &=& \frac{C_t f}{2}  \cos \theta\ \mathcal{R}_Y\,, \\
g_{hA_0N_0} = g_{hA_0\Delta_0} &=& \frac{C_t f}{2}  \cos 2\theta\ \mathcal{I}_Y\,, \\
g_{hH^+N^-/\Delta^-} =  g^\ast_{hH^-N^+/\Delta^+}&=& \frac{C_t f}{2} \left( \pm  \cos \theta\ \mathcal{R}_Y + i \cos 2 \theta\ \mathcal{I}_Y \right)\,.
\eeq

\subsubsection*{Quartic couplings}

\begin{multline}
\mathcal{L}_{\rm qua} = \frac{m_h^2}{24 v^2_{\rm SM}}\left( (3-7 \sin^2 \theta) h_1^4 + 2 (1-5 \sin^2 \theta) h_1^2 s^2  - (1+3 \sin^2 \theta) s^4\right) + \\
g_{ss\pi\pi} s^2 \pi_i \pi_j + g_{sh\pi\pi} s h \pi_i \pi_j + g_{hh\pi\pi} h^2 \pi_i \pi_j + g_{\pi\pi\pi\pi} \pi_i \pi_j \pi_l \pi_k\,.
\end{multline}

\beq
g_{ssh_2h_2} &=& -\frac{1}{12 v_{\rm SM}^2} \left( m_h^2 (1+3 \sin^2 \theta) + \frac{C_g}{8} (2 m_W^2 + m_Z^2\right)\,, \\
g_{ssA_0A_0} &=& -\frac{1}{12 v_{\rm SM}^2} \left( m_h^2 (1+3 \sin^2 \theta) + \frac{C_g}{8} (2 (1+\sin^2 \theta) m_W^2 + \cos^2 \theta m_Z^2\right)\,, \\
g_{ssN_0N_0/\Delta_0\Delta_0} &=& - \frac{m_h^2}{4 v_{\rm SM}^2} (1 +\sin^2 \theta \mp \delta \cos \theta)\,, \\
g_{ssN_0\Delta_0} &=& \frac{C_g}{16 v_{\rm SM}^2} \sin^2 \theta (2 m_W^2 - m_Z^2)\,, \\
g_{ssh_2N_0/\Delta_0} &=& - \frac{C_t}{6 \sqrt{2}} \sin \theta\ \mathcal{R}_Y\,, \\
g_{ssA_0N_0/\Delta_0}  &=& - \frac{C_t}{6 \sqrt{2}} \sin \theta \cos \theta\ \mathcal{I}_Y\,,
\eeq

\beq
g_{ssH^+H^-} &=& - \frac{1}{6 v_{\rm SM}^2} \left( m_h^2 (1+3 \sin^2 \theta) + \frac{C_g}{8} \left( 2m_W^2 + m_Z^2 (1+\sin^2 \theta) \right) \right) \,,\\
g_{ssN^+N^-/\Delta^+\Delta^-} &=& -\frac{m_h^2}{2v_{\rm SM}^2} (1+\sin^2 \theta \mp \delta \cos \theta)\,. \\
g_{ssN^+\Delta^-} &=& g_{ssN^-\Delta^+} = \frac{C_g}{16 v_{\rm SM}^2} \sin^2 \theta m_z^2\,,\\
g_{ssH^+N^-/\Delta^-} &=& g_{ssH^-N^+/\Delta^+}^\ast = -\frac{C_t}{6 \sqrt{2}} \sin \theta (\pm \mathcal{R}_Y + i \cos \theta\ \mathcal{I}_Y)\,.
\eeq

\beq
g_{shA_0 N_0/\Delta_0} &=& \pm \frac{1}{6 \sqrt{2} v_{\rm SM}^2} \left( m_h^2 (\sin^2 \theta \mp \delta \cos \theta) - \phantom{\frac{C_g}{16} } \right. \\
&& \left. \frac{C_g}{16} \left( 2 m_W^2 (3 - 4 \sin^2 \theta \mp 4 \cos \theta) - m_Z^2 (1-4 \sin^2 \theta)\right)   \right)\,, \\
g_{shN_0N_0} &=& - g_{sh\Delta_0\Delta_0} = \frac{C_t}{6} \sin \theta \cos \theta\ \mathcal{I}_Y\,.
\eeq

\beq
g_{shH^+N^-/\Delta^-} &=& g_{shH^-N^+/\Delta^+}^\ast = \pm \frac{i}{3 \sqrt{2}} \left( m_h^2 (\sin^2 \theta \mp  \delta \cos \theta) - \phantom{\frac{C_g}{16}} \right. \\
&&\left. \frac{C_g}{16} \left( m_W^2 2 (1\mp 6 \cos \theta) + m_Z^2 (3 - 4 \sin^2 \theta \pm 4 \cos \theta) \right) \right)\,, \\
g_{s h N^+ N^-/\Delta^+\Delta^-} &=& \pm \frac{C_t}{6} \sin 2 \theta \ \mathcal{I}_Y\,.
\eeq

\beq
g_{hhh_2h_2} &=& \frac{1}{12 v_{\rm SM}^2} \left( m_h^2 (1-5 \sin^2 \theta) - \frac{C_g}{8} (2 m_W^2 + m_Z^2) \right)\, \\
g_{hhA_0A_0} &=& \frac{1}{4 v_{\rm SM}^2} \left( m_h^2 (1-3 \sin^2 \theta) + \frac{C_g}{8} \cos (2\theta) (2 m_W^2 - m_Z^2) \right)\,\\
g_{hhN_0N_0/\Delta_0\Delta_0} &=& \frac{1}{12 v_{\rm SM}^2} \left( m_h^2 (2-7 \sin^2 \theta \pm 2 \delta \cos \theta) - \phantom{\frac{C_g}{16}} \right. \\
 & & \left. \frac{C_g}{16} \left( m_W^2 2(5-6 \sin^2 \theta \mp 8 \cos \theta) - m_Z^2 3 (1-2\sin^2 \theta)\right) \right)\,,  \\
 g_{hhN_0\Delta_0} &=& \frac{1}{6 v_{\rm SM}^2} \left( m_h^2 (\cos 2 \theta) - \frac{C_g}{16} \left( m_W^2 2(1-6\sin^2 \theta) - m_Z^2 3 (\cos 2 \theta)\right)\right)\,, \\
 g_{hhA_0N_0/\Delta_0} &=& - \frac{C_t}{4 \sqrt{2}} \sin 2 \theta\ \mathcal{I}_Y\,, \\
 g_{hhh_2N_0/\Delta_0} &=& - \frac{C_t}{6 \sqrt{2}} \sin \theta\ \mathcal{R}_Y\,.
\eeq

\beq
g_{hhH^+H^-} &=& \frac{1}{2 v_{\rm SM}^2} \left( m_h^2 (1-3 \sin^2 \theta) + \frac{C_g}{8} m_Z^2 \cos 2 \theta \right)\,, \\
g_{hhN^+N^-/\Delta^+\Delta^-} &=& \frac{1}{6 v_{\rm SM}^2} \left( m_h^2 (2-7 \sin^2 \theta \pm \delta \cos \theta) - \phantom{\frac{C_g}{16}} \right. \\
&& \left. \frac{C_g}{16} \left( m_W^2 2 (1\mp12 \cos \theta) +m_Z^2 (5-6 \sin^2 \theta \pm 8 \cos \theta) \right) \right)\,, \\
g_{hhH^+ N^-/\Delta^-} &=& g_{hhH^- N^+/\Delta^+}^\ast = - \frac{C_t}{6 \sqrt{2}} \sin \theta \left( \pm \mathcal{R}_Y + 3 i \cos \theta\ \mathcal{I}_Y \right)\,.
\eeq

\subsection{Yukawa couplings} \label{app:Yukawa}

The couplings of two pNGBs to fermions, defined in Eq.~(\ref{eq:NLcoup}), are listed in the following Table~\ref{table:topyukawa}.

\begin{table}[h!]
\begin{center}
\footnotesize{
\begin{tabular}{c|cccccc|c}

$\xi^N_u, \xi^N_{d,e}$  &$ h_1$ & $h_2$ & $A_0$ & $\Delta_0$ & $ N_0$ &$ s$  \\
\hline
&&&&&& & \\
$h_1$  & $Y_f\sin\theta$  & $0$  & $\pm {\bf Y_{fD}}\sin\theta  $  &$ \mp i \frac{\bf{Y_{fT}}}{\sqrt{2} }\cos\theta  $     & $ \mp i \frac{\bf{Y_{fT}}}{\sqrt{2}}\cos\theta  $ & $0$  \\

$h_2 $ & $0$   & $Y_f\sin\theta$ &  $0$ & $\pm \frac{Y_{f0}}{\sqrt{2}}  $ & $\pm \frac{Y_{f0}}{\sqrt{2}}  $&  $0$ \\

$A_0$  &$\pm \bf{Y_{fD}}\sin\theta  $   &  $0$ &  $Y_f\sin\theta$ & $-i \frac{Y_{f0}}{\sqrt{2}}\cos \theta  $  & $-i\frac{Y_{f0}}{\sqrt{2} }\cos \theta  $ &  $0$  \\
$\Delta$  &$ \mp i \frac{\bf{Y_{fT}}}{\sqrt{2}}\cos\theta  $   & $\pm \frac{Y_{f0}}{\sqrt{2}} $  &$-i \frac{Y_{f0}}{\sqrt{2}}\cos \theta  $  & $Y_f\sin\theta$  & $0$ & $ \pm \frac{\bf{Y_{fD}}}{\sqrt{2}}\sin\theta  $ \\
$N$  &$ \mp i\frac{\bf{Y_{fT}}}{\sqrt{2}}\cos\theta $   &$\pm \frac{Y_{f0}}{\sqrt{2}}$  &$-i\frac{ Y_{f0}}{\sqrt{2}}\cos \theta $  & $0$ & $Y_f\sin\theta$ & $\mp \frac{ \bf{Y_{fD}}}{\sqrt{2}}\sin\theta$   \\
$s$  &$0$   &$0$  &$0$  &$\pm \frac{ \bf{Y_{fD}}}{\sqrt{2}}\sin\theta$ &$\mp \frac{ \bf{Y_{fD}}}{\sqrt{2}}\sin\theta$ &  $Y_f\sin\theta$ \\
\hline
&&&&&& & \\
\end{tabular}

\begin{tabular}{c|ccc|c}
 $ \xi_u^+, \xi_{d,e}^-$ &$H^\mp$ & $\Delta^\mp$ & $N^\mp $  \\
\hline
&&& &\\
$h_1$  &    $ \mp i\sqrt{2} \bf{Y_{fD}}\sin\theta  $  & $ -i\bf{Y_{fT}}\cos\theta  $ & $ -i\bf{Y_{fT}}\cos\theta  $ \\
$h_2 $ &$0$ & $Y_{f0} $&$Y_{f0} $  \\
$A_0$  & $0  $  &  $\mp iY_{f0}  $ &$\pm iY_{f0}  $\\
$\Delta$  & $Y_{f0} $ & $ 0$ & $0$        \\
$N$  & $-Y_{f0}$ &  $0$ &$0$ \\
$s$  &$0$ & $\bf{Y_{fD}}\sin\theta$&$-\bf{Y_{fD}}\sin\theta$ \\
\hline
&&& &\\
\end{tabular}

\begin{tabular}{c|ccc|c}
$\xi^C_u, \xi^C_{d,e}$&$H^-$ & $\Delta^-$ & $N^- $  \\
\hline
&&& &\\
$H^+$  & $Y_f\sin\theta$ &$\mp Y_{f0}\frac{(1\mp \cos\theta)}{\sqrt{2}}$ &$\pm Y_{f0}\frac{(1\pm\cos\theta)}{\sqrt{2}}$   \\
$\Delta^+$  & $\mp Y_{f0}\frac{(1\pm\cos\theta)}{\sqrt{2}}$ & $Y_f\sin\theta$ & 0\\
$N^+$  &$\pm Y_{f0}\frac{(1\mp\cos\theta)}{\sqrt{2}}$ & $0$ &  $Y_f\sin\theta$   \\
\hline
&&& &\\
\end{tabular}
}
\caption{Couplings of two pNGBs to fermions generated by the effective Yukawas. The  first  table lists the couplings  of two neutral pNGBs; the second one  lists the couplings of one charged and one neutral pNGB;  and  the  third  one  lists the couplings of two charged pNGBs (the signs refer to up and down/leptons, respectively). } \label{table:topyukawa}
\end{center}
\end{table}

\newpage

\bibliographystyle{JHEP}

\bibliography{CompoDM}

\providecommand{\href}[2]{#2}\begingroup\raggedright\begin{thebibliography}{100}

\bibitem{Aad:2012tfa}
{\scshape ATLAS} collaboration, G.~Aad et~al., \emph{{Observation of a new
  particle in the search for the Standard Model Higgs boson with the ATLAS
  detector at the LHC}},
  \href{http://dx.doi.org/10.1016/j.physletb.2012.08.020}{\emph{Phys. Lett.}
  {\bf B716} (2012) 1--29}, [\href{http://arxiv.org/abs/1207.7214}{{\tt
  1207.7214}}].

\bibitem{Chatrchyan:2012xdj}
{\scshape CMS} collaboration, S.~Chatrchyan et~al., \emph{{Observation of a new
  boson at a mass of 125 GeV with the CMS experiment at the LHC}},
  \href{http://dx.doi.org/10.1016/j.physletb.2012.08.021}{\emph{Phys. Lett.}
  {\bf B716} (2012) 30--61}, [\href{http://arxiv.org/abs/1207.7235}{{\tt
  1207.7235}}].

\bibitem{ArkaniHamed:2002qx}
N.~Arkani-Hamed, A.~G. Cohen, E.~Katz, A.~E. Nelson, T.~Gregoire and J.~G.
  Wacker, \emph{{The Minimal moose for a little Higgs}},
  \href{http://dx.doi.org/10.1088/1126-6708/2002/08/021}{\emph{JHEP} {\bf 08}
  (2002) 021}, [\href{http://arxiv.org/abs/hep-ph/0206020}{{\tt
  hep-ph/0206020}}].

\bibitem{Schmaltz:2002wx}
M.~Schmaltz, \emph{{Physics beyond the standard model (theory): Introducing the
  little Higgs}},
  \href{http://dx.doi.org/10.1016/S0920-5632(03)01409-9}{\emph{Nucl. Phys.
  Proc. Suppl.} {\bf 117} (2003) 40--49},
  [\href{http://arxiv.org/abs/hep-ph/0210415}{{\tt hep-ph/0210415}}].

\bibitem{Chacko:2005pe}
Z.~Chacko, H.-S. Goh and R.~Harnik, \emph{{The Twin Higgs: Natural electroweak
  breaking from mirror symmetry}},
  \href{http://dx.doi.org/10.1103/PhysRevLett.96.231802}{\emph{Phys. Rev.
  Lett.} {\bf 96} (2006) 231802},
  [\href{http://arxiv.org/abs/hep-ph/0506256}{{\tt hep-ph/0506256}}].

\bibitem{Csaki:2017cep}
C.~Csaki, T.~Ma and J.~Shu, \emph{{The Maximally Symmetric Composite Higgs}},
  \href{http://arxiv.org/abs/1702.00405}{{\tt 1702.00405}}.

\bibitem{Manton:1979kb}
N.~S. Manton, \emph{{A New Six-Dimensional Approach to the Weinberg-Salam
  Model}}, \href{http://dx.doi.org/10.1016/0550-3213(79)90192-5}{\emph{Nucl.
  Phys.} {\bf B158} (1979) 141--153}.

\bibitem{Fairlie:1979at}
D.~B. Fairlie, \emph{{Higgs' Fields and the Determination of the Weinberg
  Angle}}, \href{http://dx.doi.org/10.1016/0370-2693(79)90434-9}{\emph{Phys.
  Lett.} {\bf B82} (1979) 97--100}.

\bibitem{Hosotani:1983xw}
Y.~Hosotani, \emph{{Dynamical Mass Generation by Compact Extra Dimensions}},
  \href{http://dx.doi.org/10.1016/0370-2693(83)90170-3}{\emph{Phys. Lett.} {\bf
  B126} (1983) 309--313}.

\bibitem{Weinberg:1975gm}
S.~Weinberg, \emph{{Implications of Dynamical Symmetry Breaking}},
  \href{http://dx.doi.org/10.1103/PhysRevD.13.974}{\emph{Phys. Rev.} {\bf D13}
  (1976) 974--996}.

\bibitem{Susskind:1978ms}
L.~Susskind, \emph{{Dynamics of Spontaneous Symmetry Breaking in the
  Weinberg-Salam Theory}},
  \href{http://dx.doi.org/10.1103/PhysRevD.20.2619}{\emph{Phys. Rev.} {\bf D20}
  (1979) 2619--2625}.

\bibitem{Dimopoulos:1979es}
S.~Dimopoulos and L.~Susskind, \emph{{Mass Without Scalars}},
  \href{http://dx.doi.org/10.1016/0550-3213(79)90364-X}{\emph{Nucl. Phys.} {\bf
  B155} (1979) 237--252}.

\bibitem{Eichten:1979ah}
E.~Eichten and K.~D. Lane, \emph{{Dynamical Breaking of Weak Interaction
  Symmetries}},
  \href{http://dx.doi.org/10.1016/0370-2693(80)90065-9}{\emph{Phys. Lett.} {\bf
  B90} (1980) 125--130}.

\bibitem{Farhi:128227}
E.~Farhi and L.~Susskind, \emph{{Technicolour}}, {\emph{Phys. Rep.} {\bf 74}
  (Oct, 1980) 277--321. 66 p}.

\bibitem{Peskin:1990zt}
M.~E. Peskin and T.~Takeuchi, \emph{{A New constraint on a strongly interacting
  Higgs sector}},
  \href{http://dx.doi.org/10.1103/PhysRevLett.65.964}{\emph{Phys. Rev. Lett.}
  {\bf 65} (1990) 964--967}.

\bibitem{Kaplan:1983fs}
D.~B. Kaplan and H.~Georgi, \emph{{SU(2) x U(1) Breaking by Vacuum
  Misalignment}},
  \href{http://dx.doi.org/10.1016/0370-2693(84)91177-8}{\emph{Phys. Lett.} {\bf
  B136} (1984) 183--186}.

\bibitem{Kaplan:1983sm}
D.~B. Kaplan, H.~Georgi and S.~Dimopoulos, \emph{{Composite Higgs Scalars}},
  \href{http://dx.doi.org/10.1016/0370-2693(84)91178-X}{\emph{Phys. Lett.} {\bf
  B136} (1984) 187--190}.

\bibitem{Georgi:1984af}
H.~Georgi and D.~B. Kaplan, \emph{{Composite Higgs and Custodial SU(2)}},
  \href{http://dx.doi.org/10.1016/0370-2693(84)90341-1}{\emph{Phys. Lett.} {\bf
  B145} (1984) 216--220}.

\bibitem{Holdom:1981rm}
B.~Holdom, \emph{{Raising the Sideways Scale}},
  \href{http://dx.doi.org/10.1103/PhysRevD.24.1441}{\emph{Phys. Rev.} {\bf D24}
  (1981) 1441}.

\bibitem{Yamawaki:1985zg}
K.~Yamawaki, M.~Bando and K.-i. Matumoto, \emph{{Scale Invariant Technicolor
  Model and a Technidilaton}},
  \href{http://dx.doi.org/10.1103/PhysRevLett.56.1335}{\emph{Phys. Rev. Lett.}
  {\bf 56} (1986) 1335}.

\bibitem{Bando:1986bg}
M.~Bando, K.-i. Matumoto and K.~Yamawaki, \emph{{Technidilaton}},
  \href{http://dx.doi.org/10.1016/0370-2693(86)91516-9}{\emph{Phys. Lett.} {\bf
  B178} (1986) 308--312}.

\bibitem{Dietrich:2005jn}
D.~D. Dietrich, F.~Sannino and K.~Tuominen, \emph{{Light composite Higgs from
  higher representations versus electroweak precision measurements: Predictions
  for CERN LHC}},
  \href{http://dx.doi.org/10.1103/PhysRevD.72.055001}{\emph{Phys. Rev.} {\bf
  D72} (2005) 055001}, [\href{http://arxiv.org/abs/hep-ph/0505059}{{\tt
  hep-ph/0505059}}].

\bibitem{Appelquist:2010gy}
T.~Appelquist and Y.~Bai, \emph{{A Light Dilaton in Walking Gauge Theories}},
  \href{http://dx.doi.org/10.1103/PhysRevD.82.071701}{\emph{Phys. Rev.} {\bf
  D82} (2010) 071701}, [\href{http://arxiv.org/abs/1006.4375}{{\tt
  1006.4375}}].

\bibitem{Elander:2010wd}
D.~Elander and M.~Piai, \emph{{Light scalars from a compact fifth dimension}},
  \href{http://dx.doi.org/10.1007/JHEP01(2011)026}{\emph{JHEP} {\bf 01} (2011)
  026}, [\href{http://arxiv.org/abs/1010.1964}{{\tt 1010.1964}}].

\bibitem{Foadi:2012bb}
R.~Foadi, M.~T. Frandsen and F.~Sannino, \emph{{125 GeV Higgs boson from a not
  so light technicolor scalar}},
  \href{http://dx.doi.org/10.1103/PhysRevD.87.095001}{\emph{Phys. Rev.} {\bf
  D87} (2013) 095001}, [\href{http://arxiv.org/abs/1211.1083}{{\tt
  1211.1083}}].

\bibitem{Fodor:2012ty}
Z.~Fodor, K.~Holland, J.~Kuti, D.~Nogradi, C.~Schroeder and C.~H. Wong,
  \emph{{Can the nearly conformal sextet gauge model hide the Higgs
  impostor?}},
  \href{http://dx.doi.org/10.1016/j.physletb.2012.10.079}{\emph{Phys. Lett.}
  {\bf B718} (2012) 657--666}, [\href{http://arxiv.org/abs/1209.0391}{{\tt
  1209.0391}}].

\bibitem{Fodor:2015vwa}
Z.~Fodor, K.~Holland, J.~Kuti, S.~Mondal, D.~Nogradi and C.~H. Wong,
  \emph{{Toward the minimal realization of a light composite Higgs}},
  {\emph{PoS} {\bf LATTICE2014} (2015) 244},
  [\href{http://arxiv.org/abs/1502.00028}{{\tt 1502.00028}}].

\bibitem{Holdom:1986ub}
B.~Holdom and J.~Terning, \emph{{A Light Dilaton in Gauge Theories?}},
  \href{http://dx.doi.org/10.1016/0370-2693(87)91109-9}{\emph{Phys. Lett.} {\bf
  B187} (1987) 357--361}.

\bibitem{Holdom:1987yu}
B.~Holdom and J.~Terning, \emph{{No Light Dilaton in Gauge Theories}},
  \href{http://dx.doi.org/10.1016/0370-2693(88)90783-6}{\emph{Phys. Lett.} {\bf
  B200} (1988) 338--342}.

\bibitem{Maldacena:1997re}
J.~M. Maldacena, \emph{{The Large N limit of superconformal field theories and
  supergravity}}, \href{http://dx.doi.org/10.1023/A:1026654312961}{\emph{Int.
  J. Theor. Phys.} {\bf 38} (1999) 1113--1133},
  [\href{http://arxiv.org/abs/hep-th/9711200}{{\tt hep-th/9711200}}].

\bibitem{Randall:1999vf}
L.~Randall and R.~Sundrum, \emph{{An Alternative to compactification}},
  \href{http://dx.doi.org/10.1103/PhysRevLett.83.4690}{\emph{Phys. Rev. Lett.}
  {\bf 83} (1999) 4690--4693}, [\href{http://arxiv.org/abs/hep-th/9906064}{{\tt
  hep-th/9906064}}].

\bibitem{Agashe:2004rs}
K.~Agashe, R.~Contino and A.~Pomarol, \emph{{The Minimal composite Higgs
  model}}, \href{http://dx.doi.org/10.1016/j.nuclphysb.2005.04.035}{\emph{Nucl.
  Phys.} {\bf B719} (2005) 165--187},
  [\href{http://arxiv.org/abs/hep-ph/0412089}{{\tt hep-ph/0412089}}].

\bibitem{Kaplan:1991dc}
D.~B. Kaplan, \emph{{Flavor at SSC energies: A New mechanism for dynamically
  generated fermion masses}},
  \href{http://dx.doi.org/10.1016/S0550-3213(05)80021-5}{\emph{Nucl. Phys.}
  {\bf B365} (1991) 259--278}.

\bibitem{Contino:2004vy}
R.~Contino and A.~Pomarol, \emph{{Holography for fermions}},
  \href{http://dx.doi.org/10.1088/1126-6708/2004/11/058}{\emph{JHEP} {\bf 11}
  (2004) 058}, [\href{http://arxiv.org/abs/hep-th/0406257}{{\tt
  hep-th/0406257}}].

\bibitem{Cacciapaglia:2008bi}
G.~Cacciapaglia, G.~Marandella and J.~Terning, \emph{{Dimensions of
  Supersymmetric Operators from AdS/CFT}},
  \href{http://dx.doi.org/10.1088/1126-6708/2009/06/027}{\emph{JHEP} {\bf 06}
  (2009) 027}, [\href{http://arxiv.org/abs/0802.2946}{{\tt 0802.2946}}].

\bibitem{Galloway:2010bp}
J.~Galloway, J.~A. Evans, M.~A. Luty and R.~A. Tacchi, \emph{{Minimal Conformal
  Technicolor and Precision Electroweak Tests}},
  \href{http://dx.doi.org/10.1007/JHEP10(2010)086}{\emph{JHEP} {\bf 10} (2010)
  086}, [\href{http://arxiv.org/abs/1001.1361}{{\tt 1001.1361}}].

\bibitem{Cacciapaglia:2014uja}
G.~Cacciapaglia and F.~Sannino, \emph{{Fundamental Composite (Goldstone) Higgs
  Dynamics}}, \href{http://dx.doi.org/10.1007/JHEP04(2014)111}{\emph{JHEP} {\bf
  04} (2014) 111}, [\href{http://arxiv.org/abs/1402.0233}{{\tt 1402.0233}}].

\bibitem{Ma:2015gra}
T.~Ma and G.~Cacciapaglia, \emph{{Fundamental Composite 2HDM: SU(N) with 4
  flavours}}, \href{http://dx.doi.org/10.1007/JHEP03(2016)211}{\emph{JHEP} {\bf
  03} (2016) 211}, [\href{http://arxiv.org/abs/1508.07014}{{\tt 1508.07014}}].

\bibitem{Barnard:2013zea}
J.~Barnard, T.~Gherghetta and T.~S. Ray, \emph{{UV descriptions of composite
  Higgs models without elementary scalars}},
  \href{http://dx.doi.org/10.1007/JHEP02(2014)002}{\emph{JHEP} {\bf 02} (2014)
  002}, [\href{http://arxiv.org/abs/1311.6562}{{\tt 1311.6562}}].

\bibitem{Ferretti:2013kya}
G.~Ferretti and D.~Karateev, \emph{{Fermionic UV completions of Composite Higgs
  models}}, \href{http://dx.doi.org/10.1007/JHEP03(2014)077}{\emph{JHEP} {\bf
  03} (2014) 077}, [\href{http://arxiv.org/abs/1312.5330}{{\tt 1312.5330}}].

\bibitem{Ferretti:2014qta}
G.~Ferretti, \emph{{UV Completions of Partial Compositeness: The Case for a
  SU(4) Gauge Group}},
  \href{http://dx.doi.org/10.1007/JHEP06(2014)142}{\emph{JHEP} {\bf 06} (2014)
  142}, [\href{http://arxiv.org/abs/1404.7137}{{\tt 1404.7137}}].

\bibitem{vonGersdorff:2015fta}
G.~von Gersdorff, E.~Pont\'on and R.~Rosenfeld, \emph{{The Dynamical Composite
  Higgs}}, \href{http://dx.doi.org/10.1007/JHEP06(2015)119}{\emph{JHEP} {\bf
  06} (2015) 119}, [\href{http://arxiv.org/abs/1502.07340}{{\tt 1502.07340}}].

\bibitem{Vecchi:2015fma}
L.~Vecchi, \emph{{A dangerous irrelevant UV-completion of the composite
  Higgs}}, \href{http://dx.doi.org/10.1007/JHEP02(2017)094}{\emph{JHEP} {\bf
  02} (2017) 094}, [\href{http://arxiv.org/abs/1506.00623}{{\tt 1506.00623}}].

\bibitem{Ryttov:2008xe}
T.~A. Ryttov and F.~Sannino, \emph{{Ultra Minimal Technicolor and its Dark
  Matter TIMP}},
  \href{http://dx.doi.org/10.1103/PhysRevD.78.115010}{\emph{Phys. Rev.} {\bf
  D78} (2008) 115010}, [\href{http://arxiv.org/abs/0809.0713}{{\tt
  0809.0713}}].

\bibitem{Arbey:2015exa}
A.~Arbey, G.~Cacciapaglia, H.~Cai, A.~Deandrea, S.~Le~Corre and F.~Sannino,
  \emph{{Fundamental Composite Electroweak Dynamics: Status at the LHC}},
  \href{http://dx.doi.org/10.1103/PhysRevD.95.015028}{\emph{Phys. Rev.} {\bf
  D95} (2017) 015028}, [\href{http://arxiv.org/abs/1502.04718}{{\tt
  1502.04718}}].

\bibitem{Frigerio:2012uc}
M.~Frigerio, A.~Pomarol, F.~Riva and A.~Urbano, \emph{{Composite Scalar Dark
  Matter}}, \href{http://dx.doi.org/10.1007/JHEP07(2012)015}{\emph{JHEP} {\bf
  07} (2012) 015}, [\href{http://arxiv.org/abs/1204.2808}{{\tt 1204.2808}}].

\bibitem{Marzocca:2014msa}
D.~Marzocca and A.~Urbano, \emph{{Composite Dark Matter and LHC Interplay}},
  \href{http://dx.doi.org/10.1007/JHEP07(2014)107}{\emph{JHEP} {\bf 07} (2014)
  107}, [\href{http://arxiv.org/abs/1404.7419}{{\tt 1404.7419}}].

\bibitem{Wess:1971yu}
J.~Wess and B.~Zumino, \emph{{Consequences of anomalous Ward identities}},
  \href{http://dx.doi.org/10.1016/0370-2693(71)90582-X}{\emph{Phys. Lett.} {\bf
  B37} (1971) 95--97}.

\bibitem{Witten:1983tx}
E.~Witten, \emph{{Current Algebra, Baryons, and Quark Confinement}},
  \href{http://dx.doi.org/10.1016/0550-3213(83)90064-0}{\emph{Nucl. Phys.} {\bf
  B223} (1983) 433--444}.

\bibitem{Duan:2000dy}
Z.-y. Duan, P.~S. Rodrigues~da Silva and F.~Sannino, \emph{{Enhanced global
  symmetry constraints on epsilon terms}},
  \href{http://dx.doi.org/10.1016/S0550-3213(00)00550-2}{\emph{Nucl. Phys.}
  {\bf B592} (2001) 371--390}, [\href{http://arxiv.org/abs/hep-ph/0001303}{{\tt
  hep-ph/0001303}}].

\bibitem{Antipin:2015xia}
O.~Antipin, M.~Redi, A.~Strumia and E.~Vigiani, \emph{{Accidental Composite
  Dark Matter}}, \href{http://dx.doi.org/10.1007/JHEP07(2015)039}{\emph{JHEP}
  {\bf 07} (2015) 039}, [\href{http://arxiv.org/abs/1503.08749}{{\tt
  1503.08749}}].

\bibitem{Carmona:2015haa}
A.~Carmona and M.~Chala, \emph{{Composite Dark Sectors}},
  \href{http://dx.doi.org/10.1007/JHEP06(2015)105}{\emph{JHEP} {\bf 06} (2015)
  105}, [\href{http://arxiv.org/abs/1504.00332}{{\tt 1504.00332}}].

\bibitem{Blinnikov:1983gh}
S.~I. Blinnikov and M.~Khlopov, \emph{{Possible astronomical effects of mirror
  particles}}, {\emph{Sov. Astron.} {\bf 27} (1983) 371--375}.

\bibitem{Mohapatra:2001sx}
R.~N. Mohapatra, S.~Nussinov and V.~L. Teplitz, \emph{{Mirror matter as
  selfinteracting dark matter}},
  \href{http://dx.doi.org/10.1103/PhysRevD.66.063002}{\emph{Phys. Rev.} {\bf
  D66} (2002) 063002}, [\href{http://arxiv.org/abs/hep-ph/0111381}{{\tt
  hep-ph/0111381}}].

\bibitem{Foot:2004pa}
R.~Foot, \emph{{Mirror matter-type dark matter}},
  \href{http://dx.doi.org/10.1142/S0218271804006449}{\emph{Int. J. Mod. Phys.}
  {\bf D13} (2004) 2161--2192},
  [\href{http://arxiv.org/abs/astro-ph/0407623}{{\tt astro-ph/0407623}}].

\bibitem{Kaplan:2009de}
D.~E. Kaplan, G.~Z. Krnjaic, K.~R. Rehermann and C.~M. Wells, \emph{{Atomic
  Dark Matter}},
  \href{http://dx.doi.org/10.1088/1475-7516/2010/05/021}{\emph{JCAP} {\bf 1005}
  (2010) 021}, [\href{http://arxiv.org/abs/0909.0753}{{\tt 0909.0753}}].

\bibitem{Nussinov:1985xr}
S.~Nussinov, \emph{{Technocosmology: could Technibaryon excess provide a
  ``natural'' missing mass candidate?}},
  \href{http://dx.doi.org/10.1016/0370-2693(85)90689-6}{\emph{Phys. Lett.} {\bf
  B165} (1985) 55--58}.

\bibitem{Barr:1990ca}
S.~M. Barr, R.~S. Chivukula and E.~Farhi, \emph{{Electroweak Fermion Number
  Violation and the Production of Stable Particles in the Early Universe}},
  \href{http://dx.doi.org/10.1016/0370-2693(90)91661-T}{\emph{Phys. Lett.} {\bf
  B241} (1990) 387--391}.

\bibitem{Nussinov:1992he}
S.~Nussinov, \emph{{Some estimates of interaction in matter of neutral
  technibaryons made of colored constituents}},
  \href{http://dx.doi.org/10.1016/0370-2693(92)91849-5}{\emph{Phys. Lett.} {\bf
  B279} (1992) 111--116}.

\bibitem{Bellazzini:2014yua}
B.~Bellazzini, C.~Cs\'aki and J.~Serra, \emph{{Composite Higgses}},
  \href{http://dx.doi.org/10.1140/epjc/s10052-014-2766-x}{\emph{Eur. Phys. J.}
  {\bf C74} (2014) 2766}, [\href{http://arxiv.org/abs/1401.2457}{{\tt
  1401.2457}}].

\bibitem{Mrazek:2011iu}
J.~Mrazek, A.~Pomarol, R.~Rattazzi, M.~Redi, J.~Serra and A.~Wulzer, \emph{{The
  Other Natural Two Higgs Doublet Model}},
  \href{http://dx.doi.org/10.1016/j.nuclphysb.2011.07.008}{\emph{Nucl. Phys.}
  {\bf B853} (2011) 1--48}, [\href{http://arxiv.org/abs/1105.5403}{{\tt
  1105.5403}}].

\bibitem{Gripaios:2009pe}
B.~Gripaios, A.~Pomarol, F.~Riva and J.~Serra, \emph{{Beyond the Minimal
  Composite Higgs Model}},
  \href{http://dx.doi.org/10.1088/1126-6708/2009/04/070}{\emph{JHEP} {\bf 04}
  (2009) 070}, [\href{http://arxiv.org/abs/0902.1483}{{\tt 0902.1483}}].

\bibitem{Marzocca:2012zn}
D.~Marzocca, M.~Serone and J.~Shu, \emph{{General Composite Higgs Models}},
  \href{http://dx.doi.org/10.1007/JHEP08(2012)013}{\emph{JHEP} {\bf 08} (2012)
  013}, [\href{http://arxiv.org/abs/1205.0770}{{\tt 1205.0770}}].

\bibitem{Contino:2015mha}
R.~Contino and M.~Salvarezza, \emph{{One-loop effects from spin-1 resonances in
  Composite Higgs models}},
  \href{http://dx.doi.org/10.1007/JHEP07(2015)065}{\emph{JHEP} {\bf 07} (2015)
  065}, [\href{http://arxiv.org/abs/1504.02750}{{\tt 1504.02750}}].

\bibitem{Ghosh:2015wiz}
D.~Ghosh, M.~Salvarezza and F.~Senia, \emph{{Extending the Analysis of
  Electroweak Precision Constraints in Composite Higgs Models}},
  \href{http://dx.doi.org/10.1016/j.nuclphysb.2016.11.013}{\emph{Nucl. Phys.}
  {\bf B914} (2017) 346--387}, [\href{http://arxiv.org/abs/1511.08235}{{\tt
  1511.08235}}].

\bibitem{Brower:2015owo}
R.~C. Brower, A.~Hasenfratz, C.~Rebbi, E.~Weinberg and O.~Witzel,
  \emph{{Composite Higgs model at a conformal fixed point}},
  \href{http://dx.doi.org/10.1103/PhysRevD.93.075028}{\emph{Phys. Rev.} {\bf
  D93} (2016) 075028}, [\href{http://arxiv.org/abs/1512.02576}{{\tt
  1512.02576}}].

\bibitem{Pica:2016rmv}
C.~Pica and F.~Sannino, \emph{{Anomalous Dimensions of Conformal Baryons}},
  \href{http://dx.doi.org/10.1103/PhysRevD.94.071702}{\emph{Phys. Rev.} {\bf
  D94} (2016) 071702}, [\href{http://arxiv.org/abs/1604.02572}{{\tt
  1604.02572}}].

\bibitem{Appelquist:2009ty}
T.~Appelquist, G.~T. Fleming and E.~T. Neil, \emph{{Lattice Study of Conformal
  Behavior in SU(3) Yang-Mills Theories}},
  \href{http://dx.doi.org/10.1103/PhysRevD.79.076010}{\emph{Phys. Rev.} {\bf
  D79} (2009) 076010}, [\href{http://arxiv.org/abs/0901.3766}{{\tt
  0901.3766}}].

\bibitem{Aoki:2012eq}
Y.~Aoki, T.~Aoyama, M.~Kurachi, T.~Maskawa, K.-i. Nagai, H.~Ohki et~al.,
  \emph{{Lattice study of conformality in twelve-flavor QCD}},
  \href{http://dx.doi.org/10.1103/PhysRevD.86.059903,
  10.1103/PhysRevD.86.054506}{\emph{Phys. Rev.} {\bf D86} (2012) 054506},
  [\href{http://arxiv.org/abs/1207.3060}{{\tt 1207.3060}}].

\bibitem{Cheng:2014jba}
A.~Cheng, A.~Hasenfratz, Y.~Liu, G.~Petropoulos and D.~Schaich,
  \emph{{Improving the continuum limit of gradient flow step scaling}},
  \href{http://dx.doi.org/10.1007/JHEP05(2014)137}{\emph{JHEP} {\bf 05} (2014)
  137}, [\href{http://arxiv.org/abs/1404.0984}{{\tt 1404.0984}}].

\bibitem{Fodor:2011tu}
Z.~Fodor, K.~Holland, J.~Kuti, D.~Nogradi, C.~Schroeder, K.~Holland et~al.,
  \emph{{Twelve massless flavors and three colors below the conformal window}},
  \href{http://dx.doi.org/10.1016/j.physletb.2011.07.037}{\emph{Phys. Lett.}
  {\bf B703} (2011) 348--358}, [\href{http://arxiv.org/abs/1104.3124}{{\tt
  1104.3124}}].

\bibitem{Lombardo:2014pda}
M.~P. Lombardo, K.~Miura, T.~J.~N. da~Silva and E.~Pallante, \emph{{On the
  particle spectrum and the conformal window}},
  \href{http://dx.doi.org/10.1007/JHEP12(2014)183}{\emph{JHEP} {\bf 12} (2014)
  183}, [\href{http://arxiv.org/abs/1410.0298}{{\tt 1410.0298}}].

\bibitem{Cacciapaglia:2015dsa}
G.~Cacciapaglia, H.~Cai, T.~Flacke, S.~J. Lee, A.~Parolini and H.~Ser\^{o}dio,
  \emph{{Anarchic Yukawas and top partial compositeness: the flavour of a
  successful marriage}},
  \href{http://dx.doi.org/10.1007/JHEP06(2015)085}{\emph{JHEP} {\bf 06} (2015)
  085}, [\href{http://arxiv.org/abs/1501.03818}{{\tt 1501.03818}}].

\bibitem{Matsedonskyi:2014iha}
O.~Matsedonskyi, \emph{{On Flavour and Naturalness of Composite Higgs Models}},
  \href{http://dx.doi.org/10.1007/JHEP02(2015)154}{\emph{JHEP} {\bf 02} (2015)
  154}, [\href{http://arxiv.org/abs/1411.4638}{{\tt 1411.4638}}].

\bibitem{Panico:2016ull}
G.~Panico and A.~Pomarol, \emph{{Flavor hierarchies from dynamical scales}},
  \href{http://dx.doi.org/10.1007/JHEP07(2016)097}{\emph{JHEP} {\bf 07} (2016)
  097}, [\href{http://arxiv.org/abs/1603.06609}{{\tt 1603.06609}}].

\bibitem{Sannino:2016sfx}
F.~Sannino, A.~Strumia, A.~Tesi and E.~Vigiani, \emph{{Fundamental partial
  compositeness}}, \href{http://dx.doi.org/10.1007/JHEP11(2016)029}{\emph{JHEP}
  {\bf 11} (2016) 029}, [\href{http://arxiv.org/abs/1607.01659}{{\tt
  1607.01659}}].

\bibitem{Arthur:2016dir}
R.~Arthur, V.~Drach, M.~Hansen, A.~Hietanen, C.~Pica and F.~Sannino,
  \emph{{SU(2) gauge theory with two fundamental flavors: A minimal template
  for model building}},
  \href{http://dx.doi.org/10.1103/PhysRevD.94.094507}{\emph{Phys. Rev.} {\bf
  D94} (2016) 094507}, [\href{http://arxiv.org/abs/1602.06559}{{\tt
  1602.06559}}].

\bibitem{Rantaharju:2015nep}
J.~Rantaharju, V.~Drach, A.~Hietanen, C.~Pica and F.~Sannino, \emph{{Wilson
  Fermions with Four Fermion Interactions}},  in \emph{{Proceedings, 33rd
  International Symposium on Lattice Field Theory (Lattice 2015)}}, 2015.
\newblock \href{http://arxiv.org/abs/1511.03899}{{\tt 1511.03899}}.

\bibitem{Foadi:2016nbi}
R.~Foadi, \emph{{Effect of four-fermion operators on the mass of the composite
  particles}},  \href{http://arxiv.org/abs/1601.02676}{{\tt 1601.02676}}.

\bibitem{Butler:2013kdw}
{\scshape Quark Flavor Physics Working Group} collaboration, J.~N. Butler
  et~al., \emph{{Working Group Report: Quark Flavor Physics}},  in
  \emph{{Proceedings, Community Summer Study 2013: Snowmass on the Mississippi
  (CSS2013): Minneapolis, MN, USA, July 29-August 6, 2013}}, 2013.
\newblock \href{http://arxiv.org/abs/1311.1076}{{\tt 1311.1076}}.

\bibitem{ATLAS-Monojet}
{\scshape ATLAS} collaboration, G.~Aad et~al., \emph{{Search for new phenomena
  in final states with an energetic jet and large missing transverse momentum
  in pp collisions at $\sqrt{s}=$8 TeV with the ATLAS detector}},
  \href{http://dx.doi.org/10.1140/epjc/s10052-015-3517-3,
  10.1140/epjc/s10052-015-3639-7}{\emph{Eur. Phys. J.} {\bf C75} (2015) 299},
  [\href{http://arxiv.org/abs/1502.01518}{{\tt 1502.01518}}].

\bibitem{CMS-Monojet}
{\scshape CMS} collaboration, V.~Khachatryan et~al., \emph{{Search for dark
  matter, extra dimensions, and unparticles in monojet events in proton--proton
  collisions at $\sqrt{s} = 8$ TeV}},
  \href{http://dx.doi.org/10.1140/epjc/s10052-015-3451-4}{\emph{Eur. Phys. J.}
  {\bf C75} (2015) 235}, [\href{http://arxiv.org/abs/1408.3583}{{\tt
  1408.3583}}].

\bibitem{ATLAS-MonoWZ}
{\scshape ATLAS} collaboration, G.~Aad et~al., \emph{{Search for dark matter in
  events with a hadronically decaying W or Z boson and missing transverse
  momentum in $pp$ collisions at $\sqrt{s} =$ 8 TeV with the ATLAS detector}},
  \href{http://dx.doi.org/10.1103/PhysRevLett.112.041802}{\emph{Phys. Rev.
  Lett.} {\bf 112} (2014) 041802}, [\href{http://arxiv.org/abs/1309.4017}{{\tt
  1309.4017}}].

\bibitem{CMS-MonoWZ}
{\scshape CMS} collaboration, V.~Khachatryan et~al., \emph{{Search for dark
  matter in proton-proton collisions at 8 TeV with missing transverse momentum
  and vector boson tagged jets}},
  \href{http://dx.doi.org/10.1007/JHEP12(2016)083}{\emph{JHEP} {\bf 12} (2016)
  083}, [\href{http://arxiv.org/abs/1607.05764}{{\tt 1607.05764}}].

\bibitem{ATLAS-MonoH-gaga-8T}
{\scshape ATLAS} collaboration, G.~Aad et~al., \emph{{Search for Dark Matter in
  Events with Missing Transverse Momentum and a Higgs Boson Decaying to Two
  Photons in $pp$ Collisions at $\sqrt{s}=8$ TeV with the ATLAS Detector}},
  \href{http://dx.doi.org/10.1103/PhysRevLett.115.131801}{\emph{Phys. Rev.
  Lett.} {\bf 115} (2015) 131801}, [\href{http://arxiv.org/abs/1506.01081}{{\tt
  1506.01081}}].

\bibitem{ATLAS-MonoH-bb-8T}
{\scshape ATLAS} collaboration, G.~Aad et~al., \emph{{Search for dark matter
  produced in association with a Higgs boson decaying to two bottom quarks in
  $pp$ collisions at $\sqrt{s} = 8$ TeV with the ATLAS detector}},
  \href{http://dx.doi.org/10.1103/PhysRevD.93.072007}{\emph{Phys. Rev.} {\bf
  D93} (2016) 072007}, [\href{http://arxiv.org/abs/1510.06218}{{\tt
  1510.06218}}].

\bibitem{ATLAS-MonoH-bb-13T}
{\scshape ATLAS} collaboration, M.~Aaboud et~al., \emph{{Search for dark matter
  in association with a Higgs boson decaying to $b$-quarks in $pp$ collisions
  at $\sqrt s=13$ TeV with the ATLAS detector}},
  \href{http://dx.doi.org/10.1016/j.physletb.2016.11.035}{\emph{Phys. Lett.}
  {\bf B765} (2017) 11--31}, [\href{http://arxiv.org/abs/1609.04572}{{\tt
  1609.04572}}].

\bibitem{Wolfram:1978gp}
S.~Wolfram, \emph{{Abundances of Stable Particles Produced in the Early
  Universe}}, \href{http://dx.doi.org/10.1016/0370-2693(79)90426-X}{\emph{Phys.
  Lett.} {\bf B82} (1979) 65--68}.

\bibitem{Griest:1990kh}
K.~Griest and D.~Seckel, \emph{{Three exceptions in the calculation of relic
  abundances}}, \href{http://dx.doi.org/10.1103/PhysRevD.43.3191}{\emph{Phys.
  Rev.} {\bf D43} (1991) 3191--3203}.

\bibitem{Kolb:1990vq}
E.~W. Kolb and M.~S. Turner, \emph{{The Early Universe}}, {\emph{Front. Phys.}
  {\bf 69} (1990) 1--547}.

\bibitem{Gondolo:1990dk}
P.~Gondolo and G.~Gelmini, \emph{{Cosmic abundances of stable particles:
  Improved analysis}},
  \href{http://dx.doi.org/10.1016/0550-3213(91)90438-4}{\emph{Nucl. Phys.} {\bf
  B360} (1991) 145--179}.

\bibitem{Ade:2015xua}
{\scshape Planck} collaboration, P.~A.~R. Ade et~al., \emph{{Planck 2015
  results. XIII. Cosmological parameters}},
  \href{http://dx.doi.org/10.1051/0004-6361/201525830}{\emph{Astron.
  Astrophys.} {\bf 594} (2016) A13},
  [\href{http://arxiv.org/abs/1502.01589}{{\tt 1502.01589}}].

\bibitem{Cheng:2012qr}
H.-Y. Cheng and C.-W. Chiang, \emph{{Revisiting Scalar and Pseudoscalar
  Couplings with Nucleons}},
  \href{http://dx.doi.org/10.1007/JHEP07(2012)009}{\emph{JHEP} {\bf 07} (2012)
  009}, [\href{http://arxiv.org/abs/1202.1292}{{\tt 1202.1292}}].

\bibitem{Alarcon:2011zs}
J.~M. Alarcon, J.~Martin~Camalich and J.~A. Oller, \emph{{The chiral
  representation of the $\pi N$ scattering amplitude and the pion-nucleon sigma
  term}}, \href{http://dx.doi.org/10.1103/PhysRevD.85.051503}{\emph{Phys. Rev.}
  {\bf D85} (2012) 051503}, [\href{http://arxiv.org/abs/1110.3797}{{\tt
  1110.3797}}].

\bibitem{Alarcon:2012nr}
J.~M. Alarcon, L.~S. Geng, J.~Martin~Camalich and J.~A. Oller, \emph{{The
  strangeness content of the nucleon from effective field theory and
  phenomenology}},
  \href{http://dx.doi.org/10.1016/j.physletb.2014.01.065}{\emph{Phys. Lett.}
  {\bf B730} (2014) 342--346}, [\href{http://arxiv.org/abs/1209.2870}{{\tt
  1209.2870}}].

\bibitem{Akerib:2016vxi}
{\scshape LUX} collaboration, D.~S. Akerib et~al., \emph{{Results from a search
  for dark matter in the complete LUX exposure}},
  \href{http://dx.doi.org/10.1103/PhysRevLett.118.021303}{\emph{Phys. Rev.
  Lett.} {\bf 118} (2017) 021303}, [\href{http://arxiv.org/abs/1608.07648}{{\tt
  1608.07648}}].

\bibitem{XENON1T}
{\scshape XENON} collaboration, E.~Aprile et~al., \emph{{Physics reach of the
  XENON1T dark matter experiment}},
  \href{http://dx.doi.org/10.1088/1475-7516/2016/04/027}{\emph{JCAP} {\bf 1604}
  (2016) 027}, [\href{http://arxiv.org/abs/1512.07501}{{\tt 1512.07501}}].

\bibitem{LZ-CDR}
{\scshape LZ} collaboration, D.~S. Akerib et~al., \emph{{LUX-ZEPLIN (LZ)
  Conceptual Design Report}},  \href{http://arxiv.org/abs/1509.02910}{{\tt
  1509.02910}}.

\bibitem{Cirelli:2010xx}
M.~Cirelli, G.~Corcella, A.~Hektor, G.~Hutsi, M.~Kadastik, P.~Panci et~al.,
  \emph{{PPPC 4 DM ID: A Poor Particle Physicist Cookbook for Dark Matter
  Indirect Detection}}, \href{http://dx.doi.org/10.1088/1475-7516/2012/10/E01,
  10.1088/1475-7516/2011/03/051}{\emph{JCAP} {\bf 1103} (2011) 051},
  [\href{http://arxiv.org/abs/1012.4515}{{\tt 1012.4515}}].

\bibitem{Feng:2010zp}
J.~L. Feng, M.~Kaplinghat and H.-B. Yu, \emph{{Sommerfeld Enhancements for
  Thermal Relic Dark Matter}},
  \href{http://dx.doi.org/10.1103/PhysRevD.82.083525}{\emph{Phys. Rev.} {\bf
  D82} (2010) 083525}, [\href{http://arxiv.org/abs/1005.4678}{{\tt
  1005.4678}}].

\bibitem{Abdallah:2016ygi}
{\scshape H.E.S.S.} collaboration, H.~Abdallah et~al., \emph{{Search for dark
  matter annihilations towards the inner Galactic halo from 10 years of
  observations with H.E.S.S}},
  \href{http://dx.doi.org/10.1103/PhysRevLett.117.111301}{\emph{Phys. Rev.
  Lett.} {\bf 117} (2016) 111301}, [\href{http://arxiv.org/abs/1607.08142}{{\tt
  1607.08142}}].

\bibitem{Petraki:2013wwa}
K.~Petraki and R.~R. Volkas, \emph{{Review of asymmetric dark matter}},
  \href{http://dx.doi.org/10.1142/S0217751X13300287}{\emph{Int. J. Mod. Phys.}
  {\bf A28} (2013) 1330028}, [\href{http://arxiv.org/abs/1305.4939}{{\tt
  1305.4939}}].

\bibitem{Zurek:2013wia}
K.~M. Zurek, \emph{{Asymmetric Dark Matter: Theories, Signatures, and
  Constraints}},
  \href{http://dx.doi.org/10.1016/j.physrep.2013.12.001}{\emph{Phys. Rept.}
  {\bf 537} (2014) 91--121}, [\href{http://arxiv.org/abs/1308.0338}{{\tt
  1308.0338}}].

\bibitem{Sakharov:1967dj}
A.~D. Sakharov, \emph{{Violation of CP Invariance, c Asymmetry, and Baryon
  Asymmetry of the Universe}},
  \href{http://dx.doi.org/10.1070/PU1991v034n05ABEH002497}{\emph{Pisma Zh.
  Eksp. Teor. Fiz.} {\bf 5} (1967) 32--35}.

\bibitem{Cacciapaglia:2015yra}
G.~Cacciapaglia and F.~Sannino, \emph{{An Ultraviolet Chiral Theory of the Top
  for the Fundamental Composite (Goldstone) Higgs}},
  \href{http://dx.doi.org/10.1016/j.physletb.2016.02.034}{\emph{Phys. Lett.}
  {\bf B755} (2016) 328--331}, [\href{http://arxiv.org/abs/1508.00016}{{\tt
  1508.00016}}].

\bibitem{Serra:2015xfa}
J.~Serra, \emph{{Beyond the Minimal Top Partner Decay}},
  \href{http://dx.doi.org/10.1007/JHEP09(2015)176}{\emph{JHEP} {\bf 09} (2015)
  176}, [\href{http://arxiv.org/abs/1506.05110}{{\tt 1506.05110}}].

\end{thebibliography}\endgroup

 \end{document}